\documentclass[12pt]{article}
\usepackage{epsfig,amssymb,amsmath,psfrag,multirow,epstopdf,color}
\usepackage{breqn}
\usepackage{placeins}
\usepackage{array}

\numberwithin{equation}{section}

\newcommand{\be}{\begin{equation}}
\newcommand{\ee}{\end{equation}}
\newcommand{\bea}{\begin{eqnarray}}
\newcommand{\eea}{\end{eqnarray}}
\newcommand{\eqn}[1]{eq.~\eqref{#1}}
\def\beq{\begin{equation}}
\def\eeq{\end{equation}}
\def\bsp#1\esp{\begin{split}#1\end{split}}
\def\fig#1{fig.~{\ref{#1}}}

\def\sect#1{section~{\ref{#1}}}
\def\eqn#1{eq.~(\ref{#1})}
\def\Eqn#1{Equation~(\ref{#1})}
\def\eqns#1#2{eqs.~(\ref{#1}) and~(\ref{#2})}

\def\Eqn#1{Equation~(\ref{#1})}
\def\eqn#1{eq.~(\ref{#1})}
\def\eqns#1#2{eqs.~(\ref{#1}) and~(\ref{#2})}

\def\nn{\nonumber}
\def\spa#1.#2{\left\langle#1\,#2\right\rangle}
\def\spb#1.#2{\left[#1\,#2\right]}
\def\Li{{\rm Li}}

\def\ws{{w^\ast}}

\def\to{\rightarrow}
\def\lr{\leftrightarrow}
\def\e{\epsilon}

\newcommand{\de}{\delta}
\newcommand{\lnden}{\ln|\de|}
\newcommand{\lnvn}{\ln|v|}
\newcommand{\lndene}[1]{\ln^{#1}|\de|}
\newcommand{\lnvne}[1]{\ln^{#1}|v|}

\newcommand{\gK}{{\gamma}_K}

\newcommand{\NeqFour}{{\cal N}=4}

\newfont{\scyr}{wncyr10 scaled 550}

\def\beq{\begin{equation}}
\def\eeq{\end{equation}}
\newcommand{\Vt}{\tilde{V}}
\newcommand{\Et}{\tilde{E}}
 
\begin{document}

\catcode`\@=11
\font\manfnt=manfnt
\def\Watchout{\@ifnextchar [{\W@tchout}{\W@tchout[1]}}
\def\W@tchout[#1]{{\manfnt\@tempcnta#1\relax%
  \@whilenum\@tempcnta>\z@\do{%
    \char"7F\hskip 0.3em\advance\@tempcnta\m@ne}}}
\let\foo\W@tchout
\def\dubious{\@ifnextchar[{\@dubious}{\@dubious[1]}}
\let\enddubious\endlist
\def\@dubious[#1]{%
  \setbox\@tempboxa\hbox{\@W@tchout#1}
  \@tempdima\wd\@tempboxa
  \list{}{\leftmargin\@tempdima}\item[\hbox to 0pt{\hss\@W@tchout#1}]}
\def\@W@tchout#1{\W@tchout[#1]}
\catcode`\@=12


\thispagestyle{empty}

\null\vskip-10pt \hfill
\begin{minipage}[t]{42mm}
SLAC--PUB--16467\hskip1cm \ \ \ CALT--2016--003\\
\end{minipage}
\vspace{5mm}

\begingroup\centering
{\Large\bfseries\mathversion{bold}
  All orders results for self-crossing Wilson loops\\ 
  mimicking double parton scattering\par}%
\vspace{7mm}

\begingroup\scshape\large
Lance~J.~Dixon$^{(1),(2)}$ and Ilya Esterlis$^{(1)}$ \\
\endgroup
\vspace{5mm}
\begingroup\small
$^{(1)}$\emph{SLAC National Accelerator Laboratory,
Stanford University, Stanford, CA 94309, USA} \\
$^{(2)}$\emph{Walter Burke Institute for Theoretical Physics,\\
California Institute of Technology, Pasadena, CA 91125, USA}
\endgroup

\vspace{0.4cm}
\begingroup\small
E-mails:\\
{\tt lance@slac.stanford.edu},\ \ \ {\tt ilyae@stanford.edu}
\endgroup
\vspace{0.7cm}

\textbf{Abstract}\vspace{5mm}\par
\begin{minipage}{14.7cm}
Loop-level scattering amplitudes for massless particles have singularities
in regions where tree amplitudes are perfectly smooth.
For example, a $2\to4$ gluon scattering process
has a singularity in which each incoming gluon splits into a pair of gluons,
followed by a pair of $2\to2$ collisions between the gluon pairs.
This singularity mimics double parton scattering because it occurs when the
transverse momentum of a pair of outgoing gluons vanishes.
The singularity is logarithmic at fixed order in perturbation theory.
We exploit the duality between scattering amplitudes and polygonal
Wilson loops to study six-point amplitudes in this limit to high loop order
in planar $\NeqFour$ super-Yang-Mills theory.  The singular configuration
corresponds to the limit in which a hexagonal Wilson loop
develops a self-crossing.
The singular terms are governed by an evolution equation, in which
the hexagon mixes into a pair of boxes;
the mixing back is suppressed in the planar (large $N_c$) limit.
Because the kinematic dependence
of the box Wilson loops is dictated by (dual) conformal invariance,
the complete kinematic dependence of the singular terms for the
self-crossing hexagon on the one nonsingular variable is determined
to all loop orders.  The complete logarithmic dependence on the
singular variable can be obtained through nine loops,
up to a couple of constants, using a correspondence with the multi-Regge limit.
As a byproduct, we obtain a simple formula for the leading
logs to all loop orders.
We also show that, although the MHV six-gluon amplitude is singular,
remarkably, the transcendental functions entering the
non-MHV amplitude are finite in the same limit, at least through four loops.
\end{minipage}\par
\endgroup

\newpage

\tableofcontents

\newpage

\section{Introduction}

At high-energy hadron colliders such as the Tevatron and the Large Hadron
Collider, double parton scattering can take place, in which two partons
from each incoming hadron collide with each other.  The kinematic
signature of such an event is that the final state can be split into
two subsets of constituents, which are the products of the two separate
partonic collisions.  The transverse momentum of each subset should add
up to zero, whereas for a single parton scattering with the same final
state this is not generically true.  Single-parton tree amplitudes
are smooth as one approaches the kinematics of double-parton scattering.
However, at loop level the single-parton scattering amplitude can have
a logarithmic singularity in the subset transverse
momentum, which can be identified with a Landau or
pinch singularity in the Feynman parameter
integration~\cite{Landau,BjDrell,ColemanNorton,NagySoper,EZ,Ninh,%
GauntStirling,Diehl2011yj,Dennen2015bet}.
This singularity arises because, at the loop level,
each incoming parton can split into two collinear partons,
each of which then participates in a scattering.

\begin{figure}
\begin{center}
\includegraphics[width=5.0in]{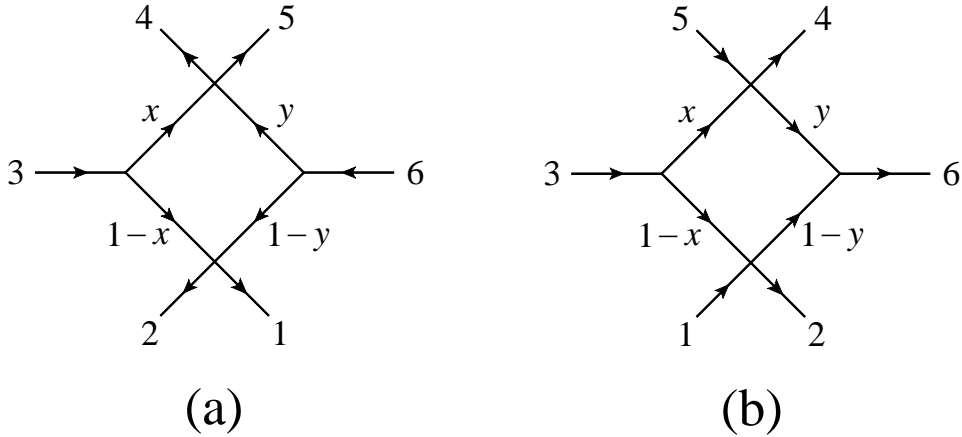}
\end{center}
\caption{(a) A $2\to4$ scattering configuration mimicking double-parton
scattering.  Incoming gluons 3 and 6 split into collinear pairs with
very small transverse momentum, and longitudinal momentum fractions
$x$ and $1-x$, and $y$ and $1-y$, respectively.
These pairs then undergo $2\to2$ scatterings into final state gluons
1, 2, 4 and 5.
(b) The analogous configuration for $3\to3$ scattering. Gluons 1, 3 and 5
are incoming.  Gluon 3 splits collinearly,
and its daughters collide with gluons 1 and 5.
These $2\to2$ collisions produce gluons 2 and 4, and two more gluons
which fuse collinearly into gluon 6.}
\label{fig:show_24_33}
\end{figure}

Figure~\ref{fig:show_24_33}(a) shows such a configuration for
massless $2\to4$ scattering, where particles 3 and 6 are incoming,
and particles 1, 2, 4 and 5 are outgoing.  This configuration is
generically singular as the vector sum of the transverse momenta of particles
1 and 2 vanishes, because of the existence of the $2\to2$ subprocesses
$(1-x)k_3 + (1-y)k_6 \to k_1+k_2$ and $xk_3 + yk_6 \to k_4+k_5$.
The lines marked $x$, $(1-x)$ ($y$, $(1-y)$) in the figure can go on-shell
in this limit; they are approximately collinear
to the initial-state particle 3 (particle 6), and they contain
the indicated fraction of its longitudinal momentum.
Similar configurations arise for $2\to(n-2)$ processes for any $n\geq6$,
although in this paper we will mostly study the six-point amplitude.

Although of less direct phenomenological relevance, there are similar
singularities for scattering processes with more than two initial
particles.  For the case of the six-point amplitude,
\fig{fig:show_24_33}(b) shows such a singularity in $3\to3$
scattering.  An initial parton splits into two collinear partons,
each of which collides with another incoming parton; two of the products
of those collisions then fuse into a single parton.
Interestingly, at least for the theory we will be considering
in this paper, the $3\to3$ process has a somewhat simpler structure in the
singular limit, and the $2\to4$ case can then be obtained from it by
an analytic continuation.

In the experimental effort to isolate hard multiple parton
interactions, the background from single-parton contributions
in the same region of phase space must be subtracted, usually
by extrapolation from data with finite values of the subset
transverse momentum.  In the case of jet final states,
the jet energy resolution typically smears the double-parton
events over a range of finite subset transverse momenta,
and the expected ``bump'' at zero may become a broad shoulder.
Hence it is of some interest to understand any non-smooth
behavior of the single-parton background.
Because the matrix element singularity is merely logarithmic,
the phase-space integration over the subset transverse momentum
is convergent.  However, the detailed shape as one approaches
zero may still be important.
It is not our purpose in this paper to perform any phenomenological
discussion of the single-parton contributions to double parton scattering
or how to separate them theoretically
(see e.g.~refs.~\cite{Cacciari2009dp,Blok2010ge,Ryskin2011kk,%
Blok2011bu,Diehl2011yj,Ryskin2012qx,Blok2013bpa,Diehl2013rca}).
Rather, we would like to use a toy model to study the behavior
of scattering amplitudes in this region to very high perturbative
order.

The toy model we have in mind is $\NeqFour$ super-Yang-Mills theory
(SYM)~\cite{Brink1976bc} in the limit of a large number of colors $N_c$.
In this limit, the theory is integrable~\cite{IntegrableReview}
and possesses a dual superconformal
symmetry~\cite{DualConformal,FourLoopNeq4,FiveLoopNeq4,AMStrong,Drummond2008vq}.
Perhaps most interestingly for the present problem,
scattering amplitudes in planar $\NeqFour$ SYM
are dual to polygonal Wilson loops with
light-like edges~\cite{AMStrong,WilsonLoopWeak,Alday2008yw,Adamo2011pv}.
Each edge of the Wilson loop corresponds to an external momentum of
the scattering amplitude, and the closure of the polygon is the
geometrical statement of overall momentum conservation.

\begin{figure}[ht]
\begin{center}
\includegraphics[width=1.5 in]{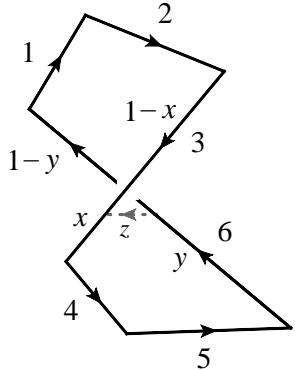}
\end{center}
\caption{Self-crossing configuration of a hexagonal Wilson loop,
regulated by a small space-like vector $\vec{z}$.}
\label{fig:sc_reg}
\end{figure}

The Wilson loop configuration that mimics
double-parton scattering is one in which the loop crosses itself,
as shown in \fig{fig:sc_reg}.  Comparing this figure
with~\fig{fig:show_24_33}, we see that the splitting of particle 3 into
two collinear intermediate particles with momentum fractions $x$ and
$1-x$ is reflected in the division of line 3 in \fig{fig:sc_reg}
into two segments labelled by $x$ and $1-x$, and similarly for
particle 6 with momentum fractions $y$ and $1-y$.
The reason lines 3 and 6 touch (in the singular limit) is
to ensure momentum conservation of the subprocesses
$(1-x)k_3 + (1-y)k_6 \to k_1+k_2$ and $xk_3 + yk_6 \to k_4+k_5$
(or their analytic continuation in the case of $3\to3$ scattering).
For convenience, we will refer to this kinematics as ``self-crossing'',
even when we discuss theories other than planar $\NeqFour$ SYM
for which there is no polygonal Wilson loop correspondence, and where
the term ``double-parton-scattering-like'' might be more appropriate.

The behavior of the self-crossing hexagonal Wilson loop in planar
$\NeqFour$ SYM has been studied by Georgiou~\cite{Georgiou} at two loops,
and by Dorn and Wuttke~\cite{DornWuttke1,DornWuttke2} at two and three loops.
This ``bosonic'' Wilson loop corresponds to the six-gluon
maximally helicity violating (MHV)
scattering amplitude, in which the gluons have the (all-outgoing)
helicity configuration~$({-}{-}{+}{+}{+}{+})$, or any permutation thereof.
The self-crossing configuration depends on one singular parameter,
$\delta \ll 1$.  This parameter is invariant under the dual conformal
symmetry possessed by Wilson loops in planar $\NeqFour$ SYM.
It serves as a proxy for the vanishing subset transverse momentum.
In \fig{fig:sc_reg}, we regulate the self-crossing singularity
by a small space-like separation vector
$\vec{z}$~\cite{Korchemsky1993hr,Korchemskaya1994qp}.
We will see that the magnitude of the separation, $\vec{z}^2$,
is proportional to $\delta$.

The self-crossing configuration
depends as well on one generic, nonsingular parameter we call $v$.
In this paper we will determine how the singular ($\ln\de$ containing)
terms in this Wilson loop (amplitude) depend on $v$ to all loop orders.
We will present the full logarithmic dependence on $\delta$
through seven loops, and at eight and nine loops up to a couple
of constants.  We'll also give the full $v$ dependence of the
nonsingular terms through five loops, neglecting terms
suppressed by powers of $\de$.

We will show that the transcendental functions entering the other six-point
helicity configuration in $\NeqFour$ SYM, called non-MHV (NMHV),
are actually nonsingular through four loops!
This result is related to an argument of Gaunt and Stirling
for one-loop QED amplitudes~\cite{GauntStirling}.

In order to provide explicit MHV results to
such a high loop order, we will make use of
the factorized singularity structure of Wilson loops that are close to
crossing~\cite{BrandtNeriSato}, in particular the analysis
and evolution equation studied by Korchemskaya and
Korchemsky~\cite{Korchemsky1993hr,Korchemskaya1994qp}.
As one approaches the singularity, the hexagonal Wilson loop mixes with
another configuration, which features two disconnected squares corresponding
to the two $2\to2$ subprocesses.  At large $N_c$, the mixing
of the two-square configuration back into the hexagon is suppressed,
and the expectation value of the two-square Wilson loop is
dictated by dual conformal invariance.  This leads to
an exact prediction for how the singular
terms in the hexagonal Wilson loop depend
on the unique nonsingular kinematic variable, $v$.

Knowing the full dependence on $v$ for the singular terms,
we can evaluate them by choosing $v$ to be anything we like.
We make use of the fact that as $v\to0$, the self-crossing limit
overlaps with the limit of multi-Regge kinematics (MRK), which
has been studied extensively in planar $\NeqFour$ SYM~\cite{Bartels2008ce,%
Bartels2008sc, Lipatov2010qg, Lipatov2010ad, Bartels2010tx, BLPCollRegge,
Dixon2011pw, Fadin2011we, Lipatov2012gk,
Dixon2012yy, Pennington2012zj, CaronHuot2013fea,
Dixon2014voa,Hatsuda2014oza,BCHS,JamesYorgos,BroedelSprenger,
Bargheer2015djt}.  In particular, an all-orders formula for the
behavior of the Wilson loop in this limit was proposed
by Basso, Caron-Huot and Sever (BCS)~\cite{BCHS}, based on integrability
and an analytic continuation from the Euclidean operator product expansion
region studied by Basso, Sever and
Vieira~\cite{Basso2013vsa,Basso2013aha,Basso2014koa,BSVIV}.
We will analyze this formula to high loop orders in the region of overlap with
the self-crossing configuration, and make use of recent high order
results by Drummond and Papathanasiou~\cite{JamesYorgos} and especially
by Broedel and Sprenger~\cite{BroedelSprenger}.
The exact $v$ dependence provides cross checks on the MRK predictions
of BCS.

Finally, we have used a very recent determination of the
full MHV amplitude at five loops~\cite{CHDvHMR65},
as well as lower-loop
results~\cite{Goncharov2010jf,Dixon2011pw,Dixon2013eka,Dixon2014voa},
to obtain the
full $v$ dependence of the nonsingular terms through this order.
We also present the nonsingular limits of the transcendental
functions entering the NMHV amplitude through
four loops, using results from
refs.~\cite{Dixon2011nj,Dixon2014iba,Dixon2015iva}.

Similar methods for controlling the singular terms
should be applicable as well to higher-point amplitudes,
but we leave that for future work.

This paper is organized as follows:
Section~\ref{prelim} reviews properties of amplitudes
and Wilson loops in planar $\NeqFour$ SYM. It describes
the self-crossing limit and explains
why the transcendental functions in
NMHV six-gluon amplitudes are expected to be nonsingular there,
while MHV amplitudes diverge.
Finally, it discusses how to frame Wilson loops to remove
their cusp divergences, and how different framings
behave in the self crossing limit.
Section~\ref{explicitresults} provides explicit results
for the MHV amplitude through five loops, after discussing
how to analytically continue into these regions.
It also looks at a few special limits ($v\to\infty$ and $v\to1$)
where the results simplify.

Section~\ref{evolveWL} discusses how the singular terms
as $\de\to0$ obey an evolution equation, one that is particularly
simple due to the large $N_c$ limit.  This equation explains
several properties of the explicit results, and allows one
to go to higher loop order using the $v\to0$ limit.
Section~\ref{SCMRKmatch} shows how the $v\to0$ limit of self-crossing
overlaps with the $w\to-1$ limit of multi-Regge-kinematics.
It also develops techniques for evaluating the Fourier-Mellin
transform in this limit, and discusses a comparison with information
from ref.~\cite{BroedelSprenger}.
Section~\ref{FinalResults} presents the final expression for
the self-crossing limit, and organizes the dependence on $\de$
into a suggestive form,
so that no pure even $\zeta$ values appear explicitly,
and particular terms with single odd $\zeta$ values
are confined to a specific dependence on $\lnden$.
Finally, in section~\ref{concout} we conclude.

We also provide multiple appendices.
In appendix~\ref{sckin} we give a detailed description of
both $2\to4$ and $3\to3$ self-crossing kinematics in terms
of the kinematics of the $2\to2$ subprocesses.
Appendix~\ref{cuspexp} gives the expansion of the light-like
cusp anomalous dimension through 10 loops.
Appendix~\ref{Evgt145} gives the full four- and five-loop
results for the MHV amplitude in the self-crossing limit,
while appendix~\ref{ENMHVvgt11234} presents results for
the nonsingular NMHV transcendental functions through four loops.
Appendix~\ref{sckin7} gives a brief description of the
self-crossing limit for the seven-point case, which will
be explored more thoroughly in future work.

Accompanying this article is an ancillary file containing
computer-readable expressions for the lengthier formulae
in this paper.

\newpage


\section{Preliminaries}
\label{prelim}

As mentioned in the introduction, scattering amplitudes
in planar $\NeqFour$ SYM are dual to polygonal Wilson loops.
For the MHV $n$-gluon amplitude, where two gluon helicities
are negative and the rest positive, the correspondence is via
\be
A_n^{\rm MHV}\ =\ A_n^{\rm MHV,\, tree} \, W_n
\ =\ A_n^{\rm BDS}(s_{i,j};\e) \, \exp[ R_n(u_{ijkl}) ] \,,
\label{AmpWLBDS}
\ee
where $A_n^{\rm MHV}$ is the partial amplitude associated
with the color factor ${\rm Tr}(T^{a_1}T^{a_2}\ldots T^{a_n})$,
$A_n^{\rm MHV,\, tree}$ is the corresponding tree-level amplitude,
$W_n$ is the Wilson $n$-gon expectation value,
$A_n^{\rm BDS}$ is the BDS ansatz~\cite{BDS},
and $R_n$ is the remainder function which corrects this ansatz
for $n\geq6$.  The tree amplitude
$A_n^{\rm MHV,\, tree}$ is given by the Parke-Taylor
formula~\cite{ParkeTaylor,ManganoParkeXu},
\be
A_n^{\rm MHV,\, tree}
\ =\ i \, \frac{{\spa{j}.{k}}^4}{\spa1.2\spa2.3\cdots\spa{n}.1} \,,
\label{PT}
\ee
where $j$ and $k$ label the two negative-helicity gluons,
in the all-outgoing helicity convention.
It is important to note that all of the dependence on $j$ and $k$
is carried by the simple prefactor ${\spa{j}.{k}}^4$.
The more complicated quantity, the bosonic Wilson loop,
carries no helicity information at all.

The BDS ansatz depends on the Mandelstam variables $s_{ij}$
and is given by
\be
A_n^{\rm BDS}\ =\ A_n^{\rm MHV,\, tree} \,
\exp\biggl[ 
\sum_{L=1}^\infty a^L \Bigl( f^{(L)}(\e) \frac{1}{2} M_n^{\rm 1-loop}(L\e) + C^{(L)} \Bigr)
\biggr] \,,
\label{BDSAnsatz}
\ee
where $M_n^{\rm 1-loop}(L\e)$ is the one-loop amplitude, normalized by the
tree amplitude, and evaluated in dimensional regularization with $D=4-2\e$,
but letting $\e\to L\e$.  The remaining quantities in \eqn{BDSAnsatz}
are constants:
\be
f^{(L)}(\e) \equiv f_0^{(L)} + \e \, f_1^{(L)} + \e^2 \, f_2^{(L)} \,,
\label{fepsdef}
\ee
where two of the constants,
\be
f_0^{(L)}\ =\ \frac{1}{4} \, \gK^{(L)} \,, \qquad
f_1^{(L)}\ =\ \frac{L}{2} \, {\cal G}_0^{(L)} \,,
\label{cuspG}
\ee
are given in terms of the (light-like) cusp anomalous dimension $\gK$
and the ``collinear'' anomalous dimension ${\cal G}_0$,
and $f_2^{(L)}$ and $C^{(L)}$ are other constants, known
analytically to three loops.
The BDS ansatz captures all the infrared singularities of the
scattering amplitude, or equivalently the ultraviolet cusp
singularities of the Wilson loop.  However, there is a simpler,
``BDS-like'' ansatz that also does this, which we will
introduce in \sect{BDSlike}.

We should note that \eqn{AmpWLBDS} is formal due to infrared divergences.
In fact the $1/\e$ pole in the logarithm of the Wilson loop is
controlled by ${\cal G}_{\rm eik}$, a quantity that differs from ${\cal G}_0$ by
a term proportional to the virtual part of the DGLAP
kernel~\cite{DHKS,Dixon2008gr}.  This difference will not matter below,
once we pass to finite quantities on both sides of the duality.

\subsection{Helicity selection rules}
\label{helsel}

Because the MHV tree amplitude~(\ref{PT}) carries all the helicity
dependence of the full MHV amplitude, we can immediately make
some all-order statements about how the self-crossing result in $\NeqFour$
SYM depends on the helicity configuration.  The spinor products $\spa{i}.{j}$
entering \eqn{PT} are all nonsingular and finite in the
generic self-crossing configuration.  Hence there are at most simple finite
factors between the different helicity configurations, and all will
have the same singularities in the self-crossing limit.
We can contrast this behavior with that of the one-loop QED six-photon amplitude
with an electron in the loop, as analyzed in ref.~\cite{GauntStirling}.
In that case, non-singular self-crossing limits of certain helicity
configurations appear for two reasons:
\begin{enumerate}
\item If two of the outgoing photons from a $2\to2$ subprocess
  have the same helicity, then the tree amplitude for $e^+e^-\to\gamma^+\gamma^+$
  vanishes for massless electrons, causing the one-loop amplitude
  to be nonsingular.
\item If the two incoming photons have opposite helicities, then the $J_z$
  of the initial state is nonzero; however, by helicity conservation for
  massless electrons, the four-electron intermediate state has $J_z=0$.
\end{enumerate}
As a result of these selection rules, the only possible singular
MHV configurations for six photons at one loop are ${+}{+}\to({-}{+})({-}{+})$,
where we use all-outgoing helicity labelings, and
separate the pairs of final-state photons for the two $2\to2$
subprocesses using parentheses.  In addition, there are no singular
NMHV configurations.

The difference between one-loop QCD or planar $\NeqFour$ SYM amplitudes
with gluons circulating in the loop, and one-loop QED amplitudes,
is that gluons can have a helicity
flip while circulating around the loop, and therefore the outgoing gluons
from a $2\to2$ subprocess can have the same helicity, in contrast to
selection rule 1.  Massless quarks, gluinos and scalars in the loop
obey rule 1, just like massless electrons.
However, rule 2, or more generally {\it $J_z$ conservation between
the initial state and the four-parton intermediate state},
is an important constraint that still needs to be applied.

\begin{figure}[ht]
\begin{center}
\includegraphics[width=6.0 in]{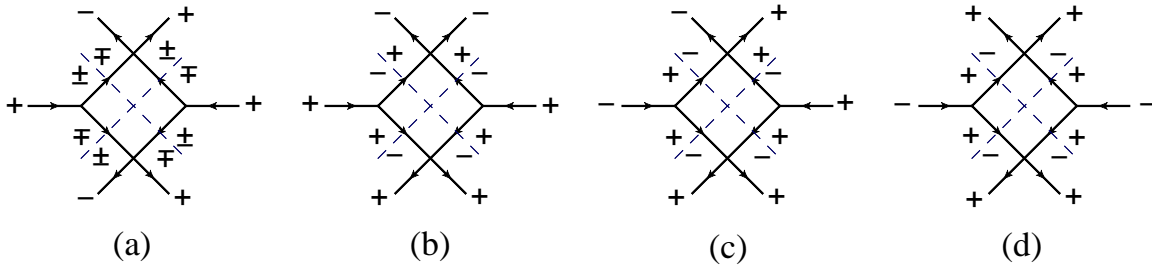}
\end{center}
\caption{Allowed helicity configurations for a singular self-crossing limit
  for the MHV six-gluon amplitude.  The cut internal lines are
  collinear with the incoming particles, as in \fig{fig:show_24_33}. 
  Only (a) is allowed for massless matter circulating in the loop; (b), (c)
  and (d) require gluons to circulate.}
\label{fig:MHVhel}
\end{figure}

Figure~\ref{fig:MHVhel} shows four different
configurations that satisfy both rules 1 and 2 for gluons
circulating in the loop.  Only the first one,
(a), for ${+}{+}\to({-}{+})({-}{+})$, appears in QED.
Configurations (b), (c) and (d) are forbidden by rule 1 in QED,
but permitted in QCD or planar $\NeqFour$ SYM amplitudes.
Correspondingly, the MHV configurations ${+}{+}\to({-}{-})({+}{+})$,\ 
${-}{+}\to({-}{+})({+}{+})$ and ${-}{-}\to({+}{+})({+}{+})$,
although non-singular in QED, are singular in QCD or
$\NeqFour$ SYM, because they obey the $J_z$ conservation rule.
Because only gluons contribute to the singularity, the singular behavior
of the one-loop QCD amplitude for these three helicity configurations
is identical to that for $\NeqFour$ SYM studied in this paper,
and therefore all three cases are simply related by the helicity
dependence of the MHV tree amplitude.

On the other hand, for ${+}{+}\to({-}{+})({-}{+})$, the QCD
and $\NeqFour$ SYM results are different, because massless
quarks, gluinos and scalars can contribute.  While the $\NeqFour$
supersymmetric sum will reconstruct a result for $\NeqFour$ SYM
that is simply related to the other MHV configurations, the 
QCD result will have a different form.

\begin{figure}[ht]
\begin{center}
\includegraphics[width=6.0 in]{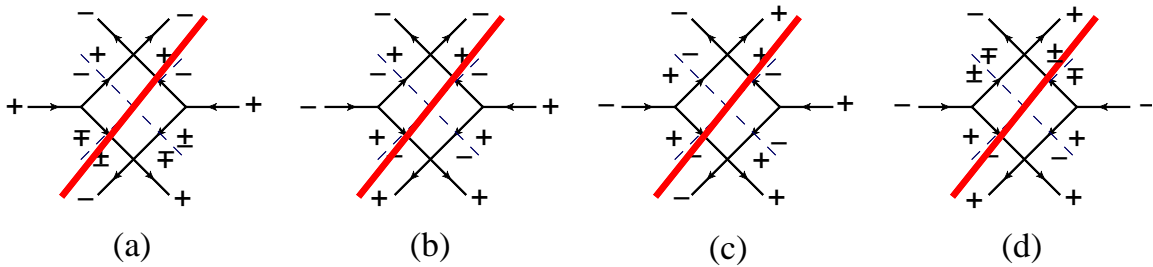}
\end{center}
\caption{NMHV six-gluon helicity configurations.  All are forbidden
  to have self-crossing singularities by the $J_z$ conservation rule
  (hence the red lines through them).
  One (QED-like) configuration for ${-}{+}\to({-}{+})({-}{+})$
  is not shown; it also violates $J_z$ conservation.}
\label{fig:NMHVhel}
\end{figure}

What about NMHV helicity configurations? Ref.~\cite{GauntStirling}
shows that there are no NMHV helicity configurations in QED (or
for any massless matter contribution) that obey rules 1 and 2.
Remarkably, even with gluons in the loop, so that
rule 1 can be relaxed, rule 2 ($J_z$ conservation)
still forbids any singular configurations.
Figure~\ref{fig:NMHVhel} shows four different helicity configurations
that would appear to factorize properly into nonvanishing $1\to2$ splittings
and $2\to2$ subprocesses.  However, all four of them are
non-singular because they violate $J_z$ conservation between
the initial and the intermediate state.
For example, in case (a), for ${+}{+}\to({-}{-})({-}{+})$,
the initial $J_z$ is zero, but the final $J_z$ is $\pm2$.
In case (b), for ${-}{+}\to({-}{-})({+}{+})$, 
the initial $J_z$ is $\pm2$, while the final $J_z$ is zero.
(There is a fifth configuration, not shown, for
${-}{+}\to({-}{+})({-}{+})$, which is of the same type allowed by rule 1;
but it violates $J_z$ conservation just as it does in QED.)

Thus, as a consequence of $J_z$ conservation,
{\it the NMHV six-gluon amplitude is nonsingular in all self-crossing limits, 
to one-loop in QCD, and in ${\cal N}=1$ or $\NeqFour$ SYM.}
Naively, the result should hold to {\it all}
orders in supersymmetric gauge theories
because supersymmetry forbids the ``non-tree-like''
$2\to2$ helicity amplitudes $({+}{+}{+}{+})$ and $({-}{+}{+}{+})$.
That four-point selection rule dictates the configurations shown in 
\fig{fig:NMHVhel}, plus the fifth, QED-like one, all of which
violate $J_z$ conservation.  At two loops in QCD, the NMHV six-gluon
amplitude will presumably become singular; it is easy to write down
helicity configurations that satisfy $J_z$ conservation
once one of the $2\to2$ amplitudes for $({-}{+}{+}{+})$ or its
parity conjugate $({+}{-}{-}{-})$ is nonzero.

The NMHV six-gluon amplitude has been computed in planar $\NeqFour$
SYM through four loops~\cite{Dixon2011nj,Dixon2014iba,Dixon2015iva}.
In appendix~\ref{ENMHVvgt11234} we provide the self-crossing limits
of the transcendental functions entering these results. 
We find that these functions are indeed completely nonsingular
in the self-crossing limit through four loops.  However,
the full NMHV amplitude contains rational function prefactors
which blow up like $1/\sqrt{\de}$ in the self-crossing limit.
Therefore the full amplitude can be singular even if the transcendental
functions are finite.  To analyze the behavior of the full amplitude
requires expanding the transcendental functions to higher order
around the self-crossing limit.  In appendix~\ref{ENMHVvgt11234}
we carry out this expansion, not in the full self-crossing limit,
but in the part that overlaps with multi-Regge-kinematics.
We find logarithmic singularities in $\de$ starting at two loops.
Thus there must be a loophole in the naive argument
for all-loop order NMHV finiteness in the self-crossing limit,
perhaps from contributions where more than one particle crosses a cut,
which can happen starting at two loops.  We leave further investigation
of this issue to future work.


\subsection{The BDS-like normalized amplitude}
\label{BDSlike}

The BDS ansatz also accounts for an anomaly in dual conformal invariance
due to the infrared (ultraviolet) divergences of the scattering amplitude
(Wilson loop).  The remainder function $R_n$ is then
invariant under dual conformal transformations.
Hence $R_n$ can only be a function of the dual conformally invariant
cross ratios,
\be
u_{ijkl}\ =\ \frac{x_{ij}^2 x_{kl}^2}{x_{ik}^2 x_{jl}^2} \,,
\label{uijkl}
\ee
where $x_{ij}^2 = (x_i - x_j)^2$, the dual coordinates $x_i^\mu$
describe the locations of the vertices of the $n$-gon, and the
scattering amplitude momenta $k_i^\mu$ are related to them by
$k_i^\mu = x_i^\mu - x_{i+1}^\mu$.  Non-trivial cross ratios require
non-adjacent vertices, since $x_{i,i+1}^2 = k_i^2 = 0$.
There are no such invariants for $n=4$ or 5, and the remainder
function first becomes nonvanishing for
$n=6$~\cite{Bartels2008ce,Bern2008ap}.

For $n=6$, the main subject of this paper, there are three independent
cross ratios,
\bea
u &=& u_1 = \frac{s_{12}\,s_{45}}{s_{123}\,s_{345}}
= \frac{x_{13}^2\,x_{46}^2}{x_{14}^2\,x_{36}^2}\,, \nn\\
\qquad v &=& u_2 = \frac{s_{23}\,s_{56}}{s_{234}\,s_{123}}
= \frac{x_{24}^2\,x_{51}^2}{x_{25}^2\,x_{41}^2}\,, \nn\\
w &=& u_3 = \frac{s_{34}\,s_{61}}{s_{345}\,s_{234}}
= \frac{x_{35}^2\,x_{62}^2}{x_{36}^2\,x_{52}^2}\,,
\label{uvw_def}
\eea
where $s_{i,i+1} = (k_i+k_{i+1})^2$, $s_{i,i+1,i+2} = (k_i+k_{i+1}+k_{i+2})^2$,
and $x_{ij}^2 \equiv (x_i - x_j)^2$.

The remainder function, $R_6(u,v,w)$, has a Euclidean branch
for which $u$, $v$ and $w$ are all positive and the function is real.
The $2\to4$ and $3\to3$ scattering configurations are physical,
Minkowski configurations, which can be obtained from the Euclidean
region by a suitable analytic
continuation~\cite{Bartels2008ce,Bartels2010tx,Georgiou,DornWuttke2}.
For $2\to4$ scattering, we are interested in the configuration with
particles 3 and 6 incoming
--- see \fig{fig:show_24_33}(a) and eqs.~(\ref{incout24})--(\ref{sub24B}).
This is achieved by letting $u\to u e^{-2\pi i}$ and leaving $v$
and $w$ positive.
For $3\to3$ scattering, we wish to take particles 1, 3 and 5 to be incoming
--- see \fig{fig:show_24_33}(b) and eqs.~(\ref{incout33})--(\ref{sub33B}).
To do this (for the case $v,w<0$), we let $u\to u e^{+2\pi i}$ $v\to v e^{\pi i}$,
$w\to w e^{\pi i}$.
(See also sections \ref{Euclto24} and \ref{from24to33vneg} below.)

Instead of considering the remainder function, which is
the (log of) the amplitude normalized by the BDS ansatz,
for our present problem it is better to normalize the 
amplitude by a ``BDS-like'' ansatz~\cite{AGM}.
The reason is that the BDS ansatz contains the one-loop
$\NeqFour$ SYM amplitude, which is also singular
in the self-crossing limit.  The BDS-like ansatz has a simpler
functional form, depending only on two-particle invariants,
and it is nonsingular in the limit.

More specifically, the BDS-like ansatz for six-gluon scattering is
\be
A_6^{\rm BDS-like}\ =\ A_6^{\rm MHV,\, tree} \,
\exp\biggl[ 
\sum_{L=1}^\infty a^L \Bigl( f^{(L)}(\e) \frac{1}{2} \hat{M}_6(L\e) + C^{(L)} \Bigr)
\biggr] \,,
\label{BDSlikeAnsatz}
\ee
where
\bea
\hat{M}_6(\e) &=& M_6^{\rm 1-loop} + Y(u,v,w) \nonumber\\
&=&  \sum_{i=1}^6 \biggl[
  - \frac{1}{\e^2} \Bigl( 1 - \e \ln(-s_{i,i+1}) \Bigr)
  - \ln(-s_{i,i+1})\ln(-s_{i+1,i+2})
  + \frac{1}{2} \ln(-s_{i,i+1})\ln(-s_{i+3,i+4}) \biggr]\nonumber\\
&&\hskip0cm\null + 6 \, \zeta_2 \,,
\label{M6hat}
\eea
with
\be
Y(u,v,w)\ =\ {\rm Li}_2(1-u) + {\rm Li}_2(1-v) + {\rm Li}_2(1-w)
+ \frac{1}{2} \Bigl( \ln^2 u + \ln^2 v + \ln^2 w \Bigr). 
\label{Ydef}
\ee
Our perturbative expansion parameter for planar SYM is
$a = g_{\textrm{YM}}^2 N_c/(8\pi^2)$, where $N_c$ is the (large)
number of colors and $g_{\textrm{YM}}$ is the Yang-Mills coupling.
Note that $\hat{M}_6(\e)$ contains only two-particle invariants
$s_{i,i+1}$, which are just the squares of the spinor products
$\spa{i,}.{i+1}$ entering the MHV tree amplitude~(\ref{PT}).
Therefore the BDS-like ansatz is completely smooth as one approaches
self-crossing kinematics, yet it still removes infrared
divergences in a way that respects dual conformal invariance,
since it only differs from the BDS ansatz by the dual conformally
invariant function $Y(u,v,w)$.

Comparing \eqns{BDSAnsatz}{BDSlikeAnsatz}, we see that
\be
A_6^{\rm BDS-like}
\ =\ A_6^{\rm BDS} \, \exp\biggl[ \frac{\gK}{8} Y(u,v,w) \biggr] \,.
\label{BDSlikevsBDS}
\ee
We define the function ${\cal E}(u,v,w)$ by
\be
A_6^{\rm MHV}\ =\ A_6^{\rm BDS-like}(s_{i,i+1},\e) \times {\cal E}(u,v,w) \,.
\label{EMHVdef}
\ee
\Eqn{BDSlikevsBDS} shows that it is related to the remainder function by 
\be
{\cal E}(u,v,w)\ =\ \exp\Bigl[
     R_6(u,v,w) - \frac{\gamma_K}{8} Y(u,v,w) \Bigr] \,.
\label{calERrelation}
\ee
Using integrability, the cusp anomalous dimension $\gK(a)$
can be computed to arbitrary loop orders~\cite{BES}. Its expansion
through 10 loops in terms of $a$ is given in appendix~\ref{cuspexp}.

Although our main focus will be on the singularities of the MHV
six-gluon amplitude, we will also present the nonsingular limits
(see \sect{helsel}) of the transcendental functions entering the NMHV
amplitude.
As described in more detail in ref.~\cite{Dixon2015iva},
the NMHV super-amplitude can be written in a BDS-like form as
\be
\bsp
&\frac{\mathcal{A}_6^{\rm NMHV}}{\mathcal{A}_6^{\rm BDS-like}}
\ =\ \frac{1}{2}\Bigl[
 [(1) + (4)] E(u,v,w) + [(2) + (5)] E(v,w,u) + [(3) + (6)] E(w,u,v)  \\
&\hskip1.8cm
+ [(1) - (4)] \Et(y_u,y_v,y_w) - [(2)-(5)] \Et(y_v,y_w,y_u)
  + [(3) - (6)] \Et(y_w,y_u,y_v) \Bigr] \,,
\label{Eform}
\esp
\ee
where $\mathcal{A}_6^{\rm NMHV}$ and $\mathcal{A}_6^{\rm BDS-like}$ are
super-amplitudes and $(1),(2),\ldots,(6)$ are shorthand notation for six
Grassmann-variable-containing dual-superconformal ``$R$'' invariants.
The coefficient functions $E$ and $\Et$ are related
to the more conventional components of the ratio function, $V$ and $\Vt$,
by
\bea
E(u,v,w) &=& V(u,v,w)
\, \exp\Bigl[ R_6(u,v,w) - \frac{\gamma_K}{8} Y(u,v,w) \Bigr] \,, 
\label{EVrelation}\\
\Et(u,v,w) &=& \Vt(u,v,w)
\, \exp\Bigl[ R_6(u,v,w) - \frac{\gamma_K}{8} Y(u,v,w) \Bigr] \,.
\label{EtVtrelation}
\eea
Using the quantities $E$ and $\Et$ also simplifies the global structure
of the NMHV amplitude~\cite{Dixon2014iba,Dixon2015iva}.
Similarly, the MHV amplitude's global structure is most simply
expressed in terms of ${\cal E}$~\cite{CHDvHMR65}.

The functions $\Et$ and $\Vt$ are odd under parity and, like all
parity-odd functions, they vanish like a power of $\delta$
as one approaches
the self-crossing limit.  However, it is very nontrivial
that the parity-even function $E(u,v,w)$ remains finite
in the limit through four loops, for all different orientations.
These finite limits are presented in appendix~\ref{ENMHVvgt11234}.


\subsection{Self-crossing kinematics}
\label{sckinsubsection}

After the analytic continuation to either $2\to4$ or $3\to3$ kinematics,
the self-crossing limit constrains
two of the cross ratios.  The precise relations between the subprocess
scattering angles and the kinematic invariants
are worked out in appendix~\ref{sckin}, strictly in the self-crossing
limit.  From momentum conservation for the $2\to2$ subprocesses,
we obtain eqs.~(\ref{22kinrelationsfirst})--(\ref{22kinrelationslast}).
Inserting these into the definitions of the cross ratios,
we have
\bea
u &=& \frac{s_{12}s_{45}}{s_{123}s_{345}}\ =\
\frac{(1-x)(1-y) s_{36}\ xy s_{36}}{x(1-y) s_{36}\ y(1-x) s_{36}}\ =\ 1,
\label{ueq1}\\
\frac{v}{w} &=& \frac{s_{23}s_{56}\ s_{345}}{s_{123}\ s_{34}s_{61}}
\ =\ \frac{(1-y)x\ y(1-x)}{x(1-y)\ y(1-x)}\ =\ 1. 
\label{veqw}
\eea
Thus there is only one nonsingular variable, $v$, characterizing
the self-crossing limit.

We regulate the self-crossing singularity by moving $u$ slightly away from 1.
We rewrite momentum conservation near the self-crossing limit as
\bea
k_1 + k_2 + (1-x)k_3 + (1-y)k_6 &=& z \\
k_4 + k_5 + xk_3 + yk_6 &=& -z,
\eea
where $z^\mu$ is a small, space-like vector $z=(0,\vec{z})$
orthogonal to $k_3$ and $k_6$ (see~\fig{fig:sc_reg}).
Then eqs.~(\ref{22kinrelationsfirst})--(\ref{22kinrelationsfourth})
all acquire an additional term of $-\vec{z}^2$ on the right-hand side.
Correcting \eqn{ueq1} for this and expanding to first order in $\vec{z}^2$,
we see that\footnote{The relation between $w$ and $v$ is corrected
at ${\cal O}(\vec{z})$.  However, we can ignore the correction
for the MHV case because
the amplitudes have no extra singularities as $w\to v$, so it
only leads to power-suppressed terms.}
\be
u\ =\ 1 - \delta, \qquad\quad w\ =\ v,
\label{ueq1minusdelta}
\ee
where
\be
\delta\ =\ \frac{\vec{z}^2}{s_{36}xy(1-x)(1-y)} \,.
\label{eq:del_eps}
\ee
Note that for $2\to4$ kinematics, $s_{36}$ is positive, so $\delta>0$;
whereas for $3\to3$ kinematics, $s_{36}$ is negative, so $\delta<0$,
and in this case we will let $\delta = -|\delta|$.

Appendix~\ref{sckin} also indicates the values of $v$ that correspond
to $2\to4$ versus $3\to3$ kinematics:
\bea
&&2\to4\ \hbox{kinematics}:\quad (u,v,w)\ =\ (1-\de,v,v),
\quad \de >0, \quad 0 < v < 1,
\label{vrange24}\\
&&3\to3\ \hbox{kinematics}:\quad (u,v,w)\ =\ (1+|\de|,v,v),
\quad v < 0\ \ \hbox{and}\ \  1 < v.
\label{vrange33}
\eea
The range in $v$ for $3\to3$ kinematics splits into
two segments because, as explained in appendix~\ref{sckin},
$v=w=\infty$ corresponds to $s_{234}=0$.
The 3-particle invariant $s_{234}$ can vanish in the interior
of phase-space only for $3\to3$ kinematics.
It corresponds to the potential factorization pole when a six-point
amplitude separates into two four-point amplitudes.  (Such a pole
is absent in supersymmetric theories in the MHV case,
due to helicity selection rules, although it can be there
in the NMHV case, where it has been studied at three and four
loops~\cite{Dixon2014iba,Dixon2015iva}.)

The point $v=1$ is not special from the point of view of the amplitude,
and we will see that ${\cal E}$ has no additional singularities there.
The Wilson loop framing (see the next section) can induce logarithmic
singularities of the form $\ln|1-v|$ at this point.
The point $v=0$ is special because, as shown in \sect{SCMRKmatch},
it overlaps with the multi-Regge limit.  We will use this correspondence
to determing the self-crossing behavior there to high loop orders.


\subsection{Framed Wilson loops}
\label{framedWL}

\begin{figure}[ht]
\begin{center}
\includegraphics[width=2.4 in]{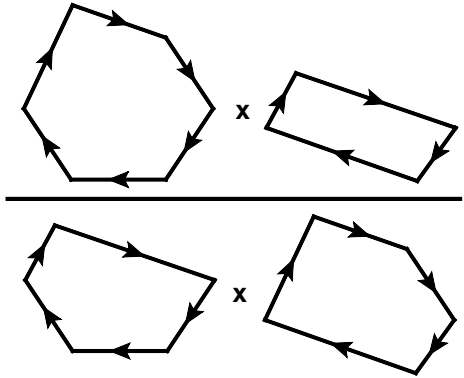}
\end{center}
\caption{The framing of a hexagonal Wilson loop, by dividing it
by two pentagons and multiplying it by a quadrilateral.}
\label{fig:fram}
\end{figure}

Before we can discuss the singular behavior of a Wilson loop
in the self-crossing limit, starting in \sect{evolveWL},
we need to regularize the singularities
that it has for any configuration, namely its cusp divergences.
A convenient way to do this is to ``frame''
the Wilson loop~\cite{Alday2010ku,Basso2013vsa},
as illustrated in \fig{fig:fram}.  Instead of considering just the hexagonal
Wilson loop, we divide it by two pentagonal Wilson loops
and then multiply back by the Wilson loop for a quadrilateral (or box,
for short).
Each pentagon has three cusps that coincide with
three of the six cusps of the hexagon, plus two new ones.
Thus dividing by the pentagons removes the cusp singularities of the hexagon,
while simultaneously introducing four additional cusps. These cusps are
then removed by multiplying by a box which shares two edges and
two cusps with each pentagon.

\begin{figure}[ht]
\begin{center}
\includegraphics[width=2.4 in]{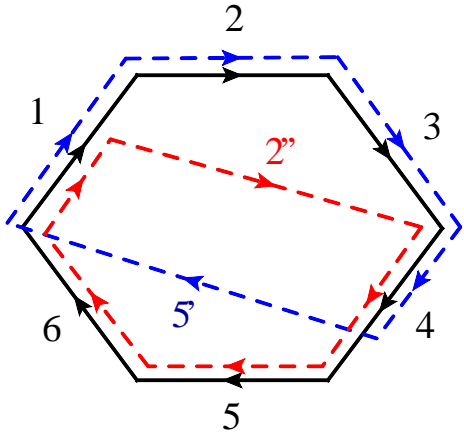}
\end{center}
\caption{Hexagonal Wilson loop with edges $k_1,\ldots,k_6$
  framed by blue and red dashed pentagons. This framing remains
  nonsingular in the self-crossing limit.}
\label{fig:framing_ns}
\end{figure}

The first pentagon is defined by selecting one of the corners of the hexagon
and creating a new edge connecting it to a light-like separated point
on a side furthest away from it, as shown
in \fig{fig:framing_ns}. The second pentagon has the same
construction, with all the labels cycled halfway around, by 3 units
modulo 6.  It might seem that there are 6 ways to do the framing:
3 pairs of opposite corners to choose, and a twofold ambiguity as to which
of the two sides is connected to the first corner.  However,
taking into account the symmetries of the self-crossing configuration,
there are really only three distinct framings.  Two of these framings
are nonsingular in the self-crossing limit, and we will use one of these.
A third framing is singular in the limit, but will still prove useful.

First we describe one of the framings
that remains nonsingular in the self-crossing limit.
Its Wilson loop will be denoted by ${\cal W}^{\rm ns}$.
We use the notation in section~5.4 of ref.~\cite{Dixon2013eka}.
The first pentagon is obtained by removing momenta $k_4,k_5,k_6$ and replacing
them with two new momenta $k_4', k_5'$ having the same sum.
The vector $k_4'$ is parallel to $k_4$, so we have:
\be
k_4 + k_5 + k_6 = k_4' + k_5', \qquad k_4' = \xi k_4.
\label{nsframeq1}
\ee
Demanding $k_5'$ be light-like fixes $\xi=s_{123}/(s_{123}-s_{56})$.
The second pentagon is obtained by replacing 
\be
k_1 + k_2 + k_3 = k_1'' + k_2'', \qquad k_1'' = \xi'' k_1,
\label{nsframeq2}
\ee
and $\xi''=s_{123}/(s_{123}-s_{23})$.
The box then has sides $k_1'',k_2'',k_4',k_5'$.
This nonsingular framing is illustrated in \fig{fig:framing_ns}.

Generically, the relation between the framed Wilson loop ${\cal W}$ and
the remainder function is given by~\cite{Dixon2013eka}
\bea
{\cal W}(u,v,w) &=& \exp\Bigl[ R_6(u,v,w) + \frac{\gK}{8} X(u,v,w) \Bigr] \,,
\label{WfromR}
\eea
where
\bea
X(u,v,w) &=&
- \Li_2(1-u) - \Li_2(1-v) - \Li_2(1-w)
\nn\\  &&\hskip0.0cm
- \ln\biggl(\frac{uv}{w(1-v)}\biggr)\ln (1-v)
- \ln u \ln w + 2 \zeta_2 \,.
\label{Xuvw}
\eea
We can rewrite this relation in terms of ${\cal E}$
using \eqn{calERrelation},
\be
{\cal W}(u,v,w)\ =\
{\cal E}(u,v,w)\, \exp\Bigl[ \frac{\gK}{8} ( X(u,v,w) + Y(u,v,w) ) \Bigr] \,.
\label{WfromE}
\ee
In ref.~\cite{Dixon2013eka} the function $X(u,v,w)$, although correct,
was assigned to the framing specified by \eqns{nsframeq1}{nsframeq2},
whereas it should have been to a flipped framing.  To say it another way,
\eqns{nsframeq1}{nsframeq2} really correspond to $X(w,v,u)$.

\begin{figure}[ht]
\begin{center}
\includegraphics[width=2.4 in]{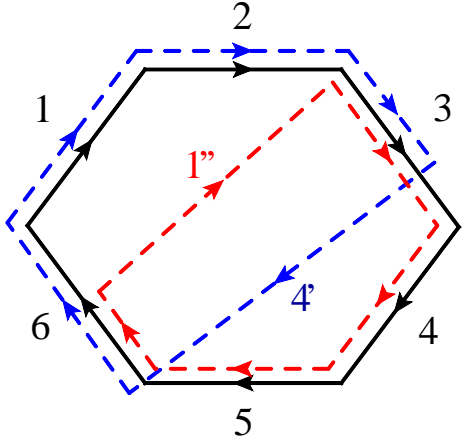}
\end{center}
\caption{A framing for the hexagonal Wilson loop
  in which the pentagons and box
  used to frame the hexagon become singular for self-crossing kinematics.}
\label{fig:framing_s}
\end{figure}

The singular framing is identical to the nonsingular framing,
except that the labels of the momenta $k_i$ or the dual coordinates
$x_i$ are lowered cyclically by one unit, $k_i\to k_{i-1}$, $x_i\to x_{i-1}$.
This framing is illustrated in \fig{fig:framing_s}.
Its Wilson loop will be denoted by ${\cal W}^{\rm \, s}$.
It corresponds to letting $u\to w\to v\to u$ in the function $X(w,v,u)$,
resulting in the function $X(v,u,w)$.  Note that there is a $\ln^2(1-v)$
in $X(u,v,w)$ in \eqn{Xuvw},
which becomes a $\ln^2(1-u)$ in $X(v,u,w)$, and is
the mathematical origin of additional $\ln^2\delta$ terms that
we will find in the self-crossing limit of $X(v,u,w)$ below.

\begin{figure}[ht]
\begin{center}
\includegraphics[width=3.7 in]{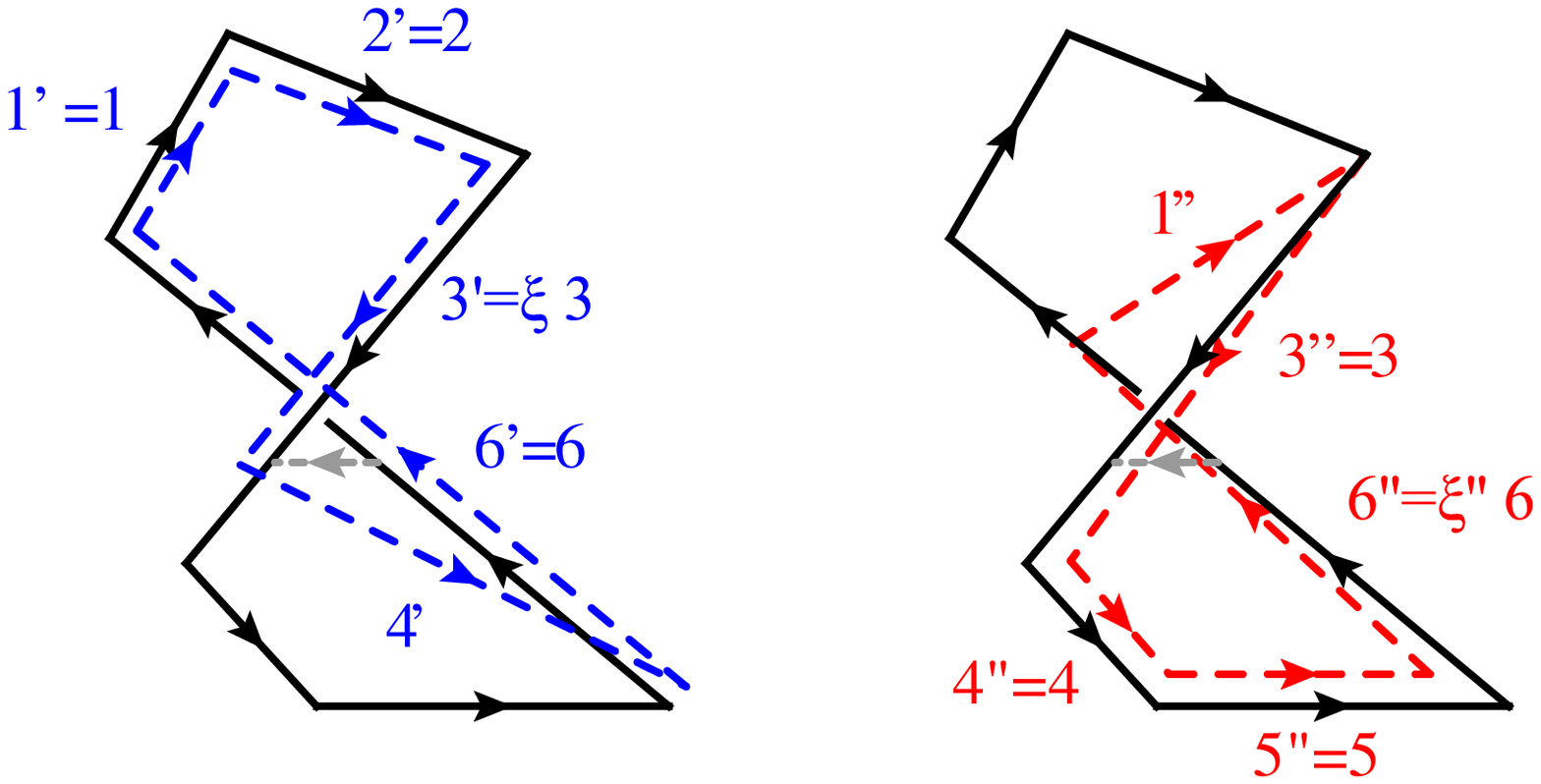}
\end{center}
\caption{The singular framing in \fig{fig:framing_s} near the
  self-crossing limit. The dashed, gray vector is the regulator $z$.
  Note that the edges of the framing Wilson loops go straight through
  the crossing point in this limit.  For $z=0$, the edges $k_6'$ and $k_4'$
  become parallel in the first (blue) framing pentagon,
  while the edges $k_1''$ and $k_3''$ become parallel in the second (red)
  pentagon.  Each framing pentagon degenerates into a box with a parallel
  segment emerging from it as $z\to0$.  This behavior induces additional,
  unwanted dependence on $\delta$.  However, it will
  simplify the $v$ dependence of the singular terms.} 
\label{fig:framing_s_sc}
\end{figure}

To see physically
why this framing is singular, note that in the blue dashed pentagon,
leg 6 is adjacent to a light-like leg (call it $4'$) which runs
from the corner between legs 5 and 6 to the middle of leg 3.
But in the self-crossing limit, the point on leg 3 which is light-like
separated from the corner between legs 5 and 6 is none other than
the self-crossing point, since that lies on both legs 3 and 6.
Therefore the blue pentagon becomes degenerate in this limit;
legs $4'$ and 6 become collinear.  See \fig{fig:framing_s_sc}.
Similarly, the red pentagon degenerates in the self-crossing limit
as legs $1^{\prime\prime}$ and 3 become collinear.  The box
is even more degenerate; it simply runs from the crossing point
out and back along leg 3, and then out and back along leg 6.
For this reason, we will find that the singularly-framed Wilson
loop has extra powers of $\ln\delta$ in its perturbative expansion.

In order to convert back and forth between $R_6$, ${\cal E}$,
${\cal W}^{\rm ns}$ and ${\cal W}^{\rm \, s}$,
we also need to record the limiting behavior of $X^{\rm ns}$,
$X^{\rm s}$ and $Y$ in the self-crossing limit for both $2\to4$ and $3\to3$
kinematics.  For the $2\to4$ self-crossing limit, we first let
$u\to ue^{-2\pi i}$, so that $\ln u \to \ln u - 2\pi i$ and
$\Li_2(1-u) \to \Li_2(1-u) + 2\pi i \ln(1-u)$.
Then we let $u\to1-\de$, $w=v$.  We obtain from $X(w,v,u)$, $X(v,u,w)$
and $Y(u,v,w)$ respectively,
\bea
X^{\rm ns}_{2\to4} &=& -2\pi i \Bigl[ \ln\de + \ln\Bigl( \frac{1-v}{v} \Bigr)\Bigr]
- 2 \Bigl( \Li_2(1-v) - \zeta_2 \Bigr)
+ \ln(1-v) \Bigl( \ln(1-v) - 2 \ln v \Bigr) \,,~~~~~~
\label{Xns24}\\
X^{\rm s}_{2\to4} &=& \ln^2\de - 2 \Bigl( \Li_2(1-v) - \zeta_2 \Bigr)
- \ln^2 v \,,
\label{Xs24}\\
Y_{2\to4} &=& 2\pi i \ln\de + 2 \Li_2(1-v) + \ln^2 v - 2 \pi^2
\nn\\
&=& - X^{\rm ns}_{2\to4} + 2\pi i \ln\left(\frac{v}{1-v}\right)
 + \ln^2\left(\frac{v}{1-v}\right) - 10 \, \zeta_2 \,.
\label{Y24}
\eea
Here we see the $\ln^2\de$ terms in $X^{\rm s}_{2\to4}$ whose
origin was mentioned earlier.

To get to $3\to3$ kinematics with $v<0$, following ref.~\cite{Bartels2010tx}
we let $\ln\de \to \lnden - i\pi$,\ \ $\ln v \to \ln|v| - i\pi$,
and then complex conjugate the result.  We find,
\bea
X^{\rm ns}_{3\to3} &=& 2\pi i \Bigl[ \lnden + L \Bigr]
+ 2 \, \Li_2(v) + \ln^2(1-v) \,,
\label{Xns33}\\
X^{\rm s}_{3\to3} &=& X^{\rm ns}_{3\to3} + \lndene{2} - L^2 \,,
\label{Xs33}\\
Y_{3\to3} &=& - X^{\rm ns}_{3\to3} + L^2 - 4 \zeta_2 \,,
\label{Y33}
\eea
where
\be
L\ \equiv\ \ln\Bigl( 1 - \frac{1}{v} \Bigr) \,.
\label{Ldef}
\ee

Considering \eqn{Y33} as well as \eqn{WfromE},
we see that the $3\to3$ self-crossing limit of ${\cal E}$ is very closely
related to that of the nonsingularly-framed Wilson loop ${\cal W}$:
\bea
{\cal E}_{3\to3} &=& {\cal W}^{\rm ns}_{3\to3} \times
 \exp\Bigl[ - \frac{\gK}{8} \Bigl( L^2 - 4 \, \zeta_2 \Bigr) \Bigr]
\label{EfromW}\\
&=& {\cal W}^{\rm \, s}_{3\to3} \times
 \exp\Bigl[ - \frac{\gK}{8} \Bigl( \lndene{2} - 4 \, \zeta_2 \Bigr) \Bigr]
\,. \label{EfromWs}
\eea
%


\section{Explicit results through five loops}
\label{explicitresults}

In this section we describe how to extract the $2\to4$ or $3\to3$
self-crossing limit of the MHV amplitude function ${\cal E}$, or equivalently
the framed hexagonal Wilson loops,
from the full remainder function $R_6^{(L)}(u,v,w)$, which has been computed
for $L=2,3,4$~\cite{Goncharov2010jf,Dixon2011pw,Dixon2013eka,Dixon2014voa}
and recently for $L=5$~\cite{CHDvHMR65}.
These results include also the nonsingular terms, those having no powers
of $\ln\de$.
We will also describe similar results for the NMHV amplitude function $E$,
which is entirely nonsingular.
Later we will examine the singular terms in the MHV case
to even higher loop order by making use of an evolution
equation for Wilson loops.

To extract the self-crossing limits,
we used properties of hexagon functions~\cite{Dixon2013eka}.
A basis for hexagon functions has been
constructed through weight eight~\cite{Dixon2015iva}.
A formula for $R_6^{(2)}$ in terms of hexagon functions was
presented already in ref.~\cite{Dixon2011nj},
and for $R_6^{(3)}$ and $R_6^{(4)}$ in ref.~\cite{Dixon2015iva}.
In practice we first found the $2\to4$ limit, and then the $3\to3$ limit
by analytic continuation from the $2\to4$ limit.
So we will describe that procedure. However, we will only present
the $3\to3$ results explicitly, because they are simpler.  The $2\to4$
results can be found by reversing the analytic continuation.


\subsection{Analytic continuation from Euclidean to $2\to4$ kinematics}
\label{Euclto24}

The analytic continuation for $2\to4$ scattering was described
in early studies of the MRK limit~\cite{Bartels2008ce}.
For the $2\to4$ scattering shown in \fig{fig:show_24_33}(a),
and given in \eqn{incout24}, the relevant invariants have the
following signs,
\be
s_{12}, s_{45}, s_{36} > 0, \qquad
s_{23}, s_{34}, s_{56}, s_{61}, s_{123}, s_{234}, s_{345} < 0.
\label{sign24}
\ee
We see that $u,v,w > 0$.  Note that $v$ and $w$ are composed
entirely of space-like (negative) invariants.  Therefore they do not need
to be analytically continued from the Euclidean region.
In contrast, $u$ is the product of two time-like invariants, divided
by two space-like ones.  The $i\varepsilon$ prescription requires
$s\to s e^{-i\pi}$ for each continuation from space-like to time-like,
hence to reach the $2\to4$ branch we take
\be
2\to4:\ \ \ u \to u \, e^{-2\pi i}, \quad v\to v, \quad w\to w.
\label{analcont24}
\ee

Next we approach the self-crossing configuration.  Like the MRK limit,
this requires $u \to 1$ from below, so we let $u=1-\de$ with $\de \ll 1$.
In the MRK limit, $v$ and $w$ both approach zero, proportional to $\de$,
but they do not have to be equal.  In the self-crossing limit,
$v$ and $w$ become equal~\cite{Georgiou,DornWuttke1,DornWuttke2}.
Thus we need to know how to analytically continue functions under
\eqn{analcont24}, followed by the limit $(u,v,w) \to (1-\de,v,v)$.
Many simplifications occur in this limit.  First of all, the quantity
$\Delta(u,v,w) = (1-u-v-w)^2 - 4 uvw$ vanishes in this limit,
$\Delta(1-\de,v,v) = -4\,\de\,v(1-v)\,+\,\de^2$.
Hexagon functions can be characterized by their ``parity'', or
how they transform under $\Delta \to -\Delta$.  Parity-odd functions
are odd under this transformation, and so they vanish like a power
of $\de$ as we approach the line $(1,v,v)$.

Hence we can restrict our attention to hexagon functions $F$ that are
even under parity.
The coproduct bootstrap for hexagon functions~\cite{Dixon2013eka}
says that we can construct their behavior on the line $(1-\de,v,v)$
iteratively in the weight, by using the differential equation
\be
\frac{d}{dv}F(1-\de,v,v)
= \frac{F^v + F^w}{v}  - \frac{F^{1-v}+F^{1-w}}{1-v} \,,
\label{diffv}
\ee
where $F^x$ denotes the $x$ component of the weight $\{n-1,1\}$ coproduct
of the weight $n$ function $F$, evaluated on the same line.
In \eqn{diffv}, we dropped terms involving parity-odd functions,
which would arise from coproducts of the form $F^{y_i}$,
because such contributions vanish on the self-crossing line. 

\Eqn{diffv} has the same structure as it does on the Euclidean branch,
although on that branch it is possible to set $\de\to0$, as there is
no singularity as $u\to1$ in this case.  This limit of the remainder function
was given through four loops in refs.~\cite{Dixon2013eka,Dixon2014voa},
and at five loops in ref.~\cite{CHDvHMR65}, as the value on the line $(u,u,1)$.
Since the remainder function is totally symmetric under exchange of its
three arguments, this line is equivalent to $(1,v,v)$ up to a relabeling.
Just as was found earlier in the Euclidean case,
the iterative solution to \eqn{diffv} lies in the space of harmonic
polylogarithms (HPLs)~\cite{Remiddi1999ew}
$H_{\vec{w}}(v)$ with argument $v$ and weight vector
$\vec{w} = (w_1,w_2,\ldots,w_n)$ with all $w_i\in \{0,1\}$.
The only difference is that on the $2\to4$ and $3\to3$ self-crossing
branches there may be imaginary parts ($i\pi$ factors)
as well as factors of $\ln\de$.  

For example, for the two-loop remainder function $R_6^{(2)}$,
we can use its $\{3,1\}$ coproducts to write its derivative~(\ref{diffv}) as
\bea
v(1-v) \, \frac{dR_6^{(2)}(1-\de,v,v)}{dv} &=& H_{2,1}^v - H_{3}^v
- \frac{1}{2} \Bigl[ H_2^u \ln v + \ln u \ln^2 v \Bigr]
 \Big|_{u\to u e^{-2\pi i}\to 1-\de}
\nn\\
&=& H_{2,1}^v - H_{3}^v - i\pi \Bigl[ \ln\de \ln v - \ln^2 v \Bigr] \,,
\label{dR62sc}
\eea
where we let $\ln u\to \ln(1-\de)-2\pi i = -2\pi i$ in the second step.
Since $dH_2^u/du = (\ln u)/(1-u) \to -2\pi i/(1-u)$, we also have
that $H_2^u\to 2\pi i\ln(1-u) = 2\pi i \ln\de$ under this analytic
continuation. The first two terms of \eqn{dR62sc} match the result found
in eq.~(7.18) of ref.~\cite{Dixon2013eka} for the behavior on the Euclidean
branch,
\be
v(1-v) \, \frac{dR_6^{(2)}(1,v,v)}{dv} \bigg|_{\rm Eucl.} = H_{2,1}^v - H_{3}^v \,.
\label{dR62Eucl}
\ee

The fact that the $v$ derivative of the remainder function is $1/[v(1-v)]$
times a transcendental function is just a reflection of the
final-entry condition~\cite{CaronHuot2011kk}. For general $(u,v,w)$
only six final entries appear:
\be
\bigg\{ \frac{u}{1-u}, \frac{v}{1-v}, \frac{w}{1-w}, y_u, y_v, y_w \bigg\}.
\label{fe6}
\ee
On the line $(1,v,v)$ (on any branch), $u$, $y_u$, $y_v$ and $y_w$
are all trivial, and $w=v$, so the set~(\ref{fe6}) collapses to
the single final entry $v/(1-v)$, and $d\ln[v/(1-v)]/dv = 1/[v(1-v)]$.
So the $v$ derivative of the remainder function must have this prefactor.
However, the $v$ derivatives of generic hexagon functions,
which are needed in intermediate steps of the iterative construction,
will not have this property.

We also need to fix a boundary condition for the integration of \eqn{diffv}.
We do so at the point $v=1$.  To obtain the values of the hexagon
functions on the $2\to4$ branch at $(1-\de,1,1)$, we first obtain them
along the Euclidean branch of the line $(u,1,1)$.  On this line, the
hexagon functions are also HPLs, with argument $u$, although the
parity-odd functions are non-vanishing here.
The remainder function is given 
on this line through four loops in refs.~\cite{Dixon2013eka,Dixon2014voa},
and at five loops in ref.~\cite{CHDvHMR65}.
(See also the ancillary files associated with ref.~\cite{Dixon2015iva}.)
For example, the two-loop remainder function is
\bea
R_6^{(2)}(u,1,1) &=& 
\frac{1}{2} \biggl[ H_{4}^u - H_{3,1}^u + 3 \, H_{2,1,1}^u
   - \frac{1}{4} (H_{2}^u)^2 + H_{1}^u ( H_{3}^u - 2 \, H_{2,1}^u )\nn\\
&&\hskip0.6cm\null
   + \frac{1}{2} ( H_{2}^u - \zeta_2 ) (H_{1}^u)^2 - 5 \zeta_4 \biggr] \,,
\label{R62_u11}
\eea
where $H_{3,1}^u = H_{0,0,1,1}(1-u)$, etc.  This ``standard'' Lyndon-basis
form of the
result has argument $(1-u)$ for all HPLs with trailing 1's in their weight
vectors.  If there had been a trailing 0, we could use a shuffle identity
to remove it, at the price of extracting a factor of $H_0(1-u)=\ln(1-u)$.
HPLs with trailing 1's are regular when their argument vanishes.  Thus
\eqn{R62_u11} is adapted to the point $u=1$, in the sense that it makes
manifest that $R_6^{(2)}(u,1,1)$ has no branch cut at $u=1$ (in the Euclidean
region), because there is no $\ln(1-u)$.

For the analytic continuation of $u$ around the origin, we need
to use HPL identities to rewrite the result in terms of a Lyndon basis
for $H_{\vec{w}}(u)$, rather than $H_{\vec{w}}(1-u)$.  We obtain,
\bea
R_6^{(2)}(u,1,1) &=& 
\frac{1}{4} \biggl\{ \ln u \biggl[ \frac{1}{3} [H_1(u)]^3
     + H_1(u) \Bigl[ H_2(u) + \zeta_2 \Bigr] + 2 \, H_3(u) + 2 \, \zeta_3 \biggr]
\nn\\
&&\hskip0.3cm\null
     - 6 \, H_4(u) + 2 \, H_{3,1}(u) - 2 \, H_{2,1,1}(u)
     - \frac{1}{2} \, [H_2(u)]^2
     - 2 \, H_1(u) \, \Bigl[ H_3(u) - H_{2,1}(u) \Bigr] \nn\\
&&\hskip0.3cm\null
     - [H_1(u)]^2 \Bigl[ H_2(u) - \zeta_2 \Bigr]
     + \zeta_2 \, H_2(u) - \frac{15}{4} \zeta_4 \biggr\} \,.
\label{R62_u11_ueq0_basis}
\eea
To get to the $2\to4$ branch we just set $\ln u \to \ln u - 2\pi i$
in \eqn{R62_u11_ueq0_basis}, because the $H_{\vec{w}}(u)$ are all regular
at $u=0$. Next we let $u=1-\de$ and take $\de\to0$, to get
\be
R_6^{(2)}(1-\de,1,1) = 
2\pi i \biggl[ \frac{1}{12} \, \ln^3\de + \frac{\zeta_2}{2} \, \ln\de
              - \zeta_3 \biggr] - \frac{5}{2} \zeta_4 \,.
\label{R62_1mde11}
\ee

Next we need to integrate up \eqn{diffv} for the generic hexagon functions,
imposing a boundary condition at $(1-\de,1,1)$.
For the case of $R_6^{(2)}$, we integrate \eqn{dR62sc} with the boundary
condition~(\ref{R62_1mde11}), obtaining
\bea
R_6^{(2)}(1-\de,v,v) &=& 
2\pi i \biggl[ \frac{1}{12} \, \ln^3\de
+ \Bigl( - \frac{1}{2} \, H_2^v - \frac{1}{4} \ln^2 v
        + \frac{\zeta_2}{2} \Bigr) \ln\de
- H_{2,1}^v + \frac{1}{6} \ln^3 v - \zeta_3 \biggr]\nn\\
&&\hskip0cm\null
+ H_4^v - H_{3,1}^v + 3 \, H_{2,1,1}^v
- \ln v ( H_3^v - H_{2,1}^v ) - \frac{1}{2} \, ( H_2^v )^2
- \frac{5}{2} \zeta_4 \,.
\label{R62sc24}
\eea
The leading $\ln^3\de$ term in this formula agrees with
the result of ref.~\cite{Georgiou}.

We have repeated this exercise for the higher-loop remainder functions
$R_6^{(3)}$, $R_6^{(4)}$, and $R_6^{(5)}$ (as well
as for the NMHV coefficient functions $E^{(1)}$,
$E^{(2)}$, $E^{(3)}$ and $E^{(4)}$).
Once we have obtained the remainder function in the self-crossing
configuration, we can find ${\cal W}^{\rm ns}_{2\to4}$,
${\cal W}^{\rm \, s}_{2\to4}$ and ${\cal E}_{2\to4}$
with the help of eqs.~(\ref{Xns24}), (\ref{Xs24}) and (\ref{Y24}).


\subsection{From $2\to4$ to $3\to3$ kinematics with $v<0$}
\label{from24to33vneg}

To get to $3\to3$ kinematics with $v<0$, following ref.~\cite{Bartels2010tx}
we let $\ln\de \to \lnden - i\pi$,\ \ $\ln v \to \ln|v| - i\pi$,
and then complex conjugate the result.
To do the analytic continuation around $v=0$, we first
rewrite the expressions in terms of a Lyndon basis for $H_{\vec{w}}(v)$.
We apply this procedure
to the remainder function, and then apply eqs.~(\ref{Xns33}),
(\ref{Xs33}) and (\ref{Y33}) to construct the results
for the framed Wilson loops and for ${\cal E}$.
In view of the simple relations~(\ref{EfromW}) and (\ref{EfromWs})
between them, it's sufficient to give one of these quantities,
and ${\cal E}$ has the simplest finite parts.

In order to represent the results for $3\to3$ kinematics
compactly, we define some compressed notation~\cite{Dixon2014voa}.
We first expand all products of HPLs using the shuffle algebra,
in order to linearize the expression in terms of HPLs.
The HPL weight vectors $\vec w$ consist entirely of $0$'s and $1$'s;
we encode them as binary numbers, but written as a subscript in decimal.
We use a superscript to record the length of the original weight vector.
For example,
\bea
H_{2}(z) H_{2,1}(z) &=& H_{0,1}(z) H _{0,1,1}(z)
\ =\ 6 H_{0,0,1,1,1}(z) + 3 H_{0,1,0,1,1}(z) + H_{0,1,1,0,1}(z)
\nn\\
&\to& 6 h^{[5]}_7 + 3 h^{[5]}_{11} + h^{[5]}_{13} \,.
\label{sampletoh}
\eea
Here the suppressed argument of $h$ is $z = 1/(1-v)$.  Note also that
\be
h_{1}^{[1]} = - \ln\biggl( 1 - \frac{1}{1-v} \biggr)
= \ln\biggl(1-\frac{1}{v}\biggr) = L \,.
\label{h11eqL}
\ee
In this notation, $X^{\rm ns}_{3\to3}$ and $Y_{3\to3}$ become
\bea
X^{\rm ns}_{3\to3} &=& 2\pi i ( \lnden + L ) - 2 h_2^{[2]} - 2 \zeta_2 \,,
\label{Xns33h}\\
Y_{3\to3} &=& - 2\pi i ( \lnden + L ) + 2 ( h_2^{[2]} + h_3^{[2]} ) - 2 \zeta_2 \,,
\label{Y33h}
\eea
and the two-loop remainder function evaluates to
\bea
R_6^{(2)}|_{3\to3} &=& 2\pi i \biggl[ - \frac{1}{12} \, \lndene{3}
\, + \, \frac{1}{2} ( h_{2}^{[2]} + h_{3}^{[2]} + \zeta_2 ) \lnden
\, + \, \frac{1}{2} \, h_{5}^{[3]} + h_{6}^{[3]} + h_{7}^{[3]}
+ \frac{1}{2} \, \zeta_2 \, L + 2 \, \zeta_3 \biggr]\nn\\
&&\hskip0cm\null
+ 3 \, \zeta_2 \, \lndene{2}\ +\ 6 \, \zeta_2 \, L \, \lnden
\, + \, 2 \, h_{8}^{[4]} + 2 \, h_{9}^{[4]} + h_{10}^{[4]} + h_{11}^{[4]}\nn\\
&&\hskip0cm\null
+ \zeta_2 \, ( h_{2}^{[2]} + 6 \, h_{3}^{[2]} ) - \zeta_3 \, L
+ \frac{33}{4} \, \zeta_4 \,.
\label{R62sc33}
\eea

The results for ${\cal E}$ in the $3\to3$ self-crossing configuration
with $v<0$ through five loops are:
\bea
{\cal E}_{3\to3}^{(0)} &=& 1 \,, \label{Evneg_0}\\
{\cal E}_{3\to3}^{(1)} &=& -\frac{1}{2} Y_{3\to3}\ =\
\pi i \, \Bigl[ \lnden \, + \, L \Bigr]
- h_{2}^{[2]} - h_{3}^{[2]} + \zeta_2 \,, \label{Evneg_1}\\
{\cal E}_{3\to3}^{(2)} &=& 2 \pi i \, \biggl[
- \frac{1}{12} \lndene{3} \, + \, \frac{1}{2} \zeta_2 \lnden \,
- \frac{1}{12} L^3 + \frac{1}{2} \zeta_2 L + 2 \zeta_3 \biggr]
\nn\\ &&\hskip0.0cm\null
+ 2 \Bigl( h_{8}^{[4]} + h_{9}^{[4]} + h_{10}^{[4]} + h_{11}^{[4]}
         + h_{12}^{[4]} + h_{13}^{[4]} \Bigr)
\nn\\ &&\hskip0.0cm\null
+ 3 \Bigl( h_{14}^{[4]} + h_{15}^{[4]} \Bigr)
+ \zeta_2 h_{2}^{[2]} - \zeta_3 L + 7 \zeta_4 \,, \label{Evneg_2}
\eea
\bea
{\cal E}_{3\to3}^{(3)} &=& 2 \pi i \, \biggl[
\frac{1}{80} \lndene{5} \, - \, \frac{1}{4} \zeta_3 \lndene{2}
  \, + \, \frac{1}{4} \zeta_4 \lnden \,
  - \, \frac{1}{4} \Bigl(
  4 h_{17}^{[5]} + 2 h_{19}^{[5]} + h_{21}^{[5]} + 2 h_{25}^{[5]} + h_{27}^{[5]}
  - 6 h_{31}^{[5]}
\nn\\ &&\hskip0.8cm\null
  - \zeta_2 h_{5}^{[3]} + \zeta_3 h_{3}^{[2]}
          + \frac{7}{4} \zeta_4 L + 42 \zeta_5 - 18 \zeta_2 \zeta_3
          \Bigr) \biggr]
\nn\\ &&\hskip0.0cm\null
- \frac{1}{2} \biggl[ 24 ( h_{32}^{[6]} + h_{33}^{[6]} )
     + 20 ( h_{34}^{[6]} + h_{35}^{[6]} + h_{36}^{[6]} + h_{37}^{[6]} )
     + 18 ( h_{38}^{[6]} + h_{39}^{[6]} ) + 20 ( h_{40}^{[6]} + h_{41}^{[6]} )
\nn\\ &&\hskip0.8cm\null
     + 19 ( h_{42}^{[6]} + h_{43}^{[6]} )
     + 18 ( h_{44}^{[6]} + h_{45}^{[6]} + h_{46}^{[6]} + h_{47}^{[6]} )
     + 24 ( h_{48}^{[6]} + h_{49}^{[6]} )
\nn\\ &&\hskip0.8cm\null
     + 22 ( h_{50}^{[6]} + h_{51}^{[6]} + h_{52}^{[6]} + h_{53}^{[6]} )
     + 21 ( h_{54}^{[6]} + h_{55}^{[6]} )
\nn\\ &&\hskip0.8cm\null
     + 24 ( h_{56}^{[6]} + h_{57}^{[6]} + h_{58}^{[6]}
     + h_{59}^{[6]} + h_{60}^{[6]} + h_{61}^{[6]} )
     + 30 ( h_{62}^{[6]} + h_{63}^{[6]} )
\nn\\ &&\hskip0.8cm\null
    + \zeta_2 ( 16 h_{8}^{[4]} + 2 h_{9}^{[4]} + 11 h_{10}^{[4]} + 4 h_{11}^{[4]}
            + 14 h_{12}^{[4]} + 7 h_{13}^{[4]} + 12 h_{14}^{[4]} + 6 h_{15}^{[4]} )
\nn\\ &&\hskip0.8cm\null
     + \zeta_3 ( 2 h_{4}^{[3]} - h_{5}^{[3]} + h_{6}^{[3]} - 6 h_{7}^{[3]} )
     + \frac{1}{4} \zeta_4 ( 81 h_{2}^{[2]} + 83 h_{3}^{[2]} )
     - 4 \zeta_5 L
\nn\\ &&\hskip0.8cm\null
     + \frac{3787}{48} \zeta_6 - \frac{5}{2} (\zeta_3)^2 \biggr]
\,, \label{Evneg_3}
\eea
\bea
{\cal E}_{3\to3}^{(4)} &=& 2 \pi i \, \biggl[
- \frac{1}{672} \lndene{7} \, - \, \frac{1}{80} \zeta_2 \lndene{5}
\, - \, \frac{1}{48} \zeta_3 \lndene{4}
\, - \, \frac{7}{24} \zeta_4 \lndene{3}
\nn\\ &&\hskip0.8cm\null
\, + \, \frac{1}{4} \Bigl( 4 \zeta_5 - 3 \zeta_2 \zeta_3 \Bigr) \lndene{2}
\, - \, \frac{1}{48} \Bigl( 13 \zeta_6 + 48 (\zeta_3)^2 \Bigr) \lnden \,
\biggr]
\,+\, \hbox{finite}\,, \label{Evneg_4_sing}\\
{\cal E}_{3\to3}^{(5)} &=& 2 \pi i \, \biggl[
  \frac{1}{6912} \lndene{9}
\, + \, \frac{1}{336} \zeta_2 \lndene{7} + \frac{5}{288} \zeta_3 \lndene{6}
\, + \, \frac{9}{80} \zeta_4 \lndene{5}
\, + \, \frac{1}{24} \Bigl( 6 \zeta_5 + 7 \zeta_2 \zeta_3 \Bigr) \lndene{4}
\nn\\ &&\hskip0.8cm\null
\, + \, \frac{1}{72} \Bigl( 115 \zeta_6 + 48 (\zeta_3)^2 \Bigr) \lndene{3}
\, + \, \frac{1}{16} \Bigl( - 55 \zeta_7 + 68 \zeta_2 \zeta_5
      + 44 \zeta_3 \zeta_4 \Bigr) \lndene{2}
\nn\\ &&\hskip0.8cm\null
\, + \, \frac{1}{72} \Bigl( 257 \zeta_8 + 810 \zeta_3 \zeta_5
    + 18 \zeta_2 (\zeta_3)^2  \Bigr) \lnden \,
\biggr]
\,+\, \hbox{finite}\,, \label{Evneg_5_sing}
\eea
where the suppressed argument of the $h_i^{[w]}(z)$ is $z=1/(1-v)$.
We also have results for the finite parts of ${\cal E}$
at four and five loops, i.e.~those terms lacking a power of $\lnden$.
However, these terms are rather lengthy,
so we will present them later, for the $v>0$ branch of
$3\to3$ kinematics instead (for which they are somewhat more compact);
the values for $v<0$ can be recovered by analytic continuation.
(We give the four- and five-loop finite parts of ${\cal E}$
for $v<0$ in the ancillary file accompanying this paper.)


\subsection{A few observations}
\label{obs}

Inspection of eqs.~(\ref{Evneg_0}) through (\ref{Evneg_5_sing})
reveals a number of remarkable features:
\begin{enumerate}
\item The $\lnden$ singularities in ${\cal E}_{3\to3}$
  are only in the imaginary part. This is not true
  for the remainder function (see \eqn{R62sc33}), and it
  is not true for ${\cal E}$ in $2\to4$ kinematics.
  Note that the same statements will be true for both
  Wilson loops
  ${\cal W}^{\rm ns}_{3\to3}$ and  ${\cal W}^{\rm \, s}_{3\to3}$,
  because according to \eqns{EfromW}{EfromWs},
  they are related to ${\cal E}_{3\to3}$ by real prefactors.
\item The singularities are totally independent of $v$.
  This statement is also true for the singularly-framed Wilson loop
  ${\cal W}^{\rm \, s}_{3\to3}$, but not for ${\cal W}^{\rm ns}_{3\to3}$,
  since it differs from ${\cal E}_{3\to3}$ by a finite, but $v$-dependent
  exponential factor.
\item In the finite (non-$\lnden$) parts, only a limited range of
  the subscripts $i$ on the functions $h_i^{[w]}$ appear, starting at
  $i=2^{w-1}$.  In the imaginary part, only odd values of $i$ appear.
  In the real part, both even and odd $i$ can appear, but in the non-$\zeta$
  terms $i=2k$ and $i=2k+1$ always appear with the same coefficients.
\end{enumerate}
The first two properties have their origin in the factorization
structure of self-crossed Wilson loops, to be discussed below.

The property that $h_i^{[w]}$ never appears for $i < 2^{w-1}$
ensures that the binary weight vector for $h_i^{[w]}$
always starts with a ``1''.  Then the $v$ derivative of
$h_i^{[w]}(z)$, with $z=1/(1-v)$, has a rational prefactor of
$dz/dv \times 1/(1-z) = -1/[v(1-v)]$.  As mentioned earlier,
this feature is just the manifestation of the final-entry
condition~\cite{CaronHuot2011kk} on the remainder function.
Inspection of \eqns{Xns33h}{Y33h} shows that it also holds for 
$X^{\rm ns}_{3\to3}$ and $Y_{3\to3}$, so the conversion to ${\cal E}$
and the Wilson loop framings do not spoil this property.

Another feature of the explicit results for ${\cal W}_{3\to3}^{(L)}$
is that the $h_i^{[w]}$ that appear in the imaginary part only have odd
values of $i$, while in the real part there is no such restriction.
An odd value of $i$, or a final value of ``1'' in the binary representation,
corresponds to a statement that the branch cuts in ${\cal W}_{3\to3}^{(L)}$
are only in $1-z = -v/(1-v)$.  This property allows for a branch cut
at $v=0$ (which definitely occurs) but forbids it at $v=\infty$.
Recall that $v=\infty$ is the position of the multi-particle pole
as $s_{234} \to 0$ inside $3\to3$ kinematics, and that helicity selection
rules forbid such a pole in the MHV case.  Although there is no pole,
there certainly can be a branch cut in this channel.
It is interesting that the branch cut behavior in $v$
is only found in the real part, not the imaginary part.

The leading $\lndene{5}$ term in the remainder function $R_6^{(3)}$,
was predicted in ref.~\cite{DornWuttke2}
to be $R_6^{(3)} \sim \mp (7\pi i/240) \lndene{5}$.
At this leading order in $\lnden$, the behavior of $R_6^{(3)}$,
${\cal E}^{(3)}$ and the nonsingularly-framed Wilson loop
${\cal W}^{\rm ns}_{3\to3}$ are identical; for example,
the cross-terms from exponentiation
can produce at most a term of the form $\pi^2 \lndene{4}$.
\Eqn{Evneg_3} disagrees with the prediction of ref.~\cite{DornWuttke2}
by a factor of $\pm6/7$.   What could cause the discrepancy?
In ref.~\cite{DornWuttke2}, the same dimensional regulator $\e$
was used to regulate the self-crossing singularity as the cusp singularity.
This may have led to difficulties in extracting the dependence
on the self-crossing separation alone. Here we have separated the self-crossing
and cusp singularities cleanly; the nonsingular framing removes the
cusp singularities completely, at least at this order in $\lnden$.
Also, we use a dual conformal measure $\de$ of the self-crossing
separation throughout the calculation.

\subsection{From $v<0$ to $v>1$}
\label{fromvnegtovpos}

Formulas~(\ref{Evneg_0}) to (\ref{Evneg_3})
describe the MHV amplitude
in $3\to3$ kinematics with $v<0$, or $s_{234} > 0$.
In this subsection we describe how to analytically continue them to
the other branch of $3\to3$ kinematics, where $s_{234} < 0$ and $v>1$.
From \eqn{eq:del_eps} we see that the sign of $\delta$ does not
change in passing between the two branches, because $s_{36}$ remains
negative.  We just need to analytically continue $v$ around $v=\infty$.
Letting $s_{234} \to s_{234} + i\varepsilon$ in \eqn{uvw_def},
we find the sign of the $i\pi$ term in the continuation:
\be
\ln\biggl(\frac{1}{1-v}\biggr)
\ \to\ \ln\biggl(\frac{1}{v-1}\biggr) + i\pi \,.
\label{vneg_to_vpos}
\ee

In order to carry out this analytic continuation of
eqs.~(\ref{Evneg_0})--(\ref{Evneg_3}), it is simplest to
return to the Lyndon basis for HPLs $H_{\vec{w}}(z)$
with argument $z=1/(1-v)$. That's because the point $v=\infty$
around which we are continuing is at $z=0$, and all HPLs with
trailing 1's in their weight vectors are regular at this point.
In the Lyndon basis, the only function that is not regular is
\be
H_0(z)\ =\ \ln\biggl(\frac{1}{1-v}\biggr) \,,
\label{H0z}
\ee
which is to be replaced using \eqn{vneg_to_vpos}.
Then we rewrite the result in a linearized basis, for compactness,
but using a different argument for the $h$ functions, $\hat{z}=1/v$.

The results for the $3\to3$ self-crossing configuration
with $v>1$ through three loops are:
\bea
{\cal E}_{3\to3}^{(0)}(v>1) &=& 1 \,, \label{Evgt1_0}\\
{\cal E}_{3\to3}^{(1)}(v>1) &=& \pi i \lnden \, + \, h_{2}^{[2]}
+ \zeta_2 \,, \label{Evgt1_1}\\
{\cal E}_{3\to3}^{(2)}(v>1) &=& 2\pi i \biggl[
- \frac{1}{12} \lndene{3} \, + \, \frac{1}{2} \zeta_2 \lnden
\, - \, h_{4}^{[3]} + 2 \zeta_3 \biggr]
\nn\\  &&\hskip0.0cm\null
- 2 h_{8}^{[4]} - h_{14}^{[4]}
+ \zeta_2 \Bigl( 5 h_{2}^{[2]} - h_{3}^{[2]} \Bigr)
+ \zeta_3 h_{1}^{[1]} + 7 \zeta_4 \,, \label{Evgt1_2}
\eea
\bea
{\cal E}_{3\to3}^{(3)}(v>1) &=& 2\pi i \biggl\{
\frac{1}{80} \lndene{5} \, - \, \frac{1}{4} \zeta_3 \lndene{2}
\,   + \, \frac{1}{4} \zeta_4 \lnden \,
\,  + \, \frac{1}{2} \Bigl[
12 h_{16}^{[5]} + 2 h_{18}^{[5]} + 2 h_{20}^{[5]} + h_{22}^{[5]}
\nn\\  &&\hskip1.0cm\null
          +      h_{26}^{[5]} + 2 h_{28}^{[5]}
          + \zeta_2 \Bigl( - 4 h_{4}^{[3]} + h_{5}^{[3]} + h_{6}^{[3]} \Bigr)
          + \zeta_3 h_{2}^{[2]} - 21 \zeta_5 + 9 \zeta_2 \zeta_3 \Bigr]
          \biggr\}
\nn\\  &&\hskip0.0cm\null
+ \frac{1}{2} \biggl[
  24 h_{32}^{[6]} + 4 h_{34}^{[6]} + 4 h_{36}^{[6]} + 2 h_{38}^{[6]}
        +  4 h_{40}^{[6]} + 3 h_{42}^{[6]} + 2 h_{44}^{[6]} + h_{46}^{[6]}
        +  2 h_{50}^{[6]}
\nn\\  &&\hskip1.0cm\null
        + 2 h_{52}^{[6]} + h_{54}^{[6]} + 4 h_{56}^{[6]}
        +      h_{58}^{[6]} + 6 h_{62}^{[6]}
\nn\\  &&\hskip1.0cm\null
+ \zeta_2 \Bigl(
- 56 h_{8}^{[4]} + 2 h_{9}^{[4]} - 7 h_{10}^{[4]} + h_{11}^{[4]}
+  2 h_{12}^{[4]} + h_{13}^{[4]} - 9 h_{14}^{[4]} + 6 h_{15}^{[4]} \Bigr)
\nn\\  &&\hskip1.0cm\null
+ \zeta_3 \Bigl( 2 h_{4}^{[3]} + 3 h_{5}^{[3]} + h_{6}^{[3]} - 4 h_{7}^{[3]} \Bigr)
        - \frac{1}{4} \zeta_4 \Bigl( 39 h_{2}^{[2]} + 62 h_{3}^{[2]} \Bigr)
\nn\\  &&\hskip1.0cm\null
        - \Bigl( 4 \zeta_5 + 6 \zeta_2 \zeta_3 \Bigr) h_{1}^{[1]}
        - \frac{3787}{48} \zeta_6 + \frac{5}{2} (\zeta_3)^2 \biggr]
\,, \label{Evgt1_3}
\eea
where the suppressed argument of the $h_i^{[w]}(\hat{z})$ is now $\hat{z}=1/v$.
The corresponding results for four and five loops are
given in \eqns{Evgt1_4}{Evgt1_5} in appendix~\ref{Evgt145}.

The singular terms for $v>1$
in eqs.~(\ref{Evgt1_0}) through (\ref{Evgt1_3}),
and in \eqns{Evgt1_4}{Evgt1_5},
are identical to those for the $v<0$ branch.
The finite terms are even simpler, thanks to the choice of $\hat{z}$ argument.
The binary vectors for $i$ in $h_i^{[w]}$ still start with
a ``1'' as a consequence of the final entry condition, because
$d\hat{z}/dv \times 1/(1-\hat{z}) = 1/[v(1-v)]$.
However, now there are only even values of $i$ for the non-$\zeta$
parts of both the imaginary and the real parts of ${\cal E}_{3\to3}(v>1)$.
An odd value of $i$ would correspond to a branch cut at $v=1$;
however, appendix~\ref{sckin} shows that this is an unremarkable
scattering configuration, and the amplitude is totally smooth there.


\subsection{$v\to\infty$ and $v\to1$ limits}

From the formulas~(\ref{Evgt1_0}) to (\ref{Evgt1_3}), (\ref{Evgt1_4})
and (\ref{Evgt1_5}), it is straightforward to take the limit $v\to\infty$,
which just amounts to setting all the $h_i^{[w]}$ to zero, since they all
vanish at $\hat{z}=1/v=0$.  Equivalently, one can set all the
$h_i^{[w]}$ to zero in the previous formulas for $v<0$,
since $z=1/(1-v)=0$ in this limit too.
Through five loops, one gets the same result as one approaches
$v=\infty$ from either the positive or the negative side, and the results
contain no $\ln v$ divergences:
\bea
{\cal E}_{3\to3}^{(0)}(v=\infty) &=& 1 \,, \label{Evinfty_0}\\
{\cal E}_{3\to3}^{(1)}(v=\infty) &=& \pi i \, \lnden \, + \, \zeta_2 \,,
\label{Evinfty_1}\\
{\cal E}_{3\to3}^{(2)}(v=\infty) &=& 2 \pi i \, \biggl[
  - \frac{1}{12} \lndene{3} \, + \, \frac{1}{2} \zeta_2 \lnden
 \, + \, 2 \zeta_3 \biggr]
  + 7 \zeta_4 \,, \label{Evinfty_2}\\
{\cal E}_{3\to3}^{(3)}(v=\infty) &=& 2 \pi i \, \biggl[ \frac{1}{80} \lndene{5}
 \, - \, \frac{1}{4} \zeta_3 \lndene{2}
 \, + \, \frac{1}{4} \zeta_4 \lnden 
 \, - \, \frac{21}{2} \zeta_5
 \, + \, \frac{9}{2} \zeta_2 \zeta_3 \biggr] \nn\\
&&\hskip0cm\null
   - \frac{3787}{96} \zeta_6
\, + \, \frac{5}{4} (\zeta_3)^2 \,, \label{Evinfty_3}\\
{\cal E}_{3\to3}^{(4)}(v=\infty) &=& 2 \pi i \, \biggl[
  - \frac{1}{672} \lndene{7}
  \, - \, \frac{1}{80} \zeta_2 \lndene{5}
  \, - \, \frac{1}{48} \zeta_3 \lndene{4}
  \, - \, \frac{7}{24} \zeta_4 \lndene{3} \nn\\
&&\hskip0.7cm\null
  \, + \, \frac{1}{4} ( 4 \zeta_5 - 3 \zeta_2 \zeta_3 ) \lndene{2} 
  \, - \, \frac{1}{48} \Bigl( 13 \zeta_6 + 48 (\zeta_3)^2 \Bigr) \lnden \nn\\
&&\hskip0.7cm\null
  + \frac{1141}{16} \zeta_7 - \frac{119}{4} \zeta_2 \zeta_5
  - \frac{17}{4} \zeta_3 \zeta_4  \biggr] \nn\\
&&\hskip0.0cm\null
  + \frac{5}{4} \zeta_{5,3} + \frac{56911}{144} \zeta_8
  + 18 \zeta_2 (\zeta_3)^2  - \frac{63}{4} \zeta_3 \zeta_5 
\,, \label{Evinfty_4}
\eea
\bea
{\cal E}_{3\to3}^{(5)}(v=\infty) &=& 2 \pi i \, \biggl[ 
  \frac{1}{6912} \lndene{9}
  \, + \, \frac{1}{336} \zeta_2 \lndene{7}
  \, + \, \frac{5}{288} \zeta_3 \lndene{6}
  \, + \, \frac{9}{80} \zeta_4 \lndene{5} \nn\\
&&\hskip0.7cm\null
  \, + \, \frac{1}{24} ( 6 \zeta_5 + 7 \zeta_2 \zeta_3 ) \lndene{4}
  \, + \, \frac{1}{72} \Bigl( 115 \zeta_6 + 48 (\zeta_3)^2 \Bigr) \lndene{3}
\nn\\ &&\hskip0.7cm\null
  \, + \, \frac{1}{16} \Bigl( - 55 \zeta_7 + 68 \zeta_2 \zeta_5
                        + 44 \zeta_3 \zeta_4 \Bigr) \lndene{2} \nn\\
&&\hskip0.7cm\null
  + \frac{1}{72} \Bigl( 257 \zeta_8 + 18 \zeta_2 (\zeta_3)^2
                      + 810 \zeta_3 \zeta_5 \Bigr) \lnden
\nn\\
&&\hskip0.7cm\null
- \frac{40369}{64} \zeta_9 + \frac{7645}{32} \zeta_2 \zeta_7
+ \frac{3119}{64} \zeta_3 \zeta_6
+ \frac{2295}{32} \zeta_4 \zeta_5 - \frac{15}{4} (\zeta_3)^3 \biggr] \nn\\
&&\hskip0cm\null
- \frac{177}{28} \zeta_{7,3}
+ \frac{1217}{40} \zeta_2 \zeta_{5,3}
- \frac{2668732849}{537600} \zeta_{10}
- \frac{2659}{32} \zeta_4 (\zeta_3)^2
\nn\\&&\hskip0cm\null
- \frac{3091}{8} \zeta_2 \zeta_3 \zeta_5
+ \frac{9179}{64} \zeta_3 \zeta_7 + \frac{20553}{224} (\zeta_5)^2 \,.
\label{Evinfty_5}
\eea
Multiple zeta values (MZV's) begin to appear at four loops:
\be
\zeta_{5,3} = \sum_{n_1>n_2>0} \frac{1}{n_1^5 n_2^3} = 0.0377076729848\ldots,
\qquad
\zeta_{7,3} = \sum_{n_1>n_2>0} \frac{1}{n_1^7 n_2^3} = 0.0084196685030\ldots.
\label{definez53z73}
\ee

Even though there is a branch cut at $v=\infty$, the residue vanishes
there, and we get the same limiting behavior as $v$ becomes large
for either sign of $v$.  The nonsingularly framed Wilson loop
${\cal W}^{\rm ns}$ has a very similar behavior to ${\cal E}$ in
this limit, because $L^2\to0$ as $v\to\infty$, so the
exponential factor in \eqn{EfromW} simply approaches
$\exp[\frac{1}{2}\zeta_2\gK]$.

The limit of ${\cal E}_{3\to3}(v)$ as $v\to1$ from above is also
smooth in $v$.  In this case the $h_i^{[w]}$ have to be evaluated
at $\hat{z}=1/v=1$, where they are given by multiple zeta values.
Since the $\lnden$ terms are exactly the same as for other $3\to3$
values of $v$, here we just present the finite (non-$\lnden$) terms:
\bea
{\cal E}_{3\to3}^{(0),{\rm fin}}(v\to1^+) &=& 1 \,, \label{Eveq1p_0}\\
{\cal E}_{3\to3}^{(1),{\rm fin}}(v\to1^+) &=& 0 \,, \label{Eveq1p_1}\\
{\cal E}_{3\to3}^{(2),{\rm fin}}(v\to1^+) &=&
2 \pi i \zeta_3 - \frac{5}{2} \zeta_4 \,, \label{Eveq1p_2}\\
{\cal E}_{3\to3}^{(3),{\rm fin}}(v\to1^+) &=&
2 \pi i \Bigl[ - 4 \zeta_5 + 2 \zeta_2 \zeta_3 \Bigr]
+ \frac{35}{24} \zeta_6 + (\zeta_3)^2 \,, \label{Eveq1p_3}\\
{\cal E}_{3\to3}^{(4),{\rm fin}}(v\to1^+) &=&
2 \pi i \biggl[ \frac{39}{2} \zeta_7
  - \frac{19}{2} \zeta_2 \zeta_5 + \frac{5}{4} \zeta_3 \zeta_4 \biggr]
+ \frac{3}{2} \zeta_{5,3} - \frac{77}{48} \zeta_8
- \frac{5}{2} \zeta_3 \zeta_5
+ \frac{15}{2} \zeta_2 (\zeta_3)^2 
\,,~~~~ \label{Eveq1p_4}\\
{\cal E}_{3\to3}^{(5),{\rm fin}}(v\to1^+) &=&
2 \pi i \biggl[
  - \frac{857}{8} \zeta_9 + \frac{205}{4} \zeta_2 \zeta_7
  - \frac{299}{48} \zeta_3 \zeta_6
  - \frac{45}{4} \zeta_4 \zeta_5 - \frac{1}{2} (\zeta_3)^3 \biggr]
\nn\\  &&\hskip0.0cm\null
  - 6 \zeta_{7,3} + \frac{15}{2} \zeta_2 \zeta_{5,3}
  - \frac{3961}{96} \zeta_{10} + \frac{29}{2} \zeta_4 (\zeta_3)^2
  - 60 \zeta_2 \zeta_3 \zeta_5
\nn\\  &&\hskip0.0cm\null
  - \frac{63}{2} \zeta_3 \zeta_7
  - \frac{111}{8} (\zeta_5)^2 
\,. \label{Eveq1p_5}
\eea
For what it's worth,
the rational numbers multiplying the $\zeta$ values in these
equations seem to be quite a bit simpler than in most other limits
of these functions.

One can also obtain the results in the limit of $2\to4$ kinematics
as $v\to1^-$ by analytic continuation.  In general, one should
let $\lnden \to \ln\delta - i\pi$ and $\ln(v-1) \to \ln(1-v) + i \pi$,
and then complex conjugate the resulting expression.
For ${\cal E}$ the step $\ln(v-1) \to \ln(1-v) + i \pi$
can be omitted since there is no $\ln(v-1)$ singularity.
These results obtained in this way agree with those
obtained by taking the limit $v\to1^-$ directly from expressions
(not shown) for ${\cal E}_{2\to4}(v)$.

In contrast to the smooth behavior of ${\cal E}$ as $v\to1$,
the limit of the nonsingularly-framed Wilson loop
${\cal W}^{\rm ns}_{3\to3}(v)$ is divergent as $v\to1^+$;
the divergence is simply due to the factor in \eqn{EfromW}
of $\exp[-\gK L^2/8] \approx \exp[-\gK \ln^2(v-1)/8]$.
If one approaches $v=1$ from the $2\to4$ side, i.e.~$v\to1^-$,
then there are additional phases in the divergence
due to the analytic continuation $\ln(v-1) \to \ln(1-v) + i \pi$.


\section{Evolving framed self-crossing Wilson loops}
\label{evolveWL}

\subsection{A simple evolution equation}

Framing the hexagon Wilson loop removes its cusp singularities
and makes ${\cal W}^{\rm ns}$ an ultraviolet (UV) finite,
dual conformal invariant function ${\cal W}^{\rm ns}(u,v,w)$.
Denote the non-framed hexagonal Wilson loop by $W_{\text{hex}} \equiv W_6$,
and denote the combination of two pentagon and one box Wilson loops
used to do the non-singular framing by $W_{\rm f}$
(the subscript ``f'' refers to the framing), so that
\be
{\cal W}^{\rm ns} \ =\ W_{\text{hex}}\times W_{\rm f}. \label{eq:fram_fns}
\ee
Note that because of UV divergences, neither $W_{\text{hex}}$ nor $W_{\rm f}$
are purely functions of the cross ratios $u,v,w$,
but their product in \eqn{eq:fram_fns} is.

Consider the self-crossing limit, in which edges $k_3$ and $k_6$ cross
(see~\fig{fig:sc_reg}). In terms of dual conformal variables
the limit is $(u,v,w) \to (1,v,v)$. We approach the limit
in the following way: First move onto the plane $w=v$.
${\cal W}^{\rm ns}$ does not acquire any divergences in this step.
Then approach the line $u=1$ by taking $u=1-\delta$:
\be
{\cal W}^{\rm ns}(u,v,w)\ \rightarrow\ {\cal W}^{\rm ns}(u,v,v)
\ \rightarrow\ {\cal W}^{\rm ns}(1-\delta,v,v).
\ee
We choose the non-singular framing $W_{\rm f}$ because the new light-like
lines in the pentagons and the box do not go near the self-crossing point.
Thus the framing Wilson loops do not acquire any additional singular behavior
in the self-crossing limit (in contrast to the singular framing defining
${\cal W}^{\rm s}$). 

Moreover, since the job of $W_{\rm f}$ is to remove cusp divergences,
any divergences in $\delta$ for ${\cal W}^{\rm ns}$ are purely due
to $W_{\text{hex}}$.  We expect these divergences to be governed by an
evolution equation of the form
\be
\frac{d}{d\ln\delta}{\cal W}^{\rm ns}(1-\delta,v,v)
\ =\ \mathcal{F}(\delta, v) \label{eq:evol_eq_gen},
\ee
for some function $\mathcal{F}$.

\begin{figure}[ht]
\begin{center}
\includegraphics[width=3.5 in]{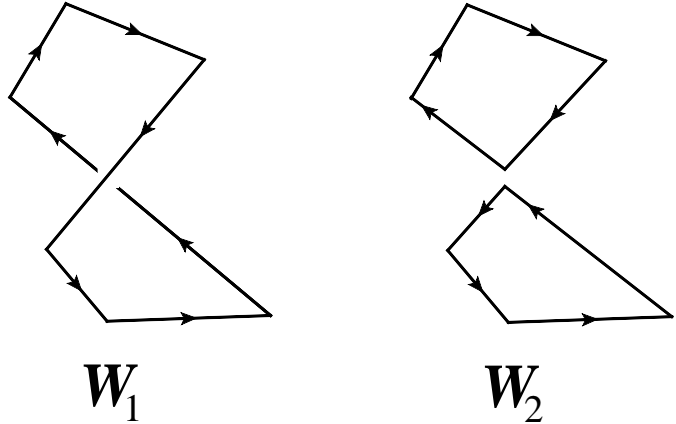}
\end{center}
\caption{Wilson loops $W_1$ and $W_2$ mix under renormalization.}
\label{fig:w1w2}
\end{figure}

To find an equation of the form~(\ref{eq:evol_eq_gen}) we follow the approach
of Korchemskaya and Korchemsky~\cite{Korchemskaya1994qp}, who studied
the crossing of two infinite Wilson lines, not necessarily light-like.
Consider two auxiliary Wilson loops $W_1$ and $W_2$,
depicted in \fig{fig:w1w2}. The only difference between $W_1$ and
$W_{\text{hex}}$ is that $W_1$ is evaluated strictly in the
self-crossing kinematics; hence $W_1$ is formally infinite and must be
renormalized.  $W_{\text{hex}}$ is finite for $\vec{z}\neq 0$
and is equal to $W_1$ for $\vec{z}=0$.
The second Wilson loop in \fig{fig:w1w2}, $W_2$, is the same
as $W_1$, except that the routing of the lines at the self-crossing
point is exchanged so that the contour forms two boxes instead
of a hexagon. It is also evaluated with the two vertices on top
of each other, after renormalization.
The renormalized quantities $W_1^r$ and $W_2^r$ are finite,
non-trivial functions of the renormalization scale $\mu$.
Under a change of renormalization scale, they
mix with each other~\cite{BrandtNeriSato}.
The functional dependence of $W_1^r$ on $\mu^2$ should be the same as the
functional dependence of $W_{\text{hex}}$ on the inverse separation
$1/\vec{z}^2$.  We will find a renormalization group (RG) equation for
$W_1^r(\mu^2)$ and then set $\mu^2=1/\vec{z}^2$ in order to obtain an
RG equation for $W_{\text{hex}}$.  This will be used, in turn, to obtain a
differential equation for ${\cal W}^{\rm ns}$.
Rewriting $\vec{z}^2$ in terms of $\delta$ will
give an equation of the form~(\ref{eq:evol_eq_gen}).

The RG equations mixing the renormalized Wilson loops $W^r_1$ and $W^r_2$
for a general theory are~\cite{BrandtNeriSato,Korchemskaya1994qp}
\be
\left(\mu\frac{\partial}{\partial\mu}
+\beta(g)\frac{\partial}{\partial g}\right) W^r_i
\ =\ - \Gamma^{ij}(\gamma,g)W^r_j - \sum_m \Gamma_{\text{cusp}}(\gamma_m,g)W_i^r
\,. \label{eq:gen_RG_eqn}
\ee
Here $\gamma$ is the crossing angle, $m$ labels the cusps of $W_1^r$,
and $\gamma_m$ are the corresponding cusp angles.
The cusp anomalous dimension for finite cusp angle $\gamma_m$
is denoted by $\Gamma_{\text{cusp}}(\gamma_m,g)$,
and $\Gamma^{ij}(\gamma,g)$ is the cross anomalous dimension matrix.

\Eqn{eq:gen_RG_eqn} holds for massive Wilson lines, for which all cusp angles
and the angle at the self-crossing point are finite.
Below we give a modified relation for the massless case.
In the subsequent equations we imagine keeping the Wilson lines massive
and taking the massless limit at the end.  Finally, in $\NeqFour$ SYM,
the beta function vanishes and \eqn{eq:gen_RG_eqn} simplifies to
\be
\mu\frac{\partial}{\partial\mu} W^r_i
\ =\ -\Gamma^{ij}(\gamma,g)W^r_j - \sum_m \Gamma_{\text{cusp}}(\gamma_m,g)W_i^r
\,. \label{eq:N=4_RG_eqn}
\ee
Let $\mathcal{W}_i^r\equiv W_i^r\times W_{\rm f}$.  As discussed above, the
(renormalized)
framing function $W_{\rm f}$ removes all six cusp divergences in
the massive case where $\gamma$ is finite.  This is true even
for $\mathcal{W}_2^r$, even though $W_{\rm f}$ is the framing
function for the hexagon and $W_2^r$ is the renormalized
Wilson loops for two boxes, because the six cusp angles match.
(In our terminology, we do not count the two cusps in $\mathcal{W}_2^r$
at the self-crossing point as cusps; we handle them separately.)
In the massless limit, where $\gamma_m$
and $\gamma$ go to infinity, things are more subtle.

Taking into account the removal of the cusp divergences when
passing from $W_i^r$ to $\mathcal{W}_i^r$, we obtain
\be
\mu\frac{\partial}{\partial\mu}\mathcal{W}_i^r
\ =\ -\Gamma^{ij}(\gamma,g)\mathcal{W}_j^r.
\label{framedRGeqn}
\ee
The components $\Gamma^{ij}(\gamma,g)$ in the large $N_c$ limit of QCD
and in the limit of large $\gamma$ can be found in
ref.~\cite{Korchemskaya1994qp}.  An important point is that,
after inserting an appropriate factor of $N_c$ into the normalization
of ${\cal W}_2^r$,
they take an upper triangular form in the large $N_c$ limit~\cite{Georgiou}.
The reason is that for a gluon to contribute to the evolution of ${\cal W}_1^r$
it should be exchanged between the two self-crossing lines, which produces
an extra factor of $N_c$, allowing ${\cal W}_1^r$ to mix into ${\cal W}_2^r$;
whereas for a gluon to cause evolution of ${\cal W}_2^r$ into ${\cal W}_1^r$,
its exchange between the two boxes in \fig{fig:w1w2}
is color suppressed, by a factor of $1/N_c$.
Since $\Gamma^{11} = \Gamma^{21} = 0$ at large $N_c$,
the evolution of both ${\cal W}_1^r$ and ${\cal W}_2^r$
is governed by ${\cal W}_2^r$, through the matrix elements
$\Gamma^{12}$ and $\Gamma^{22}$.  Furthermore,
$\Gamma^{22}$ is proportional to the cusp anomalous dimension.

\subsection{The large $\gamma$ limit}

In the limit of large $\gamma$, ref.~\cite{Korchemskaya1994qp}
finds that $\Gamma^{12}$ behaves like a constant times $i\pi$ in QCD,
while the leading behavior of the cusp anomalous dimension
is proportional to $\gamma \times \gK$, where $\gK$ is the light-like
cusp anomalous dimension.
We assume a similar form for the matrix elements holds here
(see also the discussion in ref.~\cite{Georgiou}),
\bea
\Gamma^{12} &\to&  - i\pi \, \Gamma_1(a), \label{crosslargegamma}\\
\Gamma^{22} &\to& \gamma \frac{\gK(a)}{2}
\,, \label{cusplargegamma}
\eea
as $\gamma\to\infty$, where we converted to our normalization
of the cusp anomalous dimension and perturbative
expansion parameter $a = g_{\textrm{YM}}^2 N_c/(8\pi^2)$.
Inserting these values into \eqn{framedRGeqn}, we find
\bea
\mu\frac{\partial}{\partial\mu}\mathcal{W}_1^r
&=& i\pi\Gamma_1(a)\mathcal{W}_2^r \,, \label{eq:W1_fram}\\
\mu\frac{\partial}{\partial\mu}\mathcal{W}_2^r
&=& -\gamma \frac{\gK(a)}{2}\mathcal{W}_2^r \,. \label{eq:W2_fram}
\eea

In the massless limit $\gamma\rightarrow\infty$, \eqn{eq:W2_fram}
is not defined. We can attempt to get into this limit using the method
of refs.~\cite{Korchemskaya1992je,Georgiou}.  Taking legs 3 and 6
to be massive initially, with masses $k_3^2$ and $k_6^2$, we write
\eqn{eq:W2_fram} as
\be
\mu\frac{\partial}{\partial\mu}\ln\mathcal{W}_2^r
\ \stackrel{?}{=}\ -\gamma \frac{\gK(a)}{2} \,, \label{eq:lnW2_fram}
\ee
where
\be
\gamma\ =\ \ln\biggl(\frac{s_{36}}{\sqrt{k_3^2}\sqrt{k_6^2}}\biggr) \,.
\ee
Now differentiate \eqn{eq:lnW2_fram} with respect to $s_{36}$,
and integrate back up to obtain
\be
\mu\frac{\partial}{\partial\mu}\mathcal{W}^r_2
\ \stackrel{?}{=}
\ \left(-\frac{\gK(a)}{2}\log(\mu^2s_{36})-\bar{\Gamma}(a)\right)
\mathcal{W}_2^r \,, 
\label{eq:massless_W2_diffeq}
\ee
where $\bar{\Gamma}(a)$ is an integration constant.
Notice the appearance of the non-dual-conformal quantity
$s_{36}$.  This signals a problem with the large $\gamma$ limit
for \eqn{eq:W2_fram}.  If we ignore this problem,
we can trade $\mu^2$ for $1/\vec{z}^2 \to [\delta s_{36}xy(1-x)(1-y)]^{-1}$
by the prescription described in \eqn{eq:del_eps}.
Dropping the non-dual-conformal factor $xy(1-x)(1-y)$,
\eqn{eq:massless_W2_diffeq} becomes, in $3\to3$ kinematics,
\be
\frac{d}{d\lnden}\ln{\cal W}_2\ \stackrel{?}{=}\ -\frac{\gK(a)}{4} \lnden
\, + \, \frac{\bar\Gamma}{2} \,,
\ee
which integrates to
\be
{\cal W}_2(\delta,v)\ \stackrel{?}{=}\
f(v) \, \exp\Bigl[ -\frac{\gK(a)}{8} \lndene{2}
  \, + \, \frac{\bar\Gamma(a)}{2} \lnden \Bigr] \,,
\label{W2fv}
\ee
for some function $f(v)$ of $v$ alone.  However, we will see below
that such a solution is inconsistent with explicit results. 

More important for our purposes is \eqn{eq:W1_fram}, because
it contains no explicit $\gamma$, so we expect its large $\gamma$ limit
to be reliable.  Again trading $\mu^2$ for $1/\delta$,
\eqn{eq:W1_fram} becomes
\be
\frac{1}{2\pi i}\frac{d}{d\ln\delta}{\cal W}^{\rm ns}(1-\delta,v,v)
\ =\ -\frac{1}{4} \, \Gamma_1(a) \, \mathcal{W}_2(\delta,v) \,.
\label{eq:W_hex_diff_eq}
\ee
We have changed notation
$\mathcal{W}_2^r\rightarrow{\cal W}_2$,
$\mathcal{W}_1^r\rightarrow{\cal W}^{\rm ns}$,
to emphasize that, here and below,
the renormalization scale $\mu$ has been exchanged for
$\delta$ using \eqn{eq:del_eps}, so that we now deal with finite functions
(for non-zero $\delta$) of the dual conformal variable $v$.


\subsection{The framed double box}
\label{frameddoublebox}

In order to separate the various functional dependences of ${\cal W}^{\rm ns}$,
it will be useful to rewrite $\mathcal{W}_2$ as
$\mathcal{W}_2 = W_2 \times W_{\rm f}
= W_2 \times W_{\rm f}^{\rm s}\times (W_{\rm f}/W_{\rm f}^{\rm s})$,
where $W_{\rm f}^{\rm s}$ is a new framing function, corresponding to the
singular framing in~\fig{fig:framing_s}, for which
we defined ${\cal W}^{\rm \, s} = W_{\text{hex}}\times W_{\rm f}^{\rm s}$
back in \sect{framedWL}.  (The superscript ``s'' reminds us that
this framing function is singular in the self-crossing kinematics.)

The reason for rewriting things in this way is that the framing function
$W_{\rm f}$ involves pentagons that straddle both sides of the self-crossing
geometry.  Thus they have sensitivity to the global geometry of the hexagon,
and can carry non-trivial $v$ dependence (although $W_{\rm f}$ is not dual
conformally invariant separately from $W_{\text{hex}}$).
On the other hand, the framing corresponding to $W_{\rm f}^{\rm s}$
is so degenerate
in the self-crossing limit, consisting essentially only of boxes on one
side of the crossing point or the other,  that it cannot depend on $v$
(see \fig{fig:framing_s_sc}).  For $3\to3$ kinematics,
using the exact results in \cite{DHKS} for four- and five-sided Wilson loops,
or equivalently by comparing \eqns{EfromW}{EfromWs} for
the different hexagon framings, we can relate the two framing functions:
\be
\frac{W_{\rm f}}{W_{\rm f}^{\rm s}}
\ =\ \frac{{\cal W}_2}{{\cal W}_2^{\rm \, s}}
\ =\ \exp\Bigl[ - \frac{\gK(a)}{8} \Bigl( \lndene{2} - L^2 \Bigr) \Bigr] \,,
\label{WfoverWfs}
\ee
where $L=\ln(1-1/v)$.

The evolution equation~(\ref{eq:W_hex_diff_eq}) becomes, in $3\to3$
kinematics,
\be
\frac{1}{2\pi i}\frac{d}{d\ln\delta}{\cal W}^{\rm ns}_{3\to3}(1-\delta,v,v)
\ =\ - \frac{1}{4} \, \Gamma_1(a)
\, \exp\Bigl[ - \frac{\gK(a)}{8} \Bigl( \lndene{2} - L^2 \Bigr) \Bigr]
 \, \mathcal{W}_2^{\rm \, s}(\delta) \,,
\label{eq:evol_eqn_kin_factor}
\ee
where ${\cal W}^{\rm \, s}_2 \equiv W_2 \times W_{\rm f}^{\rm s}$ is dual conformally
invariant, but too singular to depend on $v$, so we write
it as a function only of $\delta$.  Thus all non-trivial kinematic
dependence has been factored out in the exponential pre-factor
of \eqn{eq:evol_eqn_kin_factor}.

\begin{figure}[ht]
\begin{center}
\includegraphics[width=3.2 in]{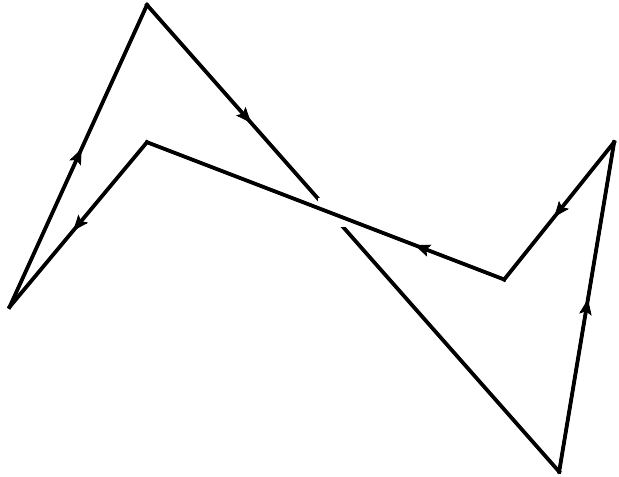}
\end{center}
\caption{The two box Wilson loops that the hexagon mixes into in $3\to3$
  kinematics both have Euclidean kinematics.}
\label{fig:eucl33}
\end{figure}

We can also use \eqn{EfromW}
to convert \eqn{eq:evol_eqn_kin_factor} to one for ${\cal E}$
in $3\to3$ kinematics,
since the relative factor between them is independent of $\delta$,
\be
\frac{1}{2\pi i}\frac{d}{d\ln\delta}{\cal E}_{3\to3}(1-\delta,v,v)
\ =\ - \frac{1}{4} \, \Gamma_1(a)
\, \exp\Bigl[ - \frac{\gK(a)}{8} \Bigl( \lndene{2} - 4\zeta_2 \Bigr) \Bigr]
 \, \mathcal{W}_2^{\rm \, s}(\delta) \,.
\label{eq:Eevol_eqn}
\ee
This equation explains the $v$ independence of the singular terms
in ${\cal E}_{3\to3}$ from 1 to 5 loops.
It also explains why the singularities are purely in the
imaginary part:   The hexagon $3\to3$ kinematics we studied
alternate between incoming and outgoing momenta, as one goes cyclically
around the loop, as shown in \fig{fig:show_24_33}(b)
and also in~\fig{fig:eucl33}.  The two boxes that this hexagon
factorizes into have Euclidean kinematics, in the sense that the
invariants connecting adjacent legs (the so-called $s$ and $t$ channels
for a scattering process) are both space-like. The only time-like invariant
is in the $u$ channel involving diagonally opposite legs, but this invariant
cannot produce cuts in the large $N_c$ limit, and so the box Wilson loops
must be real.  In other words, a factor of $\pi i$ comes from the cross
anomalous dimension, and the remaining factors in \eqn{eq:Eevol_eqn}
are real.

Suppose we took \eqn{W2fv} for the $\de$-dependence of
${\cal W}_2(\delta,v)$ seriously.  Then we could combine it with
\eqn{WfoverWfs} and the observation that ${\cal W}_2^{\rm \, s}$
is independent of $v$ to fix $f(v)$ and conclude that
\be
{\cal W}_2^{\rm \, s}(\de)\ =\
 C(a) \, \exp\Bigl[ \frac{\bar\Gamma(a)}{2} \lnden \Bigr] \,
\label{W2sbusted}
\ee
in $3\to3$ kinematics, where $C(a)$ is independent of both
$v$ and $\de$.  Then we would find that
\eqn{eq:evol_eqn_kin_factor} becomes
\be
\frac{1}{2\pi i}\frac{d}{d\ln\delta}{\cal W}^{\rm ns}_{3\to3}(1-\delta,v,v)
\ \stackrel{?}{=}\ - \frac{1}{4} \, C(a) \Gamma_1(a)
\, \exp\Bigl[ - \frac{\gK(a)}{8} \Bigl( \lndene{2} - L^2 \Bigr)
  + \frac{\bar\Gamma(a)}{2} \lnden \Bigr]
\,.
\label{eq:evol_eqn_wrong}
\ee
In \sect{FinalResults} we will find that this expression
is incompatible with explicit results, beginning at order $a^4$
and $\lndene{3}$.  The leading-logarithmic terms in \eqn{eq:evol_eqn_wrong}
(terms of order $a^L \lndene{2L-2}$) do have the correct form,
but that is a tiny part of the available data.

It might be possible to compute the singularly-framed Wilson loop
${\cal W}_2^{\rm \, s}(\de)$ directly, but we will not attempt to do so here.
Instead, in the next section,
we will use a matching between the self-crossing limit at
small $v$ and a corner of the multi-Regge limit, in order to compute
the left-hand side of \eqn{eq:evol_eqn_kin_factor}.  We leave it to
future work to evaluate ${\cal W}_2^{\rm \, s}(\de)$ directly.


\section{Matching the self-crossing and multi-Regge limits}
\label{SCMRKmatch}

We now understand the $v$-dependence of the singular terms
in the MHV amplitude
and associated Wilson loops in the self-crossing limit.
We wish to use this information to evaluate the $\ln\de$ terms to high loop
orders for all values of $v$, using \eqn{eq:evol_eqn_kin_factor} and
an evaluation of ${\cal W}^{\rm ns}_{3\to3}(v)$ at some value of
$v$ to high orders.  A convenient place to do this is for $v$
near zero, where it overlaps the multi-Regge limit.

The multi-Regge limit of six-gluon scattering in planar $\NeqFour$
SYM depends on a singular parameter $\delta$ and on a complex
parameter, conventionally called $w$.  (In order to minimize confusion
between this $w$ and the cross ratio $w$ for generic kinematics,
in this section we call the three cross ratios $(u_1,u_2,u_3)$.)
The purpose of this section is first to identify the $w\to-1$ limit
within multi-Regge kinematics with the $v\to0$ limit within
the self-crossing configuration.  Then we evaluate the all-orders MRK
formulae~\cite{BCHS} in this limit, in order to provide an expression for
the self-crossing configuration to high loop orders.

The MRK limit is defined in $2\to4$ kinematics by the analytic continuation
$u_1 \to u_1 e^{-2\pi i}$ from the Euclidean region, followed by letting
\be
u_1 = 1 - \de \,, \qquad u_2 = \frac{\de}{|1+w|^2} \,,
\qquad u_3 = \frac{\de\,|w|^2}{|1+w|^2} \,,
\label{MRK24ui}
\ee
and taking $\delta \to 0$ with $\delta$ positive.
In $3\to3$ kinematics, we first continue $u_1 \to u_1 e^{+2\pi i}$, 
$u_2\to u_2 e^{\pi i}$, $u_3\to u_3 e^{\pi i}$ from the Euclidean region,
and then take the same limit~(\ref{MRK24ui})  but with $\delta$ negative,
$\delta = -|\delta|$.

The self-crossing limit, in either $2\to4$ or
$3\to3$ kinematics, is defined by
the same analytic continuation as in the MRK case, followed by
\be
u_1 = 1 - \de \,, \qquad u_2 = v \,, \qquad u_3 = v \,.
\label{sc24ui}
\ee
By comparing \eqns{MRK24ui}{sc24ui}, we see that the overlap
region is the limit $w\to-1$ of MRK, and the limit $v\to0$ of
self-crossing kinematics.  The key relation for passing between
the two limits is
\be
v\ =\ \frac{\de}{|1+w|^2} \,, \qquad \ln v\ =\ \ln\delta - \ln|1+w|^2
\label{scMRK24relation}
\ee
for $2\to4$ kinematics, and
\be
|v|\ =\ \frac{|\de|}{|1+w|^2} \,, \qquad \ln|v|\ =\ \lnden \, - \, \ln|1+w|^2
\label{scMRK33relation}
\ee
for $3\to3$ kinematics.


\subsection{Self-crossing-MRK limit of the Wilson loop in
  $3\to3$ kinematics}

The behavior of the remainder function $R_6$ in the $2\to4$ MRK limit
is~\cite{Fadin2011we,CaronHuot2013fea}:
\bea
\exp\Bigl[ R_6 + i\pi\delta_{\textrm{MRK}} \Bigr] \Big|_{\textrm{MRK},\,2\to4}\!\!
&=& \cos\pi\omega_{ab} 
+ i  \frac{a}{2} \sum_{n=-\infty}^\infty
(-1)^n\left(\frac{w}{\ws}\right)^{\frac{n}{2}}
\int_{-\infty}^{+\infty}
\frac{d\nu}{\nu^2+\frac{n^2}{4}}|w|^{2i\nu}
\, \Phi_{\textrm{Reg}}(\nu,n) \nn\\
&&\hskip5.0cm\null
\times\left(-\frac{1}{1-u_1}\frac{|1+w|^2}{|w|}\right)^{\omega(\nu,n)}
,
\label{MRK24}
\eea
where
\bea
\omega_{ab} &=& \frac{1}{8}\,\gK(a)\,\ln|w|^2\,,\\
\delta_{\textrm{MRK}} &=& \frac{1}{8}\,\gK(a)\,\ln\frac{|w|^2}{|1+w|^4}\,,
\eea
and $\gK(a)$ is the cusp anomalous dimension.

The behavior in the $3\to3$ kinematic region is obtained
by letting $\ln(1-u_1)\to\ln(u_1-1)-i\pi = \lnden \, - i\pi$
and performing a complex conjugation~\cite{Bartels2010tx}:
\bea
\exp\Bigl[ R_6 - i\pi\delta_{\textrm{MRK}} \Bigr] \Big|_{\textrm{MRK},\,3\to3}\!\!
&=& \cos\pi\omega_{ab} 
- i  \frac{a}{2} \sum_{n=-\infty}^\infty
(-1)^n\left(\frac{w}{\ws}\right)^{\frac{n}{2}}
\int_{-\infty}^{+\infty}
\frac{d\nu}{\nu^2+\frac{n^2}{4}}|w|^{2i\nu}
\, \Phi_{\textrm{Reg}}(\nu,n) \nn\\
&&\hskip6.5cm\null
\times\left(\frac{|1+w|^2}{|\delta| \, |w|}\right)^{\omega(\nu,n)}
.
\label{MRK33}
\eea
In particular, there is no longer a phase in the last factor, so the
real part in $3\to3$ kinematics is trivial,
\be
{\rm Re} \, \exp\Bigl[ R_6 - i\pi\delta_{\textrm{MRK}} \Bigr]
        \Big|_{\textrm{MRK},\,3\to3}
\ =\ \cos\pi\omega_{ab} \,.
\label{ReR633}
\ee

According to \eqn{WfromR},
the nonsingularly-framed Wilson loop is related to $\exp[R_6]$
by multiplying by $\exp[\tfrac{\gK}{8} \, X]$.
We can use \eqn{Xns33} to evaluate $X$ in the self-crossing limit as
$v\to0$ (from the negative side):
\be
X^{\rm ns}_{3\to3}|_{v\to0}
\ =\  2\pi i \ln\Bigl( \frac{|\de|}{|v|} \Bigr)
\ =\ 2\pi i \ln|1+w|^2 \,.
\label{Xns33vto0}
\ee
Notice that as $w\to-1$, $\omega_{ab}\to0$, while
the phase $\delta_{\textrm{MRK}}$ becomes $-\gK/4 \times \ln|1+w|^2$,
so that 
\be
\exp[-i\pi\delta_{\textrm{MRK}}]
\ \to\ \exp\Bigl[ \frac{\gK}{8} X^{\rm ns}_{3\to3}|_{v\to0} \Bigr] \,.
\ee
Because these phases coincide, the $v\to0$ limit of the $3\to3$
self-crossing configuration of the non-singularly-framed Wilson loop
is purely imaginary --- apart from the trivial term `1' arising from
the limit of $\cos\pi\omega_{ab}$.
It is given by the $(w,w^*)\to-1$ limit of \eqn{MRK33}:
\be
{\cal W}^{\rm ns}_{3\to3}(v\to0)\ =\
1 \, - \, i  \frac{a}{2} \sum_{n=-\infty}^\infty
\left(\frac{w}{\ws}\right)^{\frac{n}{2}}
(-1)^n \int_{-\infty}^{+\infty}
\frac{d\nu}{\nu^2+\frac{n^2}{4}}|w|^{2i\nu}
\, \Phi_{\textrm{Reg}}(\nu,n) \, |v|^{-\omega(\nu,n)} \,.
\label{Wns33vto0}
\ee
Where $w$ and $w^*$ still appear in \eqn{Wns33vto0}, they
are needed to regularize the sum and integral.


\subsection{Evaluation of the Fourier-Mellin transform for $w\to-1$}

In this subsection we describe how to evaluate (most of) the terms in
\eqn{Wns33vto0} directly to high loop orders, in the limit $w\to-1$.
While this paper was being written, ref.~\cite{BroedelSprenger}
appeared, which provides an efficient evaluation of the MRK limit
to high loop orders for generic values of $(w,w^*)$ in terms of
SVHPLs~\cite{BrownSVHPLs}.  The two methods are complementary,
in the sense that SVHPLs are not required for the limit $w\to-1$,
so a fair amount of computational machinery can be bypassed.
On the other hand, we will see that our method misses a few terms,
for which we can use the results of ref.~\cite{BroedelSprenger}.

In ref.~\cite{BCHS}, the BFKL eigenvalue $\omega$
and impact factor $\Phi_{\textrm{Reg}}$ were computed to all orders
using integrability.  It is straightforward to obtain the perturbative
expansions of $\omega$ and $\Phi_{\textrm{Reg}}$ to high orders,
and insert them into \eqn{Wns33vto0}.
Actually, it is better to trade the integral over $\nu$ for an integral
over the rapidity $u$ which appears in ref.~\cite{BCHS}.
This saves one step because the functions $\omega$ and $\Phi_{\textrm{Reg}}$ are
initially defined in terms of $u$ rather than $\nu$.

At any fixed loop order,
$\omega$ and $\Phi_{\textrm{Reg}}$ (or alternatively the BFKL
measure) are polynomials in the function
$\psi(x) = d\ln\Gamma(x)/dx$ and its derivatives, where
$x=1\pm i u + n/2$, together with rational functions of $u$ and $n$
and Riemann zeta values.
To evaluate the integral over the real rapidity $u$, we deform
it into the complex plane, where it has an infinite sequence
of poles, at $iu=n/2+m$ for non-negative $m$.
We calculate the residues and then examine the behavior
of the double sum in $n$ and $m$.
We wish to pick up terms that contain at least one power each of
$\ln(1+w)$ and $\ln(1+w^*)$.
This divergent behavior as $(w,w^*)\to-1$ comes from the leading
behavior of the summand as $m,n\to\infty$.
We let $n=\hat{n}-m$, so that $\hat{n}\geq m$.
The residues are all functions involving
$\psi^{(p)}(m)$ or $\psi^{(p)}(1+\hat{n})$,
combined with rational terms.  However, the $\psi^{(p)}$ terms
for $p>0$ always give power-suppressed terms as $m,n\to\infty$.
Among the $\psi$ functions, we only need to keep $\psi(m)\approx \ln m$
and $\psi(1+\hat{n})\approx \ln \hat{n}$.

After dropping all power-suppressed terms, we find that the double
sum can be expressed solely in terms of 
\be
S_k(w,w^*)\ =\ \sum_{\hat{n}=1}^\infty \frac{(-w)^{\hat{n}}}{\hat{n}}
\sum_{m=1}^{\hat{n}-1} \frac{(-w^*)^m}{m} (\ln \hat{n} + \ln m + 2\gamma_E)^k \,,
\label{Sk}
\ee
for non-negative integers $k$,
where $\gamma_E$ is the Euler-Mascheroni constant, plus a similar
term with the roles of $\hat{n}$ and $m$ exchanged.
When we add the term with $\hat{n}\lr m$,
we obtain an unrestricted sum over $\hat{n}$ and $m$,
up to diagonal terms with $\hat{n}=m$, which can be dropped because
they are not divergent.
We then use the binomial theorem to decouple the two sums:
\be
S_k(w,w^*)\ =\
\sum_{p=0}^k \left({k \atop p}\right) \hat{S}_p(w) \hat{S}_{k-p}(w^*) \,,
\label{SintermsofShat}
\ee
where
\be
\hat{S}_p(w)\ =\ \sum_{n=1}^\infty \frac{(-w)^{n}}{n}
(\ln n + \gamma_E)^p \,.
\label{Shatp}
\ee

Next we extract the terms in \eqn{Shatp} containing
positive powers of $\ln(1+w)$, for generic values of $p$,
\be
\hat{S}_p(w)\ =\
\sum_{\ell=1}^{p+1} d_{p,\ell} \ln^\ell(1+w)\ +\ {\cal O}(\ln^0(1+w)),
\label{Shatpexp}
\ee
where $d_{p,\ell}$ are some constants.
We do this by taking the Mellin transform of both sides of \eqn{Shatpexp}.
Let $w=-z$, and consider
\bea
I_p &=& \int_0^1 dz \, z^{N-1} \hat{S}_p(-z) = \sum_{n=1}^\infty
\frac{(\ln n + \gamma_E)^p}{n} \int_0^1 dz \, z^{N+n-1}
\nn\\
&=& \sum_{n=1}^\infty \frac{(\ln n + \gamma_E)^p}{n(n+N)} \,.
\label{ShatMellin}
\eea
The large $N$ limit of the Mellin transform is sensitive only to
the $z\to1$ limit of the function being transformed.
It is straightforward to approximate the sum over $n$ in \eqn{ShatMellin}
by an integral, which can be evaluated in terms of classical polylogarithms.
Then the desired limit as $N\to\infty$ can be taken,
keeping only terms with positive powers of
$\ln \hat{N} \equiv \ln N + \gamma_E$ multiplying $1/N$.
The first few values are
\bea
I_0 &=& \frac{1}{N} \ln\hat{N} \,, \nn\\
I_1 &=& \frac{1}{N} \biggl[ \frac{1}{2} \ln^2\hat{N} \biggr] \,, \nn\\
I_2 &=& \frac{1}{N} \biggl[ \frac{1}{3} \ln^3\hat{N} + 2 \zeta_2 \ln\hat{N}
  \biggr] \,, \nn\\
I_3 &=& \frac{1}{N} \biggl[ \frac{1}{4} \ln^4\hat{N}
  + 3 \zeta_2 \ln^2\hat{N} \biggr] \,, \nn\\
I_4 &=& \frac{1}{N} \biggl[ \frac{1}{5} \ln^5\hat{N}
  + 4 \zeta_2 \ln^3\hat{N} + 42 \zeta_4 \ln\hat{N} \biggr] \,.
\label{SomeIp}
\eea
Notice that the $I_p$ obey
\be
\frac{d}{dN}  [ N I_p(N) ]\ =\ p \, I_{p-1}(N)\ +\ {\cal O}(\tfrac{1}{N})\,.
\label{Irecurse}
\ee
This result can established by integration by parts in $n$.  Hence
the structure of the integrals $I_p(N)$ at large $N$ is dictated by
this recursion relation, up to an integration constant, or equivalently
the rational number $r_m$ multiplying $\zeta_{2m}/N\times \ln\hat{N}$ in
$I_{2m}$.  This sequence of numbers,
\be
r_m = 2, 42, 1395, 80010, 7243425, 957535425, \frac{348670597275}{2},
41844302750250, \ldots,
\label{rm}
\ee
whose $m^{\rm th}$ term always contains a factor of $2^{2m-1}-1$,
is given in turn by another recursion relation,
\be
\frac{r_m}{2^{2m-1}-1}\ =\ \frac{m(2m-1)}{2} \times \frac{r_{m-1}}{2^{2(m-1)-1}-1} \,,
\label{rmrecurse}
\ee
which is valid for at least the first 35 terms.

We also need to perform the Mellin transform of $\ln^\ell(1+w)=\ln^\ell(1-z)$,
\be
M_\ell \equiv \int_0^1 dz \, z^{N-1} \ln^\ell(1-z)\,,
\label{Melldef}
\ee
by using the formula,
\be
\int_0^1 dz \, z^{N-1} (1-z)^{\alpha}
= \frac{\Gamma(N)\Gamma(\alpha+1)}{\Gamma(N+\alpha+1)} \,,
\label{alphaint}
\ee
differentiating $\ell$ times, then setting $\alpha=0$ and
taking $N\to\infty$.  The first few values are,
\bea
M_1 &=& - \frac{1}{N} \ln\hat{N} \,, \nn\\
M_2 &=& \frac{1}{N} \ln^2\hat{N} \,, \nn\\
M_3 &=& \frac{1}{N} \biggl[ - \ln^3\hat{N} - 3 \zeta_2 \ln\hat{N}
  \biggr] \,, \nn\\
M_4 &=& \frac{1}{N} \biggl[ \ln^4\hat{N}
  + 6 \zeta_2 \ln^2\hat{N} + 8 \zeta_3  \ln\hat{N} \biggr] \,, \nn\\
M_5 &=& \frac{1}{N} \biggl[ - \ln^5\hat{N}
  - 10 \zeta_2 \ln^3\hat{N} - 20 \zeta_3 \ln^2\hat{N}
  - \frac{135}{2} \zeta_4 \ln\hat{N} \biggr] \,,
\label{SomeMell}
\eea
again omitting the $1/N$ terms without logarithms.
Notice that $M_\ell$ obeys a very similar recursion relation to $I_i$:
\be
\frac{d}{dN}  [ N M_\ell(N) ]\ =\
- \ell \, M_{\ell-1}(N)\ +\ {\cal O}(\tfrac{1}{N})\,.
\label{Mrecurse}
\ee
This result can established by integration by parts in $z$.
It implies that
\be
\frac{d}{dN} ( N M[f] )\ =\ M[ -df/dL_w ]\ +\ {\cal O}(\tfrac{1}{N})\,,
\label{MrecurseII}
\ee
for the Mellin transform $M[f]$ of any function $f$ that is a polynomial in
$L_w \equiv \ln(1+w)$.

Using these results, we can rewrite $I_p$, the Mellin transform
of $\hat{S}_p$, as a linear combination of $M_\ell$,
the Mellin transforms of $\ln^\ell(1+w)$,
with $\zeta$ valued coefficients.  Then $\hat{S}_p$
is given by the same linear combination of powers of
$L_w$.  The first few orders are given by:
\bea
\hat{S}_0 &=& - L_w \,, \nn\\
\hat{S}_1 &=& \frac{1}{2} L_w^2 \,, \nn\\
\hat{S}_2 &=& - \frac{1}{3} L_w^3 - \zeta_2 L_w \,, \nn\\
\hat{S}_3 &=& \frac{1}{4} L_w^4 + \frac{3}{2} \zeta_2 L_w^2
  + 2 \zeta_3 L_w \,, \nn\\
\hat{S}_4 &=& - \frac{1}{5} L_w^5 - 2 \zeta_2 L_w^3 - 4 \zeta_3 L_w^2
   - \frac{27}{2} \zeta_4 L_w \,.
\label{SomeShatp}
\eea
We observe that the $\hat{S}_p$ also obey a recursion relation,
\be
\frac{d}{dL_w} \hat{S}_p(L_w)\ =\ - p \, \hat{S}_{p-1}(L_w)
\ +\ {\cal O}(1)\,,
\label{Shatrecurse}
\ee
which follows from \eqns{Irecurse}{MrecurseII}.
Hence the $\hat{S}_p$ are completely dictated by the coefficient
of just the first power of $L_w$ at each order $p$.

Inserting \eqn{SomeShatp} into \eqn{Sk}, we get similar formulas for
$S_k(w,w^*)$, which we can rewrite as polynomials in
$\ln|1+w|^2 = L_w + L_{w^*} \equiv L_{|w|^2}$.  For the first few orders, we find:
\bea
S_0 &=& \frac{1}{2} (L_{|w|^2})^2\,, \nn\\
S_1 &=& - \frac{1}{6} (L_{|w|^2})^3 \,, \nn\\
S_2 &=& \frac{1}{12} (L_{|w|^2})^4 + \zeta_2 (L_{|w|^2})^2 \,, \nn\\
S_3 &=& - \frac{1}{20} (L_{|w|^2})^5 - \zeta_2 (L_{|w|^2})^3
  - 2 \zeta_3 (L_{|w|^2})^2 \,, \nn\\
S_4 &=& \frac{1}{30} (L_{|w|^2})^6 + \zeta_2 (L_{|w|^2})^4
  + \frac{8}{3} \zeta_3 (L_{|w|^2})^3 + 21 \zeta_4 (L_{|w|^2})^2 \,.
\label{SomeSk}
\eea
Again there is a recursion relation,
\be
\frac{d}{dL_{|w|^2}} S_k(L_{|w|^2})\ =\ - k \, S_{k-1}(L_{|w|^2})
\ +\ {\cal O}(L_{|w|^2})\,,
\label{Srecurse}
\ee
and the $S_k$ are determined by the coefficient of $(L_{|w|^2})^2$ at each $k$.

After determining the $S_k$ to high orders, we insert them into the expression
for the large $(\hat{n},m)$ limit of the double sum for the MRK limit
at each loop order.  A divergence in both
$w\to-1$ and $w^*\to-1$ is required in order to be able to neglect
contributions to the double sum~(\ref{Sk}) from finite $\hat{n}$ and $m$.
Therefore this method, although computationally very efficient,
only determines the coefficients in front of $\ln^k|1+w|^2$ for $k\geq2$.
It misses the coefficients in front of $\ln|1+w|^2$ and the constant term.
On the other hand, using a complete basis of SVHPLs through weight 10,
we were able to determine the complete MRK limits through five loops,
in agreement with ref.~\cite{JamesYorgos}.  We also obtained $g^{(L)}_r(w,w^*)$,
the coefficient of $\ln^r\delta$ in the MRK limit of the $L$-loop
remainder function $R_6^{(L)}$, for $r\geq 2L-11$ for $6\leq L \leq 10$.
Using this information, we could also determine the
coefficients in front of $\ln|1+w|^2$ and the constant term
for all terms with weight 10 or fewer.

After generating such results through nine loops, as a polynomial
in $\ln|v|$ and $\ln|1+w|^2$, we used \eqn{scMRK33relation}
to rewrite it as a polynomial in $\ln|v|$ and $\lnden$.
Then we imposed the further constraint that the result was consistent
with the small $v$ limit of \eqn{eq:evol_eqn_kin_factor},
i.e.~that the $v$-dependence was precisely proportional to
$\exp[\frac{\gK}{8} \ln^2|v|]$.  This constraint fixes additional
terms.   Finally we compared with the results of
Broedel and Sprenger~\cite{BroedelSprenger},
which are complete through weight 13.
Their results were completely consistent with ours,
and fixed all constants through weight 13.


\section{Final result for singular terms}
\label{FinalResults}

In this section we present the final result for the singular terms
in the expansion of the nonsingular framing of the hexagonal
Wilson loop in the limit of $3\to3$ self-crossing kinematics,
as $|\delta|\to0$. The result can be written as,
\be
\frac{1}{2\pi i} \frac{d {\cal W}^{\rm ns}_{3\to3}}{d\lnden}\ =\
\exp\Bigl[ - \frac{\gK}{8} ( \lndene{2} - L^2 ) \Bigr]
\, g(\lnden,\gK) \,,
\label{defineg}
\ee
where $g$ depends only on $\lnden$ and the 't Hooft coupling.
We choose to write the final result as an expansion in the cusp
anomalous dimension $\gK(a)$ instead of the coupling parameter $a$.
We give the expansion of $\gK(a)$ through 10 loops in appendix~\ref{cuspexp}.

The leading-$\lnden$ terms in \eqn{defineg} all come from the factor
$\exp[-\gK \lndene{2}\,/8]$, multiplied by a factor of $\gK/8$,
and keeping only the leading term in $\gK(a) = 4a$:
\be
\frac{a}{2} \int_0^{\lnden} dx \exp\Bigl[ -\frac{a}{2} x^2 \Bigr]
\ =\ \frac{a}{2} \lnden \, - \, \frac{a^2}{12} \lndene{3}
\, + \, \frac{a^3}{80} \lndene{5} \, - \, \frac{a^4}{672} \lndene{7}
\, + \, \ldots.
\label{leadinglogs}
\ee
Furthermore,
we find that all of the purely even $\zeta$ terms in $g$, and many
of the terms with single odd $\zeta$ values, can be captured by
the following expression:
\bea
g_0(\lnden,\gK) &\equiv& \frac{\gK}{8}
\, \exp\Bigl[ \frac{\gK}{2} \gamma_E \lnden \Bigr]
\frac{\Gamma\Bigl(1 + \frac{\gK}{4} \lnden \Bigr)}
     {\Gamma\Bigl(1 - \frac{\gK}{4} \lnden \Bigr)} \nn\\
&& \hskip0cm\null
\times
\biggl\{ 1 + \sum_{k=1}^\infty \frac{1}{k!} \Bigl(-\frac{\gK}{8}\Bigr)^k
\Bigl[ \psi^{(2k-1)}\Bigl(1 + \frac{\gK}{4} \lnden \Bigr)
      - \psi^{(2k-1)}\Bigl(1 - \frac{\gK}{4} \lnden \Bigr) \Bigr]
\biggr\} \,, \nn\\
&&{~} \label{defgzero}
\eea
where $\psi^{(k)}(x)$ denotes the $k^{\rm th}$ derivative of
$\psi(x) = d\ln\Gamma(x)/dx$,
and $\gamma_E$ is the Euler-Mascheroni constant.

Writing the full function $g$ as $g_0$ plus a residual term $g_1$,
\be
g(\lnden,\gK)\ =\ g_0(\lnden,\gK) + g_1(\lnden,\gK),
\label{defgone}
\ee
we give the expansion of $g_1$, which starts at three loops:
\be
g_1(\lnden,\gK) = \frac{1}{4} 
\sum_{L=3}^\infty c^{(L)} \Bigl(\frac{\gK}{4}\Bigr)^L \,,
\label{cdef}
\ee
where
\bea
c^{(3)} &=& - 6 \zeta_3 \lnden \,, \label{c3}\\
c^{(4)} &=& ( 20 \zeta_5 - 8 \zeta_2 \zeta_3 ) \lnden
 \, - \, 2 (\zeta_3)^2 \,, \label{c4}\\
c^{(5)} &=& 5 (\zeta_3)^2 \lndene{2}
 \, - \, \Bigl( \frac{175}{2} \zeta_7 - 50 \zeta_2 \zeta_5 \Bigr) \lnden
 \,  + \, 25 \zeta_3 \zeta_5 + \zeta_2 (\zeta_3)^2 \,, \label{c5}\\
c^{(6)} &=& - \, ( 96 \zeta_3 \zeta_5 + 8 \zeta_2 (\zeta_3)^2 ) \lndene{2}
  \, + \, ( 441 \zeta_9 - 315 \zeta_2 \zeta_7 + 72 \zeta_4 \zeta_5
          - 24 (\zeta_3)^3 ) \lnden \nn\\
&&\hskip0cm\null
   - 194 \zeta_3 \zeta_7 - 99 (\zeta_5)^2 - 12 \zeta_4 (\zeta_3)^2
   - 16 \zeta_2 \zeta_3 \zeta_5  
\,, \label{c6}
\eea
\bea
c^{(7)} &=& \Bigl( \frac{164}{3} \zeta_3 \zeta_5 
                + \frac{16}{3} \zeta_2 (\zeta_3)^2 \Bigr) \lndene{4}
  \, + \,\frac{103}{3} (\zeta_3)^3 \lndene{3} \nn\\
&&\hskip0cm\null
  \, + \, \Bigl( \frac{1985}{2} \zeta_3 \zeta_7 + 496 (\zeta_5)^2
          + 40 \zeta_4 (\zeta_3)^2 + 114 \zeta_2 \zeta_3 \zeta_5 \Bigr)
        \lndene{2} \nn\\
&&\hskip0cm\null
   + \Bigl( - \frac{4851}{2} \zeta_{11} + 2058 \zeta_2 \zeta_9
      - 882 \zeta_4 \zeta_7 + 697 (\zeta_3)^2 \zeta_5
       + 11 \zeta_2 (\zeta_3)^3 \Bigr) \lnden \nn\\
&&\hskip0cm\null
   + 135 \zeta_2 \zeta_3 \zeta_7 + 108 \zeta_4 \zeta_3 \zeta_5
    + 48 \zeta_2 (\zeta_5)^2
   + 1818 \zeta_5 \zeta_7 + 1869 \zeta_3 \zeta_9 + 27 (\zeta_3)^4
\,, \label{c7}
\eea
\bea
c^{(8)} &=& - \biggl[ \frac{28}{5} \zeta_3 \zeta_5 \lndene{6}
  \, + \, \frac{46}{3} (\zeta_3)^3 \lndene{5}
  \, + \, \Bigl( \frac{2225}{3} \zeta_3 \zeta_7 + 384 (\zeta_5)^2
               + \frac{244}{3} \zeta_2 \zeta_3 \zeta_5 \Bigr) \lndene{4} \nn\\
&&\hskip0.5cm\null
  + \Bigl( \frac{3806}{3} (\zeta_3)^2 \zeta_5
         + \frac{206}{3} \zeta_2 (\zeta_3)^3 \Bigr) \lndene{3} \nn\\
&&\hskip0.5cm\null
  + \Bigl( 239 (\zeta_3)^4 + 925 \zeta_2 \zeta_3 \zeta_7
         + 400 \zeta_2 (\zeta_5)^2 + 428 \zeta_3 \zeta_4 \zeta_5
         + 11739 \zeta_3 \zeta_9 + 11156 \zeta_5 \zeta_7 \Bigr) \lndene{2} \nn\\
&&\hskip0.5cm\null
  + \Bigl( - 14157 \zeta_{13} + 13860 \zeta_2 \zeta_{11} - 8484 \zeta_4 \zeta_9
     + 540 \zeta_6 \zeta_7 + 474 \zeta_2 (\zeta_3)^2 \zeta_5
     + \frac{15537}{2} (\zeta_3)^2 \zeta_7 \nn\\
&&\hskip0.9cm\null
      + 7625 \zeta_3 (\zeta_5)^2 + 238 \zeta_4 (\zeta_3)^3 \Bigr) \lnden 
 \, + \, c^{(8)}_0 \biggr]
\,, \label{c8}
\eea
and
\bea
c^{(9)} &=& \frac{4}{3} (\zeta_3)^3 \lndene{7}
 \, + \, \Bigl( 88 (\zeta_5)^2 + \frac{16}{5} \zeta_2 \zeta_3 \zeta_5
              + 152 \zeta_3 \zeta_7 \Bigr) \lndene{6}
 \, + \, \Bigl( \frac{3244}{5} (\zeta_3)^2 \zeta_5
         + \frac{64}{3} \zeta_2 (\zeta_3)^3 \Bigr) \lndene{5} \nn\\
&&\hskip0cm\null
  + \Bigl( \frac{1043}{4} (\zeta_3)^4 + 690 \zeta_2 \zeta_3 \zeta_7
      - 48 \zeta_4 \zeta_3 \zeta_5 + 10626 \zeta_3 \zeta_9
      + 10155 \zeta_5 \zeta_7 + 300 \zeta_2 (\zeta_5)^2 \Bigr) \lndene{4} \nn\\
&&\hskip0cm\null
  + \Bigl( \frac{5639}{3} \zeta_2 (\zeta_3)^2 \zeta_5
         + \frac{203707}{12} (\zeta_3)^2 \zeta_7 + 16598 \zeta_3 (\zeta_5)^2
    + 128 \zeta_4 (\zeta_3)^3 \Bigr) \lndene{3} \nn\\
&&\hskip0cm\null
  + c^{(9)}_2  \lndene{2} \nn\\
&&\hskip0cm\null
  + \Bigl( - \frac{2760615}{32} \zeta_{15} + \frac{382239}{4} \zeta_2 \zeta_{13}
      - \frac{601425}{8} \zeta_4 \zeta_{11} + \frac{20547}{2} \zeta_6 \zeta_9
      + 84 \zeta_6 (\zeta_3)^3 \nn\\
&&\hskip0.6cm\null
      + \frac{10301}{2} \zeta_4 (\zeta_3)^2 \zeta_5
      + \frac{24569}{4} \zeta_2 (\zeta_3)^2 \zeta_7
      + 5585 \zeta_2 \zeta_3 (\zeta_5)^2 + 30885 (\zeta_5)^3 \nn\\
&&\hskip0.6cm\null
      + 101668 (\zeta_3)^2 \zeta_9
      + \frac{757873}{4} \zeta_3 \zeta_5 \zeta_7
      + 917 (\zeta_3)^5 \Bigr) \lnden \nn\\
&&\hskip0cm\null
+ c^{(9)}_0 \,, \label{c9}
\eea
where $c^{(8)}_0$, $c^{(9)}_2$ and $c^{(9)}_0$ are linear combinations
of multiple zeta values of weight 14, 14 and 16, respectively.
We have some partial information about their rational-number coefficients;
in particular,
\bea
2 \, c^{(9)}_2 - c^{(8)}_0 &=&
 339327 \zeta_3 \zeta_{11} + 307248 \zeta_5 \zeta_9
+ \frac{1174875}{8} (\zeta_7)^2 + 18648 \zeta_2 \zeta_3 \zeta_9
+ 13431 \zeta_2 \zeta_5 \zeta_7 \nn\\
&&\hskip0cm\null
+ 1102 \zeta_2 (\zeta_3)^4
+ 5058 \zeta_4 \zeta_3 \zeta_7 + 2271 \zeta_4 (\zeta_5)^2
- 1080 \zeta_6 \zeta_3 \zeta_5 + 26914 (\zeta_3)^3 \zeta_5 \,. \nn\\
&&{~} \label{wtfourteendiff}
\eea
No terms in $g_1$ have only even $\zeta$ values in them
(i.e.~pure powers of $\pi$).  Furthermore, the only terms in $g_1$ that have
a single odd $\zeta$ value are those with a single power of $\lnden$.
All the terms with a single odd $\zeta$ value and multiple powers
of $\lnden$ have been absorbed into $g_0$.

We also attempted to fit the explicit results for the function $g$
to the consequences of \eqn{eq:evol_eqn_wrong}:
\be
g(\lnden,\gK)\ \stackrel{?}{=}\ \hat{C}(\gK)
\, \exp\Bigl[ \frac{\bar\Gamma(\gK)}{2} \lnden \Bigr] \,,
\label{fitgwrong}
\ee
where $\hat{C}(\gK) = -C(a) \Gamma_1(a)/4$ and $\bar\Gamma(\gK)$
are taken to be arbitrary functions of the cusp anomalous dimension
(i.e.~of the coupling $a$).  Expanding around $\gK=0$, we find
that $\hat{C}(\gK) \propto \gK + {\cal O}(\gK^2)$,
$\bar\Gamma(\gK)\propto \gK^2 + {\cal O}(\gK^3)$, and then
it becomes impossible to fit the true $g(\lnden,\gK)$
with \eqn{fitgwrong}, beginning
with the $\gK^4$ term.  At this order, $g$ contains $\lndene{3}$
with a nonzero coefficient proportional to $\zeta_3$,
while the $\lndene{3}$ coefficient vanishes in the
ansatz~(\ref{fitgwrong}).

Despite this difficulty, the relative simplicity of the terms in $g_1$
with more than one power of $\lnden$
suggests that these terms might be on a different footing from
the linear terms in $\lnden$, and might have a relatively
simple kinematical origin.  It will be interesting to investigate
the structure of $g(\lnden,\gK)$ further in the future, and
to see what role the cross anomalous dimension in planar $\NeqFour$ SYM
might play.


\section{Conclusions and outlook}
\label{concout}

In this paper we observed that the duality between scattering
amplitudes and Wilson loops in planar $\NeqFour$ SYM
maps a configuration that mimics double-parton-scattering into a
self-crossing limit of the relevant Wilson loop.
We observed that helicity selection rules, especially $J_z$ conservation,
are related to the finiteness of the NMHV six-gluon amplitude
in $2\to4$ kinematics in QCD, ${\cal N}=1$ or $\NeqFour$ SYM at one loop.
Beyond one loop, there are logarithmic divergences, although
the transcendental function $E$ is nonsingular at the leading power of $\de$.
The transcendental function entering the MHV amplitude, in contrast,
can be and is singular.
We presented explicit results through five loops, including the
non-singular terms.  Then we studied the structure of the
singular terms, and determined their kinematic dependence exactly,
using an evolution equation for self-crossing Wilson loops in the
large $N_c$ limit. In particular, we explained the surprising
$v$-independence of the singular terms in ${\cal E}_{3\to3}$ in
\eqn{eq:Eevol_eqn}. Finally, we exploited the overlap between
the self-crossing limit and the multi-Regge limit as $v\to0$,
to determine the answer in that limit to nine loops, up to
a couple of zeta-valued constants.  The leading-logarithmic
part of this formula, \eqn{leadinglogs}, gives a simple representation
for the leading logs studied earlier~\cite{Georgiou,DornWuttke1,DornWuttke2}
to all loop orders.

We can use similar methods to study self-crossing limits of the $n$-point
amplitude for $n>6$.  For example, in appendix~\ref{sckin7}
we identify the self-crossing configuration for the seven-point case,
in which the Wilson loop factorizes into the product of a box
and a pentagon.  This configuration is a four-parameter subspace of the
general six-parameter space of dual conformal invariants.
(Naively, there are seven dual conformal invariants in the seven-point
case, related by the seven-fold cyclic symmetry; however in four dimensional
spacetime they obey a single Gram determinantal constraint.)
The same arguments given above suggest that the dependence on the
four surviving nonsingular parameters in this case can be determined
exactly in terms of the cusp anomalous dimension and a suitable one-loop
function.  The dependence on the singular parameter should be essentially
the same as in the six-point case.
This information will be useful in constructing or checking
the full seven-point MHV amplitude at higher loop orders; it is currently
known at the function level through two loops~\cite{Golden2014xqf},
and at symbol level through three loops~\cite{Drummond2014ffa}.

Beginning with the eight-point amplitude, multiple types of single
self-crossings are possible, corresponding to different partitions
of the final momenta into two sets of momenta, and there can also be
more than one self-crossing.   For the single self-crossing,
one always loses two degrees of freedom, because two separate
transverse coordinates are constrained.  For $n=6$, the three
cross ratios $(u,v,w)$ reduce to one, $v$.  For $n=7$, the six independent
$u_i$ reduce to the four $u_1$, $u_2$, $u_5$ $u_6$ given in
appendix~\ref{sckin7}.  For larger $n$ there will be
$(3n-15)-2 = 3n-17$ cross ratios left for a single self-crossing,
and correspondingly fewer for multiple self-crossings.
Even when the subprocesses have their own dual conformal cross ratios,
there will still be additional parameters, and the functional dependence
on those parameters should be calculable using factorization arguments
of the kind used here.

Similarly, studies of non-MHV amplitudes for $n>6$
in these limits should be possible, using the correspondence
with super-Wilson-loops~\cite{Mason2010yk,CaronHuot2010ek}.
The finiteness of the one-loop NMHV six-gluon amplitude is clearly special
to $n=6$: It is easy to add an additional final-state gluon with
negative helicity to the configurations in \fig{fig:MHVhel} to convert them
to valid NMHV seven-point configurations.
It would be quite interesting to combine some of the ideas and
results of this paper
with the recent general analysis of Landau singularities for $\NeqFour$ SYM
amplitudes in ref.~\cite{Dennen2015bet}.

In summary, the self-crossing limit and its relation to kinematical
configurations that mimic double-parton scattering should provide
a rich playground
for further investigations of the analytic behavior of multi-loop
amplitudes, in both planar $\NeqFour$ SYM and more general theories
such as QCD.


\vskip0.3cm
\noindent {\large\bf Acknowledgments}
\vskip0.3cm

We are grateful to Benjamin Basso, Andrei Belitsky, Johannes Broedel, Simon
Caron-Huot, John Joseph Carrasco, Claude Duhr, Matt von Hippel,
Gregory Korchemsky, Andrew McLeod, Tom Melia, Amit Sever,
Georgios Papathanasiou,
Mads S\o{}gaard, Martin Sprenger, Pedro Vieira and Jara Trnka for useful
discussions and comments.  We would particularly like to thank
Johannes Broedel and Martin Sprenger for providing us with the
$w\to-1$ limit of their high-loop order MRK expressions,
as well as Benjamin Basso, Simon Caron-Huot, Amit Sever and Pedro Vieira
for pointing out an error in the NMHV discussion in the first
version of this paper.
This research was supported by the US Department of Energy under contract
DE--AC02--76SF00515 and grant DE--SC0011632, by the Walter Burke
Institute, by the Gordon and Betty Moore Foundation through Grant
No.~776 to the Caltech Moore Center for Theoretical Cosmology and
Physics. LD thanks Caltech, the Aspen Center for Physics and the NSF
Grant \#1066293 for hospitality.  Figures in this paper were
made with {\sc Jaxodraw}~\cite{Binosi2003yf,Binosi2008ig}, based on
{\sc Axodraw}~\cite{Vermaseren1994je}. 

\newpage


\appendix

\section{Self-crossing kinematics}
\label{sckin}

In this appendix we describe the kinematics of $2\to4$ scattering
in the self-crossing or double-parton-scattering-like limit.
Then we do the same for the analogous $3\to3$ scattering configuration.

\subsection{$2\to4$ kinematics}

For the $2\to4$ scattering configuration shown in \fig{fig:show_24_33}(a)
we consider
\bea
k_3\ +\ k_6\ &\to& k_1\ + k_2\ +\ k_4\ +\ k_5, \label{incout24}\\
(1-x) k_3\ +\ (1-y) k_6 &\to& k_1\ +\ k_2, \label{sub24A}\\
x k_3\ +\ y k_6 &\to& k_4\ +\ k_5.  \label{sub24B}
\eea
Incoming gluons 3 and 6 split into collinear pairs with
momentum fractions $x$ and $1-x$, and $y$ and $1-y$, respectively.
These pairs then undergo $2\to2$ scatterings into final state gluons
1, 2, 4 and 5.

We work in the center-of-mass (CM) frame,
and take the spatial components of $k_3$ to be in the positive $z$ direction,
while those of $k_6$ are in the negative $z$ direction.
Momenta $k_1$ and $k_2$ are in the $xz$-plane,
while $k_4$ and $k_5$ are rotated out of this plane by an
azimuthal angle $\phi$.
Incoming momenta are labeled by the negative of the true momentum,
so that $\sum_{i=1}^6 k_i=0$.  Writing $k_i^\mu = (k_i^t,k_i^x,k_i^y,k_i^z)$,
we have:
\bea
k_1 &=& (E_1, E_1 \sin\theta_1, 0, -E_1 \cos\theta_1), \nn\\ 
k_2 &=& (E_2, -E_2 \sin\theta_2, 0, -E_2 \cos\theta_2), \nn\\ 
k_3 &=& (-\tfrac{1}{2}\sqrt{s_{36}}, 0, 0, -\tfrac{1}{2}\sqrt{s_{36}}), \nn\\
k_4 &=& (E_4, E_4 \sin\theta_4\cos\phi,
             E_4 \sin\theta_4\sin\phi, -E_4 \cos\theta_4), \nn\\ 
k_5 &=& (E_5, -E_5 \sin\theta_5\cos\phi,
             -E_5 \sin\theta_5\sin\phi, -E_5 \cos\theta_5), \nn\\ 
k_6 &=& (-\tfrac{1}{2}\sqrt{s_{36}}, 0, 0, \tfrac{1}{2}\sqrt{s_{36}}).
\label{sc24kin}
\eea

Momentum conservation for the $2\to2$ subprocesses in
\eqns{sub33A}{sub33B} implies that
\bea
s_{12} &=& (1-x)(1-y) s_{36} \,, \label{22kinrelationsfirst} \\
s_{45} &=& x y s_{36} \,, \\
s_{123} &=& - x (1-y) s_{36} \,, \\
s_{345} &=& - y (1-x) s_{36} \,, \label{22kinrelationsfourth} \\
x s_{34} &=& y s_{56}\,, \\
(1-x) s_{23} &=& (1-y) s_{61} \,.
\label{22kinrelationslast}
\eea
Inserting \eqn{sc24kin} into these relations, we can express
the energies $E_i$ and two of the angles in \eqn{sc24kin} in terms
of the momentum fractions $x$ and $y$, and the other two polar angles:
\bea
E_1 &=& \sqrt{s_{36}} \,
\frac{(1-x)(1-y)}{(1+\cos\theta_1)(1-x)+(1-\cos\theta_1)(1-y)} \,, \\
E_2 &=& \frac{\sqrt{s_{36}}}{2} \, 
\frac{(1+\cos\theta_1)(1-x)^2+(1-\cos\theta_1)(1-y)^2}
          {(1+\cos\theta_1)(1-x)+(1-\cos\theta_1)(1-y)} \,, \\
\cos\theta_2 &=& \frac{(1-\cos\theta_1)(1-y)^2-(1+\cos\theta_1)(1-x)^2}
           {(1+\cos\theta_1)(1-x)^2+(1-\cos\theta_1)(1-y)^2} \,, \\
\sin\theta_2 &=& \frac{2(1-x)(1-y)\sin\theta_1}
  {(1+\cos\theta_1)(1-x)^2+(1-\cos\theta_1)(1-y)^2} \,, \\
E_4 &=& \frac{\sqrt{s_{36}}}{2} \,
\frac{(1+\cos\theta_5)x^2+(1-\cos\theta_5)y^2}
   {(1+\cos\theta_5)x+(1-\cos\theta_5)y} \,, \\
E_5 &=& \sqrt{s_{36}} \,
\frac{xy}{(1+\cos\theta_5)x+(1-\cos\theta_5)y} \,, \\
\cos\theta_4 &=& \frac{(1-\cos\theta_5)y^2-(1+\cos\theta_5)x^2}
     {(1+\cos\theta_5)x^2+(1-\cos\theta_5)y^2} \,, \\
\sin\theta_4 &=& \frac{2xy\sin\theta_5}
     {(1+\cos\theta_5)x^2+(1-\cos\theta_5)y^2} \,.
\label{Ethsol24}
\eea

It's convenient to trade the angles $\theta_1$ and $\theta_5$ in
the overall CM frame for the polar angles in the CM frames
for the respective $2\to2$ subprocesses~(\ref{sub24A}) and (\ref{sub24B}),
which we call $\theta_A$ and $\theta_B$, respectively:
\bea
-\frac{(1-x)s_{23}}{s_{12}}\ =\ \frac{1-\cos\theta_A}{2}
 &=&  \frac{(1-y)(1-\cos\theta_1)}{(1+\cos\theta_1)(1-x)+(1-\cos\theta_1)(1-y)}
 \,, \\
-\frac{x s_{34}}{s_{45}}\ =\ \frac{1-\cos\theta_B}{2}
 &=& \frac{y(1-\cos\theta_5)}{(1+\cos\theta_5)x+(1-\cos\theta_5)y} \,.
\label{relate24_AB_15}
\eea
Solving for $\theta_1$ and $\theta_5$, we have,
\bea
\cos\theta_1 &=& 1
- \frac{2(1-x)(1-c_A)}{(1-c_A)(1-x)+(1+c_A)(1-y)} \,, \\
\sin\theta_1 &=&
\frac{2 s_A \sqrt{(1-x)(1-y)}}{(1-c_A)(1-x)+(1+c_A)(1-y)} \,, \\
\cos\theta_5 &=& 1
- \frac{2 x (1-c_B)}{(1-c_B)x+(1+c_B)y} \,, \\
\sin\theta_5 &=&
\frac{2 s_B \sqrt{xy}}{(1-c_B)x+(1+c_B)y} \label{th15elim24} \,,
\eea
where $c_{A,B} = \cos\theta_{A,B}$ and $s_{A,B} = \sin\theta_{A,B}$.

Of the four invariants appearing in the cross ratio
$v = s_{23} s_{56} /(s_{234} s_{123})$, three are simply related
to the invariants for the individual $2\to2$ subprocesses.  For
example, $(1-x)s_{23}=(1-y)s_{16}$ is the $t$-channel invariant for
subprocess $A$,
so it is simple to relate it to the $u$-channel invariant $(1-x)s_{13}$,
the $s$-channel invariant $s_{12}$ and
the CM scattering angle $\theta_A$. The only
invariant that straddles two subprocesses is $s_{234} = s_{23} + s_{34} + s_{24}$,
and it does so only through $s_{24}$.  Thus the dependence of $v$
on the azimuthal angle enters only through $s_{24}$.

It is convenient to normalize all invariants by $s_{36}$.  Computing
$s_{24}/s_{36}$ from \eqn{sc24kin} and using \eqns{Ethsol24}{th15elim24}
to express the result in terms of $x$, $y$, $\theta_A$, $\theta_B$ and $\phi$,
we find that
\be
\frac{s_{24}}{s_{36}} = \frac{1}{4} \Bigl[
 x(1-y)(1-c_A)(1+c_B) + y(1-x)(1+c_A)(1-c_B) 
+ 2 \sqrt{xy(1-x)(1-y)} s_A s_B \cos\phi \Bigr] \,.
\label{sc24_s24_s36}
\ee
The cross ratio $v$ can be written in terms of $s_{24}/s_{36}$ as,
\be
v = - \frac{(1-c_A)(1-c_B)}
{4 \Bigl[ s_{24}/s_{36} - \tfrac{1}{2}(1-c_A)(1-y)
                      - \tfrac{1}{2}(1-c_B) y \Bigr] } \,.
\ee
After inserting \eqn{sc24_s24_s36}, we find that
\bea
\frac{1}{v} - 1 &=& \frac{(1\!+\!c_A)(1\!-\!c_B)xy
                        + (1\!-\!c_A)(1\!+\!c_B)(1-x)(1-y)
    - 2 \sqrt{xy(1-x)(1-y)} s_A s_B \cos\phi}{(1-c_A)(1-c_B)}
\nn\\
&=& \frac{1}{(1-c_A)(1-c_B)}
 \biggl\{
   \Bigl[ \sqrt{(1+c_A)(1-c_B)xy} - \sqrt{(1-c_A)(1+c_B)(1-x)(1-y)} \Bigr]^2
\nn\\&&\hskip3cm\null
    + 2 \sqrt{xy(1-x)(1-y)} s_A s_B (1-\cos\phi) \biggr\} \,.
 \label{vinvm1_24_AB_2}
\eea
The second form~(\ref{vinvm1_24_AB_2}) makes manifest that
for $2\to4$ scattering,
\be
\frac{1}{v} - 1\ \geq\ 0, \qquad \hbox{or} \qquad
0\ \leq\ v\ \leq\ 1.
\label{vineq24}
\ee

The $v=0$ limit is only achieved when the denominator of
\eqn{vinvm1_24_AB_2} vanishes.  This happens when
one of the two $2\to2$ subprocesses becomes collinear,
either $\theta_A$ or $\theta_B \to 0$.
The $v=1$ limit requires the numerator of \eqn{vinvm1_24_AB_2} to vanish,
which implies that the event is planar,
\be
v\ =\ 1\quad \Leftrightarrow\quad \phi\ =\ 0, \quad
y\ =\ \frac{(1-c_A)(1+c_B)(1-x)}{(1+c_A)(1-c_B) x + (1-c_A)(1+c_B)(1-x)} \,.
\label{sc24veq1}
\ee
%


\subsection{$3\to3$ kinematics}

For the $3\to3$ scattering configuration shown in \fig{fig:show_24_33}(b)
we consider
\bea
k_1\ +\ k_3\ +\ k_5 &\to& k_2\ +\ k_4\ +\ k_6, \label{incout33}\\
k_1\ +\ (1-x) k_3 &\to& k_2\ +\ (1-y) k_6, \label{sub33A}\\
k_5\ +\ x k_3 &\to& k_4\ +\ y k_6.  \label{sub33B}
\eea
Gluon 3 splits into two partons, one of which collides with
gluon 1 and the other with gluon 5.  The products of those two collisions
are gluons 2 and 4, and two more gluons.  The latter two gluons
then merge into gluon 6.

Because $k_3$ is incoming and $k_6$ is outgoing, $s_{36}$ is negative.
We can choose a ``brick-wall frame'' for these two momenta,
in which $k_3$ is in the positive $z$ direction, while
$k_6$ is in the negative $z$ direction, with $k_6^z = -k_3^z$.
As in the $2\to4$ case, we take momenta $k_1$ and $k_2$ to lie in the
$xz$-plane, while $k_4$ and $k_5$ are rotated out of this plane by an
azimuthal angle $\phi$.
We parametrize the momenta as:
\bea
k_1 &=& (-E_1, E_1 \sin\theta_1, 0, E_1 \cos\theta_1), \nn\\ 
k_2 &=& (E_2, -E_2 \sin\theta_2, 0, -E_2 \cos\theta_2), \nn\\ 
k_3 &=& (-\tfrac{1}{2}\sqrt{-s_{36}}, 0, 0, -\tfrac{1}{2}\sqrt{-s_{36}}), \nn\\
k_4 &=& (E_4, E_4 \sin\theta_4\cos\phi,
              E_4 \sin\theta_4\sin\phi, -E_4 \cos\theta_4), \nn\\ 
k_5 &=& (-E_5, -E_5 \sin\theta_5\cos\phi,
              -E_5 \sin\theta_5\sin\phi, E_5 \cos\theta_5), \nn\\ 
k_6 &=& (\tfrac{1}{2}\sqrt{-s_{36}}, 0, 0, -\tfrac{1}{2}\sqrt{-s_{36}}).
\label{sc33kin}
\eea

Momentum conservation for the $2\to2$ subprocesses in
\eqns{sub33A}{sub33B} implies that kinematic relations in
eqs.~(\ref{22kinrelationsfirst})--(\ref{22kinrelationslast}) still hold.
Inserting \eqn{sc33kin} into these relations, we again express
the energies $E_i$ and two of the angles in \eqn{sc33kin} in terms
of the momentum fractions $x$ and $y$, and the other two polar angles:
\bea
E_1 &=& \sqrt{-s_{36}} \,
\frac{(1-x)(1-y)}{(1+\cos\theta_1)(1-x)-(1-\cos\theta_1)(1-y)} \,, \\
E_2 &=& \frac{\sqrt{-s_{36}}}{2} \, 
\frac{(1+\cos\theta_1)(1-x)^2+(1-\cos\theta_1)(1-y)^2}
          {(1+\cos\theta_1)(1-x)-(1-\cos\theta_1)(1-y)} \,, \\
\cos\theta_2 &=& \frac{(1-\cos\theta_1)(1-y)^2-(1+\cos\theta_1)(1-x)^2}
           {(1+\cos\theta_1)(1-x)^2+(1-\cos\theta_1)(1-y)^2} \,, \\
\sin\theta_2 &=& \frac{2(1-x)(1-y)\sin\theta_1}
  {(1+\cos\theta_1)(1-x)^2+(1-\cos\theta_1)(1-y)^2} \,, \\
E_4 &=& \frac{\sqrt{-s_{36}}}{2} \,
\frac{(1+\cos\theta_5)x^2+(1-\cos\theta_5)y^2}
   {(1+\cos\theta_5)x-(1-\cos\theta_5)y} \,, \\
E_5 &=& \sqrt{-s_{36}} \,
\frac{xy}{(1+\cos\theta_5)x-(1-\cos\theta_5)y} \,, \\
\cos\theta_4 &=& \frac{(1-\cos\theta_5)y^2-(1+\cos\theta_5)x^2}
     {(1+\cos\theta_5)x^2+(1-\cos\theta_5)y^2} \,, \\
\sin\theta_4 &=& \frac{2xy\sin\theta_5}
     {(1+\cos\theta_5)x^2+(1-\cos\theta_5)y^2} \,.
\label{Ethsol33}
\eea

Again we trade the angles $\theta_1$ and $\theta_5$ in
the brick-wall frame for the polar angles $\theta_A$ and $\theta_B$
in the CM frames for the respective $2\to2$
subprocesses~(\ref{sub33A}) and (\ref{sub33B}).
In this case, the relations are:
\bea
-\frac{s_{23}}{s_{13}}\ =\ -\frac{(1-y) s_{16}}{(1-x) s_{13}}
\ =\ \frac{1-c_A}{2}
 &=&  \frac{1-y}{1-x} \frac{1-\cos\theta_1}{1+\cos\theta_1} \,, \\
-\frac{s_{34}}{s_{35}}\ =\ -\frac{y s_{65}}{x s_{35}}
\ =\ \frac{1-c_B}{2}
 &=& \frac{y}{x} \frac{1-\cos\theta_5}{1+\cos\theta_5} \,.
\label{relate33_AB_15}
\eea
Solving for $\theta_1$ and $\theta_5$, we have,
\bea
\cos\theta_1 &=&
\frac{1-y - \tfrac{1}{2}(1-x)(1-c_A)}
{1-y + \tfrac{1}{2}(1-x)(1-c_A)} \,, \\
\sin\theta_1 &=&
\frac{\sqrt{2 (1-x) (1-y) (1-c_A)}}
{1-y + \tfrac{1}{2}(1-x)(1-c_A)} \,, \\
\cos\theta_5 &=& 
\frac{y - \tfrac{1}{2}x(1-c_B)}{y + \tfrac{1}{2}x(1-c_B)} \,, \\
\sin\theta_5 &=&
\frac{\sqrt{2 x y (1-c_B)}}{y + \tfrac{1}{2}x(1-c_B)}
\label{th15elim33} \,.
\eea

As in the $2\to4$ case, the dependence of $v$
on the azimuthal angle enters only through $s_{24}$,
which appears in $v = s_{23} s_{56} /(s_{234} s_{123})$ through
$s_{234} = s_{23} + s_{34} + s_{24}$.
We compute $s_{24}/s_{36}$ from \eqn{sc33kin} and use \eqns{Ethsol33}{th15elim33}
to express the result in terms of $x$, $y$, $\theta_A$, $\theta_B$ and $\phi$:
\bea
\frac{s_{24}}{s_{36}} &=& - \frac{2}{(1+c_A)(1+c_B)}
 \Bigl[ x (1-y) (1-c_A) + y (1-x) (1-c_B)
\nn\\ &&\hskip3.5cm\null
   + 2 \sqrt{x y (1-x) (1-y) (1-c_A) (1-c_B)}
       \cos\phi \Bigr] \,.
\label{sc33_s24_s36}
\eea
The expression for $v$ in terms of $s_{24}/s_{36}$ for $3\to3$ scattering is,
\be
v = - \frac{1-c_A}{1+c_A}
\frac{1-c_B}{1+c_B}
 \biggl[ \frac{s_{24}}{s_{36}}
  + \frac{1-c_A}{1+c_A} (1-y)
  + \frac{1-c_B}{1+c_B} y \biggr]^{-1} \,.
\ee
We insert \eqn{sc33_s24_s36} for $s_{24}/s_{36}$, and find that
\bea
1-\frac{1}{v} &=& 
2 \, \frac{(1-c_A) (1-x) (1-y) + (1-c_B) x y
  - 2 \sqrt{x y (1-x) (1-y) (1-c_A) (1-c_B)} \cos\phi}
   {(1-c_A)(1-c_B)}
\nn\\
&=& 
\frac{2}{(1-c_A)(1-c_B)}
\biggl\{ \Bigl[ \sqrt{(1-c_A)(1-x)(1-y)} - \sqrt{(1-c_B) x y} \Bigr]^2
\nn\\
&&\hskip3cm\null
  + 2 \sqrt{x y (1-x) (1-y) (1-c_A) (1-c_B)}(1-\cos\phi) \biggr\} \,.
\label{omv33inv}
\eea

From \eqn{omv33inv} we can see that for $3\to3$ scattering,
\be
1-\frac{1}{v}\ \geq\ 0,
\label{vineq33}
\ee
which is the complement of the region~(\ref{vineq24}) for $2\to4$
scattering.
The numerator of \eqn{omv33inv} is minimized, as a function of $\phi$,
at $\cos\phi=1$, or $\phi=0$, that is, for a
planar scattering configuration.
At this minimum, the numerator becomes a
perfect square, which can only equal zero, i.e.~$v=1$,
for
\be
v\ =\ 1\quad \Leftrightarrow\quad \phi\ =\ 0, \quad
y\ =\ \frac{(1-x)(1-c_A)}{(1-x)(1-c_A) + x (1-c_B) } \,.
\label{sc33veq1}
\ee

There are two branches that solve the inequality~(\ref{vineq33}):
$v<0$ and $v\geq1$.  The fact that there are two branches
is related to the fact that the $v\to\infty$ limit corresponds to
multiparticle factorization.  More generally, as discussed in
ref.~\cite{Dixon2014iba}, multi-particle factorization involves taking
two cross ratios large at the same rate, with the third cross ratio held fixed.

\begin{figure}
\begin{center}
\includegraphics[width=5.5in]{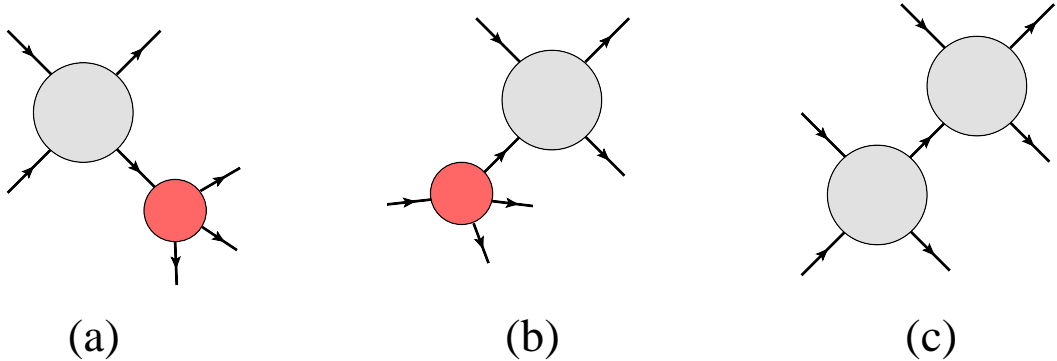}
\end{center}
\caption{In $2\to4$ scattering, multi-particle factorization only
occurs at the boundary of phase space, where a triple collinear
splitting takes place, either in the final state (a), or in 
the initial state (b).  In $3\to3$ scattering, in contrast, multi-particle
factorization can appear in the middle of the phase space; both amplitudes
in (c) are generic $2\to2$ scattering amplitudes.}
\label{fig:fact_24_33}
\end{figure}

For $2\to4$ scattering, multi-particle factorization can only occur
at the boundary of phase space, where there is a triple-collinear ($1\to3$)
splitting, either in the final state, as shown in \fig{fig:fact_24_33}(a),
or in the initial state, as depicted in \fig{fig:fact_24_33}(b).
On the other hand, a multi-particle factorization
configuration appears in the middle of the phase space for $3\to3$ scattering,
as illustrated in \fig{fig:fact_24_33}(c).

This configuration, with $v=\infty$, has
\be
\cos\phi\ =\ - \frac{(1-c_A)(1-c_B) - 2(1-c_A)(1-x)(1-y) - 2(1-c_B)xy}
   {4\sqrt{xy(1-x)(1-y)(1-c_A)(1-c_B)}} \,.
\label{cosphivinfinite}
\ee
For the MHV configuration studied in this paper, there is no pole in
the multi-particle factorization limit, because one cannot factorize
an MHV six-point amplitude into two MHV four-point amplitudes.  So
we should not expect any additional singularity as $v\to\infty$.
Indeed, eqs.~(\ref{Evinfty_0})--(\ref{Evinfty_5})
show that ${\cal E}$ has no $\ln v$ singularity in this limit.
The Wilson loop ${\cal W}^{\rm ns}$ is also smooth there,
since it becomes equal to ${\cal E}\exp[-\frac{1}{2}\zeta_2\gK]$
as $v\to\infty$.


\section{Perturbative expansion of cusp anomalous dimension}
\label{cuspexp}

The cusp anomalous dimension is known to all loop
orders~\cite{BES}. Here we give the expansion in the coupling
parameter $a$ through 10 loops:
\bea
\gK(a) &=& 4 a - 4 \zeta_2 a^2  + 22 \zeta_4 a^3
- \Bigl( \frac{219}{2} \zeta_6 + 4 (\zeta_3)^2 \Bigr) a^4
+ \Bigl( \frac{1774}{3} \zeta_8 + 8 \zeta_2 (\zeta_3)^2 + 40 \zeta_3 \zeta_5
  \Bigr) a^5 \nn\\
&&\hskip0cm\null
  - \Bigl( \frac{136883}{40} \zeta_{10} + 48 \zeta_4 (\zeta_3)^2
  + 80 \zeta_2 \zeta_3 \zeta_5 + 102 (\zeta_5)^2
  + 210 \zeta_3 \zeta_7 \Bigr) a^6 \nn\\
&&\hskip0cm\null
  + \Bigl( \frac{115201335}{5528} \zeta_{12} + 235 \zeta_6 (\zeta_3)^2
  + 4 (\zeta_3)^4 + 492 \zeta_4 \zeta_3 \zeta_5 + 204 \zeta_2 (\zeta_5)^2
  + 420 \zeta_2 \zeta_3 \zeta_7  \nn\\
&&\hskip0.6cm\null
  + 1092 \zeta_5 \zeta_7
  + 1176 \zeta_3 \zeta_9 \Bigr) a^7 \nn\\
&&\hskip0cm\null
  - \Bigl( \frac{295221817}{2240} \zeta_{14} + 1287 \zeta_4 (\zeta_5)^2
  + 12 \zeta_2 (\zeta_3)^4 + 2352 \zeta_2 \zeta_3 \zeta_9
  + \frac{5955}{2} (\zeta_7)^2 \nn\\
&&\hskip0.6cm\null
  + 2184 \zeta_2 \zeta_5 \zeta_7 + 2472 \zeta_6 \zeta_3 \zeta_5
  + 2625 \zeta_4 \zeta_3 \zeta_7 + 6930 \zeta_3 \zeta_{11}
  + 6216 \zeta_5 \zeta_9 + \frac{2453}{2} \zeta_8 (\zeta_3)^2 \nn\\
&&\hskip0.6cm\null
  + 80 (\zeta_3)^3 \zeta_5 \Bigr) a^8 \nn\\
&&\hskip0cm\null
  + \Bigl( \frac{74468151565}{86808} \zeta_{16} + 5955 (\zeta_7)^2 \zeta_2
  + 84 (\zeta_3)^4 \zeta_4 + 13860 \zeta_{11} \zeta_2 \zeta_3
  + 6633 \zeta_6 (\zeta_5)^2 \nn\\
&&\hskip0.6cm\null
  + 37125 \zeta_5 \zeta_{11} + 420 \zeta_7 (\zeta_3)^3
  + \frac{27537}{4} \zeta_{10} (\zeta_3)^2 + 14022 \zeta_4 \zeta_5 \zeta_7
  + 240 (\zeta_3)^3 \zeta_2 \zeta_5  \nn\\
&&\hskip0.6cm\null
  + 34425 \zeta_7 \zeta_9 + \frac{39527}{3} \zeta_8 \zeta_3 \zeta_5
  + 616 (\zeta_5)^2 (\zeta_3)^2 + 14868 \zeta_4 \zeta_3 \zeta_9
  + 12432 \zeta_5 \zeta_2 \zeta_9  \nn\\
&&\hskip0.6cm\null
  + 13395 \zeta_6 \zeta_3 \zeta_7 + 42471 \zeta_{13} \zeta_3 \Bigr) a^9 \nn\\
&&\hskip0cm\null
  - \Bigl( \frac{40084328285043}{7018720} \zeta_{18}
  + \frac{311805}{8} (\zeta_7)^2 \zeta_4 + \frac{1665675}{8} \zeta_7 \zeta_{11}
  + \frac{920205}{4} \zeta_{13} \zeta_5  \nn\\
&&\hskip0.6cm\null
  + 36062 \zeta_8 (\zeta_5)^2 + \frac{28328314}{691} \zeta_{12} (\zeta_3)^2
  + 434 (\zeta_3)^4 \zeta_6 + \frac{2147145}{8} \zeta_{15} \zeta_3
  + 2352 \zeta_9 (\zeta_3)^3  \nn\\
&&\hskip0.6cm\null
  + 2160 (\zeta_5)^3 \zeta_3 + \frac{375564}{5} \zeta_{10} \zeta_3 \zeta_5
  + 1704 \zeta_4 (\zeta_3)^3 \zeta_5 + 1260 (\zeta_3)^3 \zeta_7 \zeta_2 \nn\\
&&\hskip0.6cm\null
  + 1836 (\zeta_3)^2 (\zeta_5)^2 \zeta_2 + 6594 \zeta_5 (\zeta_3)^2 \zeta_7
  + 74250 \zeta_{11} \zeta_2 \zeta_5
  + \frac{176715}{2} \zeta_{11} \zeta_4 \zeta_3 \nn\\
&&\hskip0.6cm\null
  + 84942 \zeta_{13} \zeta_2 \zeta_3 + 68850 \zeta_7 \zeta_2 \zeta_9
  + 80829 \zeta_9 \zeta_4 \zeta_5 + 76671 \zeta_6 \zeta_3 \zeta_9
  + \frac{146901}{2} \zeta_6 \zeta_5 \zeta_7  \nn\\
&&\hskip0.6cm\null
  + 72230 \zeta_8 \zeta_3 \zeta_7
  + \frac{403165}{4} (\zeta_9)^2 + 4 (\zeta_3)^6 \Bigr) a^{10} \nn\\
&&\hskip0.6cm\null
  + {\cal O}(a^{11}) \,.
\label{cusphighorder}
\eea

\newpage

\section{Four- and five-loop results for the MHV amplitude
in the $3\to3$ self-crossing limit for $v>1$}
\label{Evgt145}

In this appendix we give the full four- and five-loop results for 
the MHV amplitude, normalized by the BDS-like ansatz,
${\cal E} \equiv A_6^{\rm MHV}/A_6^{\rm BDS-like}$, 
in the $3\to3$ self-crossing limit for $v>1$.  We use the 
same notation used at lower loops in
eqs.~(\ref{Evgt1_0}), (\ref{Evgt1_1}), 
(\ref{Evgt1_2}) and (\ref{Evgt1_3});
the argument of the $h_i^{[w]}(z)$ is now $z=1/v$.
The four-loop result is
\bea
{\cal E}_{3\to3}^{(4)}(v>1) &=& 2 \pi i \, \biggl\{
- \frac{1}{672} \lndene{7}
\, - \, \frac{1}{80} \zeta_2 \lndene{5}
\, - \, \frac{1}{48} \zeta_3 \lndene{4}
\, - \, \frac{7}{24} \zeta_4 \lndene{3}
\nn\\  &&\hskip0.7cm\null
\, + \, \frac{1}{4} ( 4 \zeta_5 - 3 \zeta_2 \zeta_3 ) \lndene{2}
\, - \,\frac{1}{48} \Bigl( 13 \zeta_6 + 48 (\zeta_3)^2 \Bigr) \lnden \, 
\nn\\  &&\hskip0.7cm\null
  - \frac{1}{2} \biggl[
            120 h_{64}^{[7]} + 24 h_{66}^{[7]} + 24 h_{68}^{[7]} + 3 h_{70}^{[7]}
           + 24 h_{72}^{[7]} + 8 h_{74}^{[7]}
 + 4 h_{76}^{[7]} + h_{78}^{[7]}
\nn\\  &&\hskip1.5cm\null
           + 24 h_{80}^{[7]} + 9 h_{82}^{[7]} + 9 h_{84}^{[7]} + h_{86}^{[7]}
           + 6 h_{88}^{[7]} + 2 h_{90}^{[7]} + h_{92}^{[7]} + 2 h_{94}^{[7]}
           + 6 h_{98}^{[7]}
\nn\\  &&\hskip1.5cm\null
 + 6 h_{100}^{[7]} + h_{102}^{[7]}
           + 6 h_{104}^{[7]} + 2 h_{106}^{[7]} + h_{108}^{[7]} + 2 h_{110}^{[7]}
           + 12 h_{112}^{[7]} + 3 h_{114}^{[7]}
\nn\\  &&\hskip1.5cm\null
 + 3 h_{116}^{[7]} + 2 h_{118}^{[7]}
                           + 2 h_{122}^{[7]} + 4 h_{124}^{[7]}
\nn\\  &&\hskip1.5cm\null
           + \zeta_2 \Bigl(
                     - 36 h_{16}^{[5]} + 3 h_{17}^{[5]} - 4 h_{18}^{[5]} + h_{19}^{[5]}
                     - 3 h_{20}^{[5]} + h_{21}^{[5]} - h_{22}^{[5]} + 2 h_{23}^{[5]}
\nn\\  &&\hskip2.3cm\null
 + 6 h_{24}^{[5]} + h_{25}^{[5]} - h_{26}^{[5]} + 2 h_{27}^{[5]}
                     - 3 h_{28}^{[5]} + 2 h_{29}^{[5]} + 2 h_{30}^{[5]} \Bigr)
\nn\\  &&\hskip1.5cm\null
           + \zeta_3 \Bigl(
             22 h_{8}^{[4]} + 11 h_{9}^{[4]} + 8 h_{10}^{[4]} + h_{11}^{[4]}
                   + 5 h_{12}^{[4]} + h_{13}^{[4]} + h_{14}^{[4]} \Bigr)
\nn\\  &&\hskip1.5cm\null
           - \frac{1}{4} \zeta_4 \Bigl(
             39 h_{4}^{[3]} - 4 h_{5}^{[3]} - 4 h_{6}^{[3]} \Bigr)
           + \Bigl( 14 \zeta_5 - 5 \zeta_2 \zeta_3 \Bigr) h_{2}^{[2]}
\nn\\  &&\hskip1.5cm\null
           - \frac{1141}{8} \zeta_7 + \frac{119}{2} \zeta_2 \zeta_5
           + \frac{17}{2} \zeta_3 \zeta_4 \biggr] \biggl\}
\nn\\  &&\hskip0.0cm\null
- \frac{1}{2} \biggl\{
 240 h_{128}^{[8]} + 48 h_{130}^{[8]} + 48 h_{132}^{[8]} + 4 h_{134}^{[8]}
+ 48 h_{136}^{[8]} + 16 h_{138}^{[8]} + 6 h_{140}^{[8]} + 3 h_{142}^{[8]}
\nn\\  &&\hskip0.7cm\null
+ 48 h_{144}^{[8]} + 16 h_{146}^{[8]} + 16 h_{148}^{[8]} + h_{150}^{[8]}
+ 8 h_{152}^{[8]} + 3 h_{154}^{[8]} + 2 h_{156}^{[8]} + 5 h_{158}^{[8]}
\nn\\  &&\hskip0.7cm\null
+ 48 h_{160}^{[8]} + 18 h_{162}^{[8]} + 18 h_{164}^{[8]} + 2 h_{166}^{[8]}
+ 18 h_{168}^{[8]} + 5 h_{170}^{[8]} + 2 h_{172}^{[8]} + 5 h_{174}^{[8]}
\nn\\  &&\hskip0.7cm\null
+ 12 h_{176}^{[8]} + 4 h_{178}^{[8]} + 4 h_{180}^{[8]} + 4 h_{182}^{[8]}
+ 2 h_{184}^{[8]} + 4 h_{186}^{[8]} + 4 h_{188}^{[8]} + 6 h_{190}^{[8]}
\nn\\  &&\hskip0.7cm\null
               + 12 h_{194}^{[8]} + 12 h_{196}^{[8]} + 2 h_{198}^{[8]}
+ 12 h_{200}^{[8]} + 4 h_{202}^{[8]} + 2 h_{204}^{[8]} + 4 h_{206}^{[8]}
\nn\\  &&\hskip0.7cm\null
+ 12 h_{208}^{[8]} + 4 h_{210}^{[8]} + 4 h_{212}^{[8]} + 4 h_{214}^{[8]}
+ 2 h_{216}^{[8]} + 4 h_{218}^{[8]} + 4 h_{220}^{[8]} + 6 h_{222}^{[8]}
\nn\\  &&\hskip0.7cm\null
+ 24 h_{224}^{[8]} + 6 h_{226}^{[8]} + 6 h_{228}^{[8]} + 4 h_{230}^{[8]}
+ 6 h_{232}^{[8]} + 5 h_{234}^{[8]} + 5 h_{238}^{[8]} + 4 h_{236}^{[8]}
 \nn\\  &&\hskip0.7cm\null
              + 4 h_{242}^{[8]} + 4 h_{244}^{[8]} + 5 h_{246}^{[8]}
+ 8 h_{248}^{[8]} + 5 h_{250}^{[8]}                + 30 h_{254}^{[8]} \nn
\eea
\bea
&&\null
  + \zeta_2 \Bigl( 
         - 552 h_{32}^{[6]} + 4 h_{33}^{[6]} - 104 h_{34}^{[6]} + 3 h_{35}^{[6]}
         - 104 h_{36}^{[6]} + h_{37}^{[6]} - 17 h_{38}^{[6]} + 5 h_{39}^{[6]}
\nn\\  &&\hskip0.8cm\null
         - 102 h_{40}^{[6]} + 2 h_{41}^{[6]} - 40 h_{42}^{[6]} + 5 h_{43}^{[6]}
         - 26 h_{44}^{[6]} + 4 h_{45}^{[6]} - h_{46}^{[6]} + 6 h_{47}^{[6]}
\nn\\  &&\hskip0.8cm\null
         + 12 h_{48}^{[6]} + 2 h_{49}^{[6]} - 26 h_{50}^{[6]} + 4 h_{51}^{[6]}
         - 26 h_{52}^{[6]} + 4 h_{53}^{[6]} - h_{54}^{[6]} + 6 h_{55}^{[6]}
\nn\\  &&\hskip0.8cm\null
         - 54 h_{56}^{[6]} + 4 h_{57}^{[6]} - 10 h_{58}^{[6]} + 5 h_{59}^{[6]}
         + 4 h_{60}^{[6]} + 5 h_{61}^{[6]}  - 15 h_{62}^{[6]} + 30 h_{63}^{[6]} \Bigr)
\nn\\  &&\hskip0.0cm\null
+ \zeta_3 \Bigl( 44 h_{16}^{[5]} + 23 h_{17}^{[5]} + 15 h_{18}^{[5]} - h_{19}^{[5]}
       + 16 h_{20}^{[5]} + 3 h_{21}^{[5]}             - 2 h_{23}^{[5]}
\nn\\  &&\hskip0.8cm\null
       + 10 h_{24}^{[5]} + 2 h_{25}^{[5]}             - 2 h_{27}^{[5]}
       + 2 h_{28}^{[5]} + h_{29}^{[5]} - h_{30}^{[5]} - 20 h_{31}^{[5]} \Bigr) 
\nn\\  &&\hskip0.0cm\null
- \frac{1}{4} \zeta_4 \Bigl( 558 h_{8}^{[4]} + 142 h_{9}^{[4]}
             + 267 h_{10}^{[4]} + 22 h_{11}^{[4]}
             + 232 h_{12}^{[4]} + 22 h_{13}^{[4]} + 67 h_{14}^{[4]} + 58 h_{15}^{[4]}
                      \Bigr)
\nn\\  &&\hskip0.0cm\null
+ \zeta_5 \Bigl( 28 h_{4}^{[3]} + 21 h_{5}^{[3]} + 8 h_{6}^{[3]} - 16 h_{7}^{[3]} \Bigr) 
- \zeta_2 \zeta_3 \Bigl( 98 h_{4}^{[3]} + 54 h_{5}^{[3]}
                      + 25 h_{6}^{[3]} + 10 h_{7}^{[3]} \Bigr)
\nn\\  &&\hskip0.0cm\null
- \frac{1}{48} \zeta_6 \Bigl( 8477 h_{2}^{[2]} + 3687 h_{3}^{[2]}\Bigr ) 
+ \frac{3}{2} (\zeta_3)^2 \Bigl( 4 h_{2}^{[2]} + h_{3}^{[2]} \Bigr)
\nn\\  &&\hskip0.0cm\null
+ \frac{1}{2} \Bigl( 26 \zeta_7 - 228 \zeta_2 \zeta_5
                     - 175 \zeta_3 \zeta_4 \Bigr) h_{1}^{[1]}
- \frac{5}{2} \zeta_{5,3} - \frac{56911}{72} \zeta_8 
+ \frac{63}{2} \zeta_3 \zeta_5 - 36 \zeta_2 (\zeta_3)^2 \biggr\}
\,, \label{Evgt1_4}
\eea
and the five-loop one is
\bea
{\cal E}_{3\to3}^{(5)}(v>1) &=& 2 \pi i \, \biggl\{
\frac{1}{6912} \lndene{9}
\, + \, \frac{1}{336} \zeta_2 \lndene{7}
\, + \, \frac{5}{288} \zeta_3 \lndene{6}
\, + \, \frac{9}{80} \zeta_4 \lndene{5}
\nn\\  &&\hskip0.7cm\null
\, + \, \frac{1}{24} ( 6 \zeta_5 + 7 \zeta_2 \zeta_3 ) \lndene{4}
\, + \, \frac{1}{72} \Bigl( 115 \zeta_6 + 48 (\zeta_3)^2 \Bigr) \lndene{3}
\nn\\  &&\hskip0.7cm\null
\, + \, \frac{1}{16} ( - 55 \zeta_7 + 68 \zeta_2 \zeta_5
                  + 44 \zeta_3 \zeta_4 ) \lndene{2}
\nn\\  &&\hskip0.7cm\null
  + \frac{1}{72} \Bigl( 257 \zeta_8 + 810 \zeta_3 \zeta_5
                 + 18 \zeta_2 (\zeta_3)^2 \Bigr) \lnden \,
\nn\\  &&\hskip0.7cm\null
  + \frac{1}{4} \biggl[ 3360 h_{256}^{[9]} + 720 h_{258}^{[9]} + 720 h_{260}^{[9]}
 + 20 h_{262}^{[9]}
\nn\\  &&\hskip1.5cm\null
           + 720 h_{264}^{[9]} + 192 h_{266}^{[9]} + 32 h_{268}^{[9]} + 20 h_{270}^{[9]}
           + 720 h_{272}^{[9]} + 196 h_{274}^{[9]}
\nn\\  &&\hskip1.5cm\null
 + 196 h_{276}^{[9]} + 9 h_{278}^{[9]}
           + 48 h_{280}^{[9]} + 21 h_{282}^{[9]} + 16 h_{284}^{[9]} + 11 h_{286}^{[9]}
           + 720 h_{288}^{[9]}
\nn\\  &&\hskip1.5cm\null
 + 204 h_{290}^{[9]} + 204 h_{292}^{[9]} + 10 h_{294}^{[9]}
           + 204 h_{296}^{[9]} + 56 h_{298}^{[9]} + 10 h_{300}^{[9]} + 15 h_{302}^{[9]}
\nn\\  &&\hskip1.5cm\null
           + 72 h_{304}^{[9]} + 26 h_{306}^{[9]} + 26 h_{308}^{[9]} + 11 h_{310}^{[9]}
           + 12 h_{312}^{[9]} + 11 h_{314}^{[9]} + 14 h_{316}^{[9]}
\nn\\  &&\hskip1.5cm\null
 + 6 h_{318}^{[9]}
           + 720 h_{320}^{[9]} + 228 h_{322}^{[9]} + 228 h_{324}^{[9]} + 16 h_{326}^{[9]}
           + 228 h_{328}^{[9]} + 66 h_{330}^{[9]}
\nn\\  &&\hskip1.5cm\null
 + 16 h_{332}^{[9]} + 18 h_{334}^{[9]} 
           + 228 h_{336}^{[9]} + 66 h_{338}^{[9]} + 66 h_{340}^{[9]} + 17 h_{342}^{[9]}
           + 16 h_{344}^{[9]}
\nn\\  &&\hskip1.5cm\null
 + 18 h_{346}^{[9]} + 18 h_{348}^{[9]} + 10 h_{350}^{[9]}
           + 120 h_{352}^{[9]} + 40 h_{354}^{[9]} + 40 h_{356}^{[9]} + 16 h_{358}^{[9]}
\nn\\  &&\hskip1.5cm\null
           + 40 h_{360}^{[9]} + 22 h_{362}^{[9]} + 14 h_{364}^{[9]} + 11 h_{366}^{[9]}
           + 12 h_{368}^{[9]} + 14 h_{370}^{[9]} + 14 h_{372}^{[9]}
\nn\\  &&\hskip1.5cm\null
 + 11 h_{374}^{[9]}
           + 20 h_{376}^{[9]} + 13 h_{378}^{[9]} + 6 h_{380}^{[9]} + 18 h_{382}^{[9]}
                           + 120 h_{386}^{[9]} + 120 h_{388}^{[9]}
\nn\\  &&\hskip1.5cm\null
 + 19 h_{390}^{[9]}
           + 120 h_{392}^{[9]} + 43 h_{394}^{[9]} + 18 h_{396}^{[9]} + 20 h_{398}^{[9]}
           + 120 h_{400}^{[9]} + 42 h_{402}^{[9]}
\nn
\eea
\bea
&&\null
 + 42 h_{404}^{[9]} + 18 h_{406}^{[9]}
           + 16 h_{408}^{[9]} + 19 h_{410}^{[9]} + 20 h_{412}^{[9]} + 10 h_{414}^{[9]}
           + 120 h_{416}^{[9]} + 40 h_{418}^{[9]} + 40 h_{420}^{[9]}
\nn\\  &&\hskip0.0cm\null
  + 16 h_{422}^{[9]}
  + 40 h_{424}^{[9]} + 22 h_{426}^{[9]} + 14 h_{428}^{[9]} + 11 h_{430}^{[9]}
           + 12 h_{432}^{[9]} + 14 h_{434}^{[9]} + 14 h_{436}^{[9]} + 11 h_{438}^{[9]}
\nn\\  &&\hskip0.0cm\null
           + 20 h_{440}^{[9]} + 13 h_{442}^{[9]} + 6 h_{444}^{[9]}
 + 18 h_{446}^{[9]}
           + 240 h_{448}^{[9]} + 60 h_{450}^{[9]} + 60 h_{452}^{[9]} + 14 h_{454}^{[9]}
           + 60 h_{456}^{[9]}
\nn\\  &&\hskip0.0cm\null
           + 24 h_{458}^{[9]}
 + 16 h_{460}^{[9]} + 8 h_{462}^{[9]}
           + 60 h_{464}^{[9]} + 26 h_{466}^{[9]} + 26 h_{468}^{[9]} + 12 h_{470}^{[9]}
           + 20 h_{472}^{[9]} + 14 h_{474}^{[9]}
\nn\\  &&\hskip0.0cm\null
           + 8 h_{476}^{[9]} + 18 h_{478}^{[9]}
                           + 20 h_{482}^{[9]}
 + 20 h_{484}^{[9]} + 12 h_{486}^{[9]}
           + 20 h_{488}^{[9]} + 14 h_{490}^{[9]} + 8 h_{492}^{[9]} + 18 h_{494}^{[9]} 
\nn\\  &&\hskip0.0cm\null
           + 24 h_{496}^{[9]} + 8 h_{498}^{[9]} + 8 h_{500}^{[9]}
 + 18 h_{502}^{[9]}
           + 16 h_{504}^{[9]} + 22 h_{506}^{[9]} + 20 h_{508}^{[9]}
\nn\\  &&\hskip0.0cm\null
           + \zeta_2 \Bigl( 
               - 960 h_{64}^{[7]}  + 20 h_{65}^{[7]} - 168 h_{66}^{[7]} + 20 h_{67}^{[7]}
               - 164 h_{68}^{[7]} + 9 h_{69}^{[7]} - 3 h_{70}^{[7]} + 11 h_{71}^{[7]}
               - 156 h_{72}^{[7]}
\nn\\  &&\hskip0.8cm\null
                + 10 h_{73}^{[7]} - 46 h_{74}^{[7]} + 15 h_{75}^{[7]}
               - 10 h_{76}^{[7]} + 11 h_{77}^{[7]} + 5 h_{78}^{[7]} + 6 h_{79}^{[7]}
               - 132 h_{80}^{[7]} + 16 h_{81}^{[7]} - 48 h_{82}^{[7]}
\nn\\  &&\hskip0.8cm\null
               + 18 h_{83}^{[7]} - 48 h_{84}^{[7]} + 17 h_{85}^{[7]}
               + 10 h_{86}^{[7]} + 10 h_{87}^{[7]}
               - 20 h_{88}^{[7]} + 16 h_{89}^{[7]} + 2 h_{90}^{[7]} + 11 h_{91}^{[7]}
               + 8 h_{92}^{[7]}
\nn\\  &&\hskip0.8cm\null
               + 11 h_{93}^{[7]} + 3 h_{94}^{[7]} + 18 h_{95}^{[7]}
               + 120 h_{96}^{[7]} + 19 h_{97}^{[7]} - 17 h_{98}^{[7]} + 20 h_{99}^{[7]}
               - 18 h_{100}^{[7]} + 18 h_{101}^{[7]} + 11 h_{102}^{[7]}
\nn\\  &&\hskip0.8cm\null
               + 10 h_{103}^{[7]}
               - 20 h_{104}^{[7]} + 16 h_{105}^{[7]} + 2 h_{106}^{[7]} + 11 h_{107}^{[7]}
               + 8 h_{108}^{[7]} + 11 h_{109}^{[7]} + 3 h_{110}^{[7]} + 18 h_{111}^{[7]}
\nn\\  &&\hskip0.8cm\null
               - 60 h_{112}^{[7]}
              + 14 h_{113}^{[7]} - 6 h_{114}^{[7]} + 8 h_{115}^{[7]}
               - 4 h_{116}^{[7]} + 12 h_{117}^{[7]} + 4 h_{118}^{[7]} + 18 h_{119}^{[7]}
\nn\\  &&\hskip0.8cm\null
               + 20 h_{120}^{[7]} + 12 h_{121}^{[7]}
               + 4 h_{122}^{[7]} + 18 h_{123}^{[7]}
               - 4 h_{124}^{[7]} + 18 h_{125}^{[7]} + 14 h_{126}^{[7]} \Bigr)
\nn\\  &&\hskip0.0cm\null 
+ \zeta_3 \Bigl( 760 h_{32}^{[6]} + 332 h_{33}^{[6]} + 200 h_{34}^{[6]} + 15 h_{35}^{[6]}
                  + 208 h_{36}^{[6]} + 87 h_{37}^{[6]} + 14 h_{38}^{[6]} + 2 h_{39}^{[6]}
                  + 228 h_{40}^{[6]}
\nn\\  &&\hskip0.8cm\null 
 + 98 h_{41}^{[6]} + 55 h_{42}^{[6]} + 8 h_{43}^{[6]}
                  + 24 h_{44}^{[6]} + 19 h_{45}^{[6]} + 5 h_{46}^{[6]} + 2 h_{47}^{[6]}
                  + 108 h_{48}^{[6]} + 48 h_{49}^{[6]} + 26 h_{50}^{[6]}
\nn\\  &&\hskip0.8cm\null 
 + 8 h_{51}^{[6]}
                  + 24 h_{52}^{[6]} + 19 h_{53}^{[6]} + 5 h_{54}^{[6]} + 2 h_{55}^{[6]}
                  + 44 h_{56}^{[6]} + 24 h_{57}^{[6]} + 17 h_{58}^{[6]} + 2 h_{59}^{[6]}
                  + 4 h_{60}^{[6]}
\nn\\  &&\hskip0.8cm\null 
 + 2 h_{61}^{[6]} - 2 h_{62}^{[6]} + 24 h_{63}^{[6]} \Bigr)
\nn\\  &&\hskip0.0cm\null 
           + \frac{1}{4} \zeta_4 \Bigl(
              - 1380 h_{16}^{[5]} + 30 h_{17}^{[5]} - 296 h_{18}^{[5]} + 4 h_{19}^{[5]}
              - 302 h_{20}^{[5]} + 11 h_{21}^{[5]} - 28 h_{22}^{[5]} + 42 h_{23}^{[5]}
              + 12 h_{24}^{[5]}
\nn\\  &&\hskip1.3cm\null 
 + 11 h_{25}^{[5]} - 28 h_{26}^{[5]} + 42 h_{27}^{[5]}
              - 74 h_{28}^{[5]} + 42 h_{29}^{[5]} + 42 h_{30}^{[5]}
  - 20 h_{31}^{[5]} \Bigr)
\nn\\  &&\hskip0.0cm\null 
           + \frac{1}{2} \zeta_5 \Bigl( 
            1104 h_{8}^{[4]} + 516 h_{9}^{[4]} + 355 h_{10}^{[4]} + 76 h_{11}^{[4]}
          + 194 h_{12}^{[4]} + 76 h_{13}^{[4]} + 56 h_{14}^{[4]} + 56 h_{15}^{[4]} \Bigr) 
\nn\\  &&\hskip0.0cm\null 
           - \zeta_2 \zeta_3 \Bigl( 
             200 h_{8}^{[4]} + 87 h_{9}^{[4]} + 62 h_{10}^{[4]} + 10 h_{11}^{[4]}
            + 32 h_{12}^{[4]} + 10 h_{13}^{[4]} + 9 h_{14}^{[4]} - 4 h_{15}^{[4]} \Bigr)
\nn\\  &&\hskip0.0cm\null 
    - \frac{1}{48} \zeta_6 \Bigl( 10540 h_{4}^{[3]} - 75 ( h_{5}^{[3]} + h_{6}^{[3]} )
                         + 372 h_{7}^{[3]} \Bigr)
           + \frac{1}{2} (\zeta_3)^2 \Bigl( 196 h_{4}^{[3]} + 37 h_{5}^{[3]}
                           + 21 h_{6}^{[3]} - 8 h_{7}^{[3]} \Bigr)
\nn\\  &&\hskip0.0cm\null 
       + \frac{1}{16} \zeta_7 \Bigl( 6280 h_{2}^{[2]} + 183 h_{3}^{[2]} \Bigr) 
       - \frac{1}{2} \zeta_2 \zeta_5 \Bigl( 284 h_{2}^{[2]} - 5 h_{3}^{[2]} \Bigr) 
       - \frac{1}{4} \zeta_3 \zeta_4 \Bigl( 355 h_{2}^{[2]} - 22 h_{3}^{[2]} \Bigr)
\nn\\  &&\hskip0.0cm\null 
           - \frac{1}{2} \Bigl( \zeta_{5,3} + \frac{217}{24} \zeta_8
                  + 15 \zeta_2 (\zeta_3)^2 - 17 \zeta_3 \zeta_5 \Bigr) h_{1}^{[1]}
\nn\\  &&\hskip0.0cm\null 
       - \frac{40369}{16} \zeta_9 + \frac{7645}{8} \zeta_2 \zeta_7
       + \frac{3119}{16} \zeta_3 \zeta_6
       + \frac{2295}{8} \zeta_4 \zeta_5 - 15 (\zeta_3)^3 \biggr] \biggr\}
\nn
\eea
\bea 
&&\null 
+ \frac{1}{8} \biggl\{ 
 13440 h_{512}^{[10]} + 2880 h_{514}^{[10]} + 2880 h_{516}^{[10]} - 160 h_{518}^{[10]}
+ 2880 h_{520}^{[10]} + 768 h_{522}^{[10]} + 80 h_{524}^{[10]}
\nn\\  &&\hskip0.0cm\null 
 + 152 h_{526}^{[10]} + 2880 h_{528}^{[10]} + 768 h_{530}^{[10]}
 + 768 h_{532}^{[10]} - 24 h_{534}^{[10]}
+ 128 h_{536}^{[10]} + 72 h_{538}^{[10]} + 80 h_{540}^{[10]}
\nn\\  &&\hskip0.0cm\null 
 + 48 h_{542}^{[10]}
+ 2880 h_{544}^{[10]} + 784 h_{546}^{[10]} + 784 h_{548}^{[10]} - 16 h_{550}^{[10]}
+ 784 h_{552}^{[10]} + 212 h_{554}^{[10]} + 36 h_{556}^{[10]}
\nn\\  &&\hskip0.0cm\null 
+ 80 h_{558}^{[10]}
+ 192 h_{560}^{[10]} + 84 h_{562}^{[10]} + 84 h_{564}^{[10]} + 44 h_{566}^{[10]}
+ 64 h_{568}^{[10]} + 44 h_{570}^{[10]} + 44 h_{572}^{[10]}
+ 17 h_{574}^{[10]}
\nn\\  &&\hskip0.0cm\null 
+ 2880 h_{576}^{[10]} + 816 h_{578}^{[10]} + 816 h_{580}^{[10]} - 16 h_{582}^{[10]}
+ 816 h_{584}^{[10]} + 224 h_{586}^{[10]} + 40 h_{588}^{[10]} + 74 h_{590}^{[10]}
\nn\\  &&\hskip0.0cm\null 
+ 816 h_{592}^{[10]} + 224 h_{594}^{[10]} + 224 h_{596}^{[10]} + 34 h_{598}^{[10]}
+ 40 h_{600}^{[10]} + 50 h_{602}^{[10]} + 60 h_{604}^{[10]} + 21 h_{606}^{[10]}
+ 288 h_{608}^{[10]} 
\nn\\  &&\hskip0.0cm\null 
+ 104 h_{610}^{[10]} + 104 h_{612}^{[10]} + 48 h_{614}^{[10]}
+ 104 h_{616}^{[10]} + 58 h_{618}^{[10]} + 44 h_{620}^{[10]} + 29 h_{622}^{[10]}
+ 48 h_{624}^{[10]} + 44 h_{626}^{[10]}
\nn\\  &&\hskip0.0cm\null 
+ 44 h_{628}^{[10]} + 30 h_{630}^{[10]}
+ 56 h_{632}^{[10]} + 34 h_{634}^{[10]} + 24 h_{636}^{[10]} + 80 h_{638}^{[10]}
+ 2880 h_{640}^{[10]} + 912 h_{642}^{[10]} + 912 h_{644}^{[10]}
\nn\\  &&\hskip0.0cm\null 
+ 912 h_{648}^{[10]} + 264 h_{650}^{[10]} + 64 h_{652}^{[10]} + 88 h_{654}^{[10]}
+ 912 h_{656}^{[10]} + 264 h_{658}^{[10]} + 264 h_{660}^{[10]} + 44 h_{662}^{[10]}
+ 64 h_{664}^{[10]}
\nn\\  &&\hskip0.0cm\null 
+ 64 h_{666}^{[10]} + 72 h_{668}^{[10]} + 36 h_{670}^{[10]}
+ 912 h_{672}^{[10]} + 264 h_{674}^{[10]} + 264 h_{676}^{[10]} + 44 h_{678}^{[10]}
+ 264 h_{680}^{[10]} + 118 h_{682}^{[10]}
\nn\\  &&\hskip0.0cm\null 
 + 68 h_{684}^{[10]} + 50 h_{686}^{[10]}
+ 64 h_{688}^{[10]} + 72 h_{690}^{[10]} + 72 h_{692}^{[10]} + 46 h_{694}^{[10]}
+ 72 h_{696}^{[10]} + 50 h_{698}^{[10]} + 40 h_{700}^{[10]}
\nn\\  &&\hskip0.0cm\null 
+ 84 h_{702}^{[10]}
+ 480 h_{704}^{[10]} + 160 h_{706}^{[10]} + 160 h_{708}^{[10]} + 64 h_{710}^{[10]}
+ 160 h_{712}^{[10]} + 88 h_{714}^{[10]} + 64 h_{716}^{[10]} + 42 h_{718}^{[10]}
\nn\\  &&\hskip0.0cm\null 
+ 160 h_{720}^{[10]} + 88 h_{722}^{[10]} + 88 h_{724}^{[10]} + 48 h_{726}^{[10]}
+ 56 h_{728}^{[10]} + 50 h_{730}^{[10]} + 44 h_{732}^{[10]} + 81 h_{734}^{[10]}
+ 48 h_{736}^{[10]}
\nn\\  &&\hskip0.0cm\null 
+ 56 h_{738}^{[10]} + 56 h_{740}^{[10]} + 36 h_{742}^{[10]}
+ 56 h_{744}^{[10]} + 50 h_{746}^{[10]} + 44 h_{748}^{[10]} + 80 h_{750}^{[10]}
+ 80 h_{752}^{[10]} + 52 h_{754}^{[10]}
\nn\\  &&\hskip0.0cm\null 
+ 52 h_{756}^{[10]} + 78 h_{758}^{[10]}
+ 24 h_{760}^{[10]} + 78 h_{762}^{[10]} + 72 h_{764}^{[10]} + 180 h_{766}^{[10]}
                 + 480 h_{770}^{[10]} + 480 h_{772}^{[10]} + 48 h_{774}^{[10]}
\nn\\  &&\hskip0.0cm\null 
+ 480 h_{776}^{[10]} + 172 h_{778}^{[10]} + 76 h_{780}^{[10]} + 86 h_{782}^{[10]}
+ 480 h_{784}^{[10]} + 172 h_{786}^{[10]} + 172 h_{788}^{[10]} + 66 h_{790}^{[10]}
+ 72 h_{792}^{[10]}
\nn\\  &&\hskip0.0cm\null 
+ 78 h_{794}^{[10]} + 80 h_{796}^{[10]} + 38 h_{798}^{[10]}
+ 480 h_{800}^{[10]} + 168 h_{802}^{[10]} + 168 h_{804}^{[10]} + 64 h_{806}^{[10]}
+ 168 h_{808}^{[10]} + 98 h_{810}^{[10]}
\nn\\  &&\hskip0.0cm\null 
+ 72 h_{812}^{[10]} + 44 h_{814}^{[10]}
+ 64 h_{816}^{[10]} + 76 h_{818}^{[10]} + 76 h_{820}^{[10]} + 44 h_{822}^{[10]}
+ 80 h_{824}^{[10]} + 52 h_{826}^{[10]} + 40 h_{828}^{[10]}
\nn\\  &&\hskip0.0cm\null 
+ 62 h_{830}^{[10]}
+ 480 h_{832}^{[10]} + 160 h_{834}^{[10]} + 160 h_{836}^{[10]} + 64 h_{838}^{[10]}
+ 160 h_{840}^{[10]} + 88 h_{842}^{[10]} + 64 h_{844}^{[10]} + 42 h_{846}^{[10]}
\nn\\  &&\hskip0.0cm\null 
+ 160 h_{848}^{[10]} + 88 h_{850}^{[10]} + 88 h_{852}^{[10]} + 48 h_{854}^{[10]}
+ 56 h_{856}^{[10]} + 50 h_{858}^{[10]} + 44 h_{860}^{[10]} + 81 h_{862}^{[10]}
+ 48 h_{864}^{[10]} 
\nn\\  &&\hskip0.0cm\null 
+ 56 h_{866}^{[10]} + 56 h_{868}^{[10]} + 36 h_{870}^{[10]}
+ 56 h_{872}^{[10]} + 50 h_{874}^{[10]} + 44 h_{876}^{[10]} + 80 h_{878}^{[10]}
+ 80 h_{880}^{[10]} + 52 h_{882}^{[10]}
\nn\\  &&\hskip0.0cm\null
+ 52 h_{884}^{[10]} + 78 h_{886}^{[10]}
+ 24 h_{888}^{[10]} + 78 h_{890}^{[10]} + 72 h_{892}^{[10]} + 180 h_{894}^{[10]}
+ 960 h_{896}^{[10]} + 240 h_{898}^{[10]} + 240 h_{900}^{[10]}
\nn\\  &&\hskip0.0cm\null+ 64 h_{902}^{[10]}
+ 240 h_{904}^{[10]} + 96 h_{906}^{[10]} + 56 h_{908}^{[10]} + 32 h_{910}^{[10]}
+ 240 h_{912}^{[10]} + 96 h_{914}^{[10]} + 96 h_{916}^{[10]} + 32 h_{918}^{[10]}
\nn\\  &&\hskip0.0cm\null
+ 64 h_{920}^{[10]} + 36 h_{922}^{[10]} + 32 h_{924}^{[10]} + 50 h_{926}^{[10]}
+ 240 h_{928}^{[10]} + 104 h_{930}^{[10]} + 104 h_{932}^{[10]} + 36 h_{934}^{[10]}
+ 104 h_{936}^{[10]}
\nn\\  &&\hskip0.0cm\null
+ 62 h_{938}^{[10]} + 48 h_{940}^{[10]} + 79 h_{942}^{[10]}
+ 80 h_{944}^{[10]} + 56 h_{946}^{[10]} + 56 h_{948}^{[10]} + 76 h_{950}^{[10]}
+ 32 h_{952}^{[10]} + 76 h_{954}^{[10]}
\nn\\  &&\hskip0.0cm\null 
 + 72 h_{956}^{[10]} + 150 h_{958}^{[10]}
                 + 80 h_{962}^{[10]} + 80 h_{964}^{[10]} + 64 h_{966}^{[10]}
+ 80 h_{968}^{[10]} + 56 h_{970}^{[10]} + 48 h_{972}^{[10]} + 64 h_{974}^{[10]}
\nn\\  &&\hskip0.0cm\null
+ 80 h_{976}^{[10]} + 56 h_{978}^{[10]} + 56 h_{980}^{[10]} + 84 h_{982}^{[10]}
+ 32 h_{984}^{[10]} + 76 h_{986}^{[10]} + 72 h_{988}^{[10]} + 144 h_{990}^{[10]}
+ 96 h_{992}^{[10]}
\nn\\  &&\hskip0.0cm\null
+ 32 h_{994}^{[10]} + 32 h_{996}^{[10]} + 40 h_{998}^{[10]}
+ 32 h_{1000}^{[10]} + 76 h_{1002}^{[10]} + 72 h_{1004}^{[10]} + 140 h_{1006}^{[10]}
+ 64 h_{1008}^{[10]} + 88 h_{1010}^{[10]}
\nn\\  &&\hskip0.0cm\null 
+ 88 h_{1012}^{[10]} + 140 h_{1014}^{[10]}
+ 80 h_{1016}^{[10]} + 140 h_{1018}^{[10]} + 840 h_{1022}^{[10]}
\nn\\  &&\hskip0.0cm\null 
\nn
\eea
\bea
&&\null
+ \zeta_2 \Bigl(
      - 30720 h_{128}^{[8]} - 160 h_{129}^{[8]} - 6432 h_{130}^{[8]} + 152 h_{131}^{[8]}
        - 6432 h_{132}^{[8]} - 24 h_{133}^{[8]} - 248 h_{134}^{[8]}
\nn\\  &&\hskip0.8cm\null 
 + 48 h_{135}^{[8]}
        - 6416 h_{136}^{[8]} - 16 h_{137}^{[8]} - 1748 h_{138}^{[8]} + 80 h_{139}^{[8]}
        - 396 h_{140}^{[8]} + 44 h_{141}^{[8]} - 116 h_{142}^{[8]}
\nn\\  &&\hskip0.8cm\null 
 + 17 h_{143}^{[8]}
        - 6384 h_{144}^{[8]} - 16 h_{145}^{[8]} - 1816 h_{146}^{[8]} + 74 h_{147}^{[8]}
        - 1816 h_{148}^{[8]} + 34 h_{149}^{[8]} - 50 h_{150}^{[8]}
\nn\\  &&\hskip0.8cm\null 
 + 21 h_{151}^{[8]}
        - 616 h_{152}^{[8]} + 48 h_{153}^{[8]} - 202 h_{154}^{[8]} + 29 h_{155}^{[8]}
        - 76 h_{156}^{[8]} + 30 h_{157}^{[8]} - 106 h_{158}^{[8]}
\nn\\  &&\hskip0.8cm\null 
 + 80 h_{159}^{[8]}
        - 6288 h_{160}^{[8]}              - 2016 h_{162}^{[8]} + 88 h_{163}^{[8]}
        - 2016 h_{164}^{[8]} + 44 h_{165}^{[8]} - 96 h_{166}^{[8]} + 36 h_{167}^{[8]}
\nn\\  &&\hskip0.8cm\null 
        - 2016 h_{168}^{[8]} + 44 h_{169}^{[8]} - 542 h_{170}^{[8]} + 50 h_{171}^{[8]}
        - 88 h_{172}^{[8]} + 46 h_{173}^{[8]} - 130 h_{174}^{[8]} + 84 h_{175}^{[8]}
\nn\\  &&\hskip0.8cm\null 
        - 1040 h_{176}^{[8]} + 64 h_{177}^{[8]} - 312 h_{178}^{[8]} + 42 h_{179}^{[8]}
        - 312 h_{180}^{[8]} - 90 h_{182}^{[8]} + 48 h_{181}^{[8]} + 81 h_{183}^{[8]}
\nn\\  &&\hskip0.8cm\null 
        - 64 h_{184}^{[8]} + 36 h_{185}^{[8]} - 90 h_{186}^{[8]} + 80 h_{187}^{[8]}
        - 148 h_{188}^{[8]} + 78 h_{189}^{[8]} + 18 h_{190}^{[8]} + 180 h_{191}^{[8]}
\nn\\  &&\hskip0.8cm\null 
        + 480 h_{192}^{[8]} + 48 h_{193}^{[8]} - 1028 h_{194}^{[8]} + 86 h_{195}^{[8]}
        - 1028 h_{196}^{[8]} + 66 h_{197}^{[8]} - 102 h_{198}^{[8]} + 38 h_{199}^{[8]}
\nn\\  &&\hskip0.8cm\null 
        - 1032 h_{200}^{[8]} + 64 h_{201}^{[8]} - 322 h_{202}^{[8]} + 44 h_{203}^{[8]}
        - 84 h_{204}^{[8]} + 44 h_{205}^{[8]} - 148 h_{206}^{[8]} + 62 h_{207}^{[8]}
\nn\\  &&\hskip0.8cm\null 
        - 1040 h_{208}^{[8]} + 64 h_{209}^{[8]} - 312 h_{210}^{[8]} + 42 h_{211}^{[8]}
        - 312 h_{212}^{[8]} + 48 h_{213}^{[8]} - 90 h_{214}^{[8]} + 81 h_{215}^{[8]}
\nn\\  &&\hskip0.8cm\null 
        - 64 h_{216}^{[8]} + 36 h_{217}^{[8]} - 90 h_{218}^{[8]} + 80 h_{219}^{[8]}
        - 148 h_{220}^{[8]} + 78 h_{221}^{[8]} + 18 h_{222}^{[8]} + 180 h_{223}^{[8]}
\nn\\  &&\hskip0.8cm\null 
        - 2160 h_{224}^{[8]} + 64 h_{225}^{[8]} - 504 h_{226}^{[8]} + 32 h_{227}^{[8]}
        - 504 h_{228}^{[8]} + 32 h_{229}^{[8]} - 124 h_{230}^{[8]} + 50 h_{231}^{[8]}
\nn\\  &&\hskip0.8cm\null 
        - 496 h_{232}^{[8]} + 36 h_{233}^{[8]} - 198 h_{234}^{[8]} + 79 h_{235}^{[8]}
        - 144 h_{236}^{[8]} + 76 h_{237}^{[8]} - 4 h_{238}^{[8]} + 150 h_{239}^{[8]}
\nn\\  &&\hskip0.8cm\null 
        + 80 h_{240}^{[8]} + 64 h_{241}^{[8]} - 144 h_{242}^{[8]} + 64 h_{243}^{[8]}
        - 144 h_{244}^{[8]} + 84 h_{245}^{[8]} - 4 h_{246}^{[8]} + 144 h_{247}^{[8]}
\nn\\  &&\hskip0.8cm\null 
        - 208 h_{248}^{[8]} + 40 h_{249}^{[8]} - 4 h_{250}^{[8]} + 140 h_{251}^{[8]}
      - 72 h_{252}^{[8]} + 140 h_{253}^{[8]} - 60 h_{254}^{[8]} + 840 h_{255}^{[8]} \Bigr)
\nn\\  &&\hskip0.0cm\null 
+ \zeta_3 \Bigl(
         3040 h_{64}^{[7]} + 1304 h_{65}^{[7]} + 792 h_{66}^{[7]} + 16 h_{67}^{[7]}
        + 800 h_{68}^{[7]} + 308 h_{69}^{[7]} + 40 h_{70}^{[7]} + 27 h_{71}^{[7]}
\nn\\  &&\hskip0.8cm\null 
        + 832 h_{72}^{[7]} + 334 h_{73}^{[7]} + 190 h_{74}^{[7]} + 19 h_{75}^{[7]}
        + 56 h_{76}^{[7]} + 43 h_{77}^{[7]} + 14 h_{78}^{[7]} - 36 h_{79}^{[7]}
\nn\\  &&\hskip0.8cm\null
        + 912 h_{80}^{[7]} + 376 h_{81}^{[7]} + 220 h_{82}^{[7]} + 20 h_{83}^{[7]}
        + 220 h_{84}^{[7]} + 118 h_{85}^{[7]} + 26 h_{86}^{[7]} - 24 h_{87}^{[7]}
\nn\\  &&\hskip0.8cm\null
        + 96 h_{88}^{[7]} + 70 h_{89}^{[7]} + 40 h_{90}^{[7]} - 25 h_{91}^{[7]}
        + 20 h_{92}^{[7]} - 24 h_{93}^{[7]} - 26 h_{94}^{[7]} - 96 h_{95}^{[7]}
\nn\\  &&\hskip0.8cm\null
        + 432 h_{96}^{[7]} + 182 h_{97}^{[7]} + 106 h_{98}^{[7]} + 38 h_{99}^{[7]}
        + 104 h_{100}^{[7]} + 80 h_{101}^{[7]} + 32 h_{102}^{[7]} + 2 h_{103}^{[7]}
\nn\\  &&\hskip0.8cm\null
        + 96 h_{104}^{[7]} + 70 h_{105}^{[7]} + 40 h_{106}^{[7]} - 25 h_{107}^{[7]}
        + 20 h_{108}^{[7]} - 24 h_{109}^{[7]} - 26 h_{110}^{[7]} - 96 h_{111}^{[7]}
\nn\\  &&\hskip0.8cm\null
        + 176 h_{112}^{[7]}  + 104 h_{113}^{[7]} + 64 h_{114}^{[7]} - 10 h_{115}^{[7]}
        + 68 h_{116}^{[7]} - 3 h_{117}^{[7]} - 20 h_{118}^{[7]} - 70 h_{119}^{[7]}
\nn\\  &&\hskip0.8cm\null
        + 16 h_{120}^{[7]}               - 28 h_{122}^{[7]} - 64 h_{123}^{[7]}
        - 8 h_{124}^{[7]} - 60 h_{125}^{[7]} - 52 h_{126}^{[7]} - 560 h_{127}^{[7]} \Bigr) 
\nn\\  &&\hskip0.0cm\null
- \zeta_4 \Bigl(
   9060 h_{32}^{[6]} + 520 h_{33}^{[6]} + 2766 h_{34}^{[6]} + 330 h_{35}^{[6]}
 + 2936 h_{36}^{[6]} + 179 h_{37}^{[6]} + \frac{785}{2} h_{38}^{[6]} + 220 h_{39}^{[6]}
\nn\\  &&\hskip0.8cm\null
 + 3422 h_{40}^{[6]} + 289 h_{41}^{[6]} + \frac{1899}{2} h_{42}^{[6]} + 304 h_{43}^{[6]}
 + 588 h_{44}^{[6]} + \frac{449}{2} h_{45}^{[6]} + \frac{405}{2} h_{46}^{[6]}
 - 33 h_{47}^{[6]}
\nn\\  &&\hskip0.8cm\null
 + 2388 h_{48}^{[6]} + 344 h_{49}^{[6]} + \frac{1261}{2} h_{50}^{[6]} + 364 h_{51}^{[6]}
 + 588 h_{52}^{[6]} + \frac{449}{2} h_{53}^{[6]} + \frac{405}{2} h_{54}^{[6]}
 - 33 h_{55}^{[6]}
\nn\\  &&\hskip0.8cm\null
 + 794 h_{56}^{[6]} + 291 h_{57}^{[6]} + \frac{735}{2} h_{58}^{[6]} + 38 h_{59}^{[6]}
       + 358 h_{60}^{[6]} + 54 h_{61}^{[6]} + 15 h_{62}^{[6]} - 350 h_{63}^{[6]} \Bigr) 
\nn
\eea
\bea
&&\hskip0.0cm\null
+ \zeta_5 \Bigl(
         2208 h_{16}^{[5]} + 1097 h_{17}^{[5]} + 657 h_{18}^{[5]} + 33 h_{19}^{[5]}
        + 710 h_{20}^{[5]} + 266 h_{21}^{[5]} + 50 h_{22}^{[5]} - 30 h_{23}^{[5]}
\nn\\  &&\hskip0.8cm\null
        + 388 h_{24}^{[5]} + 138 h_{25}^{[5]} + 50 h_{26}^{[5]} - 30 h_{27}^{[5]}
        + 112 h_{28}^{[5]} + 57 h_{29}^{[5]}                 - 448 h_{31}^{[5]} \Bigr)
\nn\\  &&\hskip0.0cm\null
- \zeta_2 \zeta_3 \Bigl(
        6880 h_{16}^{[5]} + 2940 h_{17}^{[5]} + 1896 h_{18}^{[5]} + 227 h_{19}^{[5]}
         + 2072 h_{20}^{[5]} + 856 h_{21}^{[5]} + 212 h_{22}^{[5]} + 70 h_{23}^{[5]}
\nn\\  &&\hskip0.8cm\null
        + 992 h_{24}^{[5]} + 416 h_{25}^{[5]} + 212 h_{26}^{[5]} + 70 h_{27}^{[5]}
        + 388 h_{28}^{[5]} + 259 h_{29}^{[5]} + 68 h_{30}^{[5]} - 24 h_{31}^{[5]} \Bigr) 
\nn\\  &&\hskip0.0cm\null
+ \zeta_6 \Bigl(
  - \frac{28465}{3} h_{8}^{[4]} - \frac{3819}{4} h_{9}^{[4]}
  - \frac{23921}{8} h_{10}^{[4]} - \frac{2239}{24} h_{11}^{[4]}
  - \frac{6863}{4} h_{12}^{[4]} - \frac{2239}{24} h_{13}^{[4]}
\nn\\  &&\hskip0.8cm\null
  - \frac{19765}{24} h_{14}^{[4]} - \frac{5}{12} h_{15}^{[4]} \Bigr)
\nn\\   &&\hskip0.0cm\null
+ (\zeta_3)^2 \Bigl( 392 h_{8}^{[4]} + \frac{103}{2} h_{9}^{[4]}
         + 111 h_{10}^{[4]} + \frac{41}{2} h_{11}^{[4]}
         + 42 h_{12}^{[4]} + \frac{41}{2} h_{13}^{[4]}
         + 28 h_{14}^{[4]} + 34 h_{15}^{[4]} \Bigr)
\nn\\   &&\hskip0.0cm\null
+ \zeta_7 \Bigl( 1570 h_{4}^{[3]} + \frac{6959}{8} h_{5}^{[3]}
          + \frac{2319}{8} h_{6}^{[3]} - \frac{3777}{16} h_{7}^{[3]} \Bigr)
\nn\\   &&\hskip0.0cm\null
- \zeta_2 \zeta_5 \Bigl( 4984 h_{4}^{[3]} + 2367 h_{5}^{[3]}
                 + 921 h_{6}^{[3]} + \frac{659}{2} h_{7}^{[3]} \Bigr)
\nn\\   &&\hskip0.0cm\null
- \zeta_3 \zeta_4 \Bigl( 2435 h_{4}^{[3]} + 1056 h_{5}^{[3]}
                  + \frac{601}{2} h_{6}^{[3]} + 647 h_{7}^{[3]} \Bigr)
- \frac{9}{10} \zeta_{5,3} h_{3}^{[2]} 
\nn\\   &&\hskip0.0cm\null
- \zeta_8 \Bigl( \frac{2461055}{288} h_{2}^{[2]}
               + \frac{4148393}{1440} h_{3}^{[2]} \Bigr)
- \zeta_2 (\zeta_3)^2 \Bigl( 886 h_{2}^{[2]} + 165 h_{3}^{[2]} \Bigr)
+ \zeta_3 \zeta_5 \Bigl( 578 h_{2}^{[2]} + 120 h_{3}^{[2]} \Bigr)
\nn\\   &&\hskip0.0cm\null
+ \Bigl( \frac{46013}{24} \zeta_9 - \frac{107245}{16} \zeta_2 \zeta_7
  - \frac{17259}{4} \zeta_3 \zeta_6
  - \frac{15539}{4} \zeta_4 \zeta_5 + \frac{143}{3} (\zeta_3)^3 \Bigr) h_{1}^{[1]} 
\nn\\   &&\hskip0.0cm\null
- \frac{354}{7} \zeta_{7,3} + \frac{1217}{5} \zeta_2 \zeta_{5,3}
- \frac{2668732849}{67200} \zeta_{10}
- \frac{2659}{4} \zeta_4 (\zeta_3)^2 - 3091 \zeta_2 \zeta_3 \zeta_5
\nn\\   &&\hskip0.0cm\null
+ \frac{9179}{8} \zeta_3 \zeta_7  + \frac{20553}{28} (\zeta_5)^2 \biggr\}
\,. \label{Evgt1_5}
\eea

\newpage


\section{NMHV results in the $3\to3$ self-crossing limit through four loops}
\label{ENMHVvgt11234}

In this appendix we first give the results for 
the functions $E$ entering the NMHV amplitude decomposition~(\ref{Eform}),
in the $3\to3$ self-crossing limit for $v>1$.
As predicted by the arguments in \sect{helsel}, they
are all nonsingular in this limit.
We use the same notation as in the MHV case;
the argument of the $h_i^{[w]}(z)$ is $z=1/v$.
Because $E(u,v,w)=E(w,v,u)$, there are two
independent orientations of $E$ to quote at each loop order.
We let $E$ refer to the limit $E(1+|\de|,v,v)$, and $E'$ to the limit
$E(v,1+|\de|,v)$. 

The results through three loops are
\bea
E_{3\to3}^{(0)}(v>1) &=& 1
\,, \label{ENMHVvgt1_0}\\
E_{3\to3}^{\prime\,(0)}(v>1) &=& 1
\,, \label{ENMHVprimevgt1_0}\\
E_{3\to3}^{(1)}(v>1) &=& 5 \zeta_2
\,, \label{ENMHVvgt1_1}\\
E_{3\to3}^{\prime\,(1)}(v>1) &=& - 2 h_{0}^{[2]} + 5 \zeta_2
- 2 \pi i h_{0}^{[1]}
\,, \label{ENMHVprimevgt1_1}\\
E_{3\to3}^{(2)}(v>1) &=& \frac{1}{2} \Bigl[
4 h_{8}^{[4]} + h_{10}^{[4]} - 9 \zeta_2 h_{2}^{[2]}
+ 2 \zeta_3 h_{1}^{[1]} - \frac{29}{4} \zeta_4
+ 2 \pi i ( 2 h_{4}^{[3]} - 3 \zeta_3 ) \Bigr]
\,, \label{ENMHVvgt1_2}\\
E_{3\to3}^{\prime\,(2)}(v>1) &=& \frac{1}{2} \Bigl[
24 h_{0}^{[4]} + 4 h_{2}^{[4]} + 4 h_{4}^{[4]}
+ 2 h_{6}^{[4]} + 4 h_{8}^{[4]} + h_{10}^{[4]}
\nn\\ &&\hskip0.7cm\null
- \zeta_2 ( 56 h_{0}^{[2]} - 2 h_{1}^{[2]} + 9 h_{2}^{[2]} )
+ 2 \zeta_3 ( h_{0}^{[1]} + h_{1}^{[1]}) - \frac{39}{4} \zeta_4
\nn\\ &&\hskip0.7cm\null
+ 2 \pi i ( 12 h_{0}^{[3]} + 2 h_{2}^{[3]} + 2 h_{4}^{[3]}
       - 4 \zeta_2 h_{0}^{[1]} + \zeta_3 ) \Bigr]
\,, \label{ENMHVprimevgt1_2}\\
E_{3\to3}^{(3)}(v>1) &=&
- 24 h_{32}^{[6]} - 5 ( h_{34}^{[6]} + h_{36}^{[6]} + h_{40}^{[6]} ) - h_{42}^{[6]}
+ \frac{1}{2} \zeta_2 ( 110 h_{8}^{[4]} + 23 h_{10}^{[4]} )
\nn\\ &&\hskip0.0cm\null
- \zeta_3 ( 5 h_{4}^{[3]} +  2 h_{5}^{[3]} )
+ 15 \zeta_4 h_{2}^{[2]} - ( 8 \zeta_5 - 23 \zeta_2 \zeta_3 ) h_{1}^{[1]}
+ \frac{8729}{96} \zeta_6 - \frac{3}{2} (\zeta_3)^2
\nn\\ &&\hskip0.0cm\null
- \pi i \Bigl[ 24 h_{16}^{[5]} + 5 ( h_{18}^{[5]} + h_{20}^{[5]} )
- 7 \zeta_2 h_{4}^{[3]} + 5 \zeta_3 h_{2}^{[2]}
- 25 \zeta_5 + 8 \zeta_2 \zeta_3 \Bigr]
\,, \label{ENMHVvgt1_3}\\
E_{3\to3}^{\prime\,(3)}(v>1) &=&
- 120 h_{0}^{[6]} - 24 ( h_{2}^{[6]} + h_{4}^{[6]} ) - 2 h_{6}^{[6]}
- 24 h_{8}^{[6]} - 7 h_{10}^{[6]} - 2 h_{12}^{[6]} - h_{14}^{[6]}
- 24 h_{16}^{[6]} - 7 h_{18}^{[6]}
\nn\\ &&\hskip0.0cm\null
- 7 h_{20}^{[6]} - h_{22}^{[6]}
- 2 h_{24}^{[6]} - h_{26}^{[6]} - h_{28}^{[6]} - 3 h_{30}^{[6]}
- 24 h_{32}^{[6]} - 5 h_{34}^{[6]} - 5 h_{36}^{[6]} - 5 h_{40}^{[6]} - h_{42}^{[6]}
\nn\\ &&\hskip0.0cm\null
+ \zeta_2 \Bigl( 276 h_{0}^{[4]} - 2 h_{1}^{[4]}
    + 53 h_{2}^{[4]} - h_{3}^{[4]} + 53 h_{4}^{[4]} - h_{5}^{[4]}
    + 4 h_{6}^{[4]} - 3 h_{7}^{[4]} + 55 h_{8}^{[4]}
\nn\\ &&\hskip0.7cm\null
 + \frac{23}{2} h_{10}^{[4]} \Bigr)
- \zeta_3 ( 22 h_{0}^{[3]} + 11 h_{1}^{[3]} + 6 h_{2}^{[3]} - 2 h_{3}^{[3]}
        + 5 h_{4}^{[3]} + 2 h_{5}^{[3]} )
\nn\\ &&\hskip0.0cm\null
+ \frac{1}{4} \zeta_4 ( 279 h_{0}^{[2]} + 31 h_{1}^{[2]} + 60 h_{2}^{[2]} )
- 2 \zeta_5 ( 7 h_{0}^{[1]} + 4 h_{1}^{[1]} )
+ \zeta_2 \zeta_3 ( 49 h_{0}^{[1]} + 23 h_{1}^{[1]} )
\nn\\ &&\hskip0.0cm\null
+ \frac{8477}{96} \zeta_6 - 3 (\zeta_3)^2
- \pi i \Bigl[ 120 h_{0}^{[5]} + 24 ( h_{2}^{[5]} + h_{4}^{[5]} ) + 2 h_{6}^{[5]}
  + 24 h_{8}^{[5]} + 7 h_{10}^{[5]} + 2 h_{12}^{[5]}
\nn\\ &&\hskip0.7cm\null
  + h_{14}^{[5]}
   + 24 h_{16}^{[5]} + 5 h_{18}^{[5]} + 5 h_{20}^{[5]}
   - \zeta_2 ( 36 h_{0}^{[3]} - 2 h_{1}^{[3]} + 5 h_{2}^{[3]}
   - h_{3}^{[3]} + 7 h_{4}^{[3]} )
\nn\\ &&\hskip0.7cm\null
   + \zeta_3 ( 22 h_{0}^{[2]} + 11 h_{1}^{[2]} + 5 h_{2}^{[2]} )
   - \frac{39}{4} \zeta_4 h_{0}^{[1]} + 14 \zeta_5 - 5 \zeta_2 \zeta_3 \Bigr]
\,. \label{ENMHVprimevgt1_3}
\eea
The four-loop results are
\bea
E_{3\to3}^{(4)}(v>1) &=& \frac{1}{8} \Bigl\{
2880 h_{128}^{[8]} + 624 ( h_{130}^{[8]} + h_{132}^{[8]} )
- 48 h_{134}^{[8]} + 624 h_{136}^{[8]} + 152 h_{138}^{[8]} + 28 h_{142}^{[8]}
\nn\\ &&\hskip0.4cm\null
+ 624 h_{144}^{[8]} + 152 ( h_{146}^{[8]} + h_{148}^{[8]} )
- 4 h_{150}^{[8]} + 8 h_{154}^{[8]} + 16 h_{156}^{[8]} + 6 h_{158}^{[8]}
\nn\\ &&\hskip0.4cm\null
+ 624 h_{160}^{[8]} + 152 h_{162}^{[8]}
+ 152 h_{164}^{[8]} - 4 h_{166}^{[8]} + 152 h_{168}^{[8]}
+ 42 h_{170}^{[8]} + 8 h_{172}^{[8]}
\nn\\ &&\hskip0.4cm\null
 + 3 h_{174}^{[8]} + 8 h_{178}^{[8]} + 8 h_{180}^{[8]} + 16 h_{184}^{[8]} - 18 h_{190}^{[8]}
\nn\\ &&\hskip0.4cm\null
- \zeta_2 \Bigl(
  6576 h_{32}^{[6]} + 48 h_{33}^{[6]} + 1408 h_{34}^{[6]} - 28 h_{35}^{[6]}
       + 1408 h_{36}^{[6]} + 4 h_{37}^{[6]} - 8 h_{38}^{[6]} - 6 h_{39}^{[6]}
\nn\\ &&\hskip1.1cm\null
       + 1408 h_{40}^{[6]} + 4 h_{41}^{[6]} + 338 h_{42}^{[6]} - 3 h_{43}^{[6]}
       - 8 h_{44}^{[6]} + 40 h_{46}^{[6]} + 18 h_{47}^{[6]} \Bigr)
\nn\\ &&\hskip0.4cm\null
+ \zeta_3 \Bigl( 672 h_{16}^{[5]} + 276 h_{17}^{[5]} + 156 h_{18}^{[5]} - 6 h_{19}^{[5]}
       + 156 h_{20}^{[5]} + 73 h_{21}^{[5]} + 8 h_{22}^{[5]} + 18 h_{23}^{[5]} \Bigr)
\nn\\ &&\hskip0.4cm\null
- \zeta_4 \Bigl( 1986 h_{8}^{[4]} - 21 h_{9}^{[4]}
+ \frac{1053}{2} h_{10}^{[4]} + 102 h_{11}^{[4]} \Bigr)
+ \zeta_5 \Bigl( 496 h_{4}^{[3]} + 225 h_{5}^{[3]} \Bigr)
\nn\\ &&\hskip0.4cm\null
- \zeta_2 \zeta_3 \Bigl( 1524 h_{4}^{[3]} + 627 h_{5}^{[3]} \Bigr)
- \Bigl( \frac{16151}{8} \zeta_6 - 86 (\zeta_3)^2 \Bigr) h_{2}^{[2]}
\nn\\ &&\hskip0.4cm\null
+ \Bigl( \frac{12135}{16} \zeta_7 - \frac{4483}{2} \zeta_2 \zeta_5
      - 957 \zeta_3 \zeta_4 \Bigr) h_{1}^{[1]}
\nn\\ &&\hskip0.4cm\null
- \frac{9}{5} \zeta_{5,3} - \frac{12574711}{1440} \zeta_8
+ 374 \zeta_3 \zeta_5 - 566 \zeta_2 (\zeta_3)^2
\nn\\ &&\hskip0.4cm\null
+ 2 \pi i \Bigl[
  1440 h_{64}^{[7]} + 312 ( h_{66}^{[7]} + h_{68}^{[7]} + h_{72}^{[7]} )
  + 76 h_{74}^{[7]} + 8 h_{78}^{[7]} + 312 h_{80}^{[7]}
\nn\\ &&\hskip1.5cm\null
  + 76 ( h_{82}^{[7]} + h_{84}^{[7]} )
   + 4 ( h_{86}^{[7]} + h_{90}^{[7]} ) + 8 h_{92}^{[7]}
\nn\\ &&\hskip1.5cm\null
   - \zeta_2 ( 408 h_{16}^{[5]} + 80 h_{18}^{[5]}
   - 8 h_{19}^{[5]} + 80 h_{20}^{[5]}
           - 4 h_{21}^{[5]} - 4 h_{22}^{[5]} )
\nn\\ &&\hskip1.5cm\null
   + \zeta_3 ( 336 h_{8}^{[4]} + 144 h_{9}^{[4]} + 78 h_{10}^{[4]} )
   - 153 \zeta_4 h_{4}^{[3]} + ( 248 \zeta_5 - 90 \zeta_2 \zeta_3 ) h_{2}^{[2]}
\nn\\ &&\hskip1.5cm\null
   + 12 (\zeta_3)^2 h_{1}^{[1]}
   - \frac{4041}{4} \zeta_7 + 330 \zeta_2 \zeta_5
   + \frac{275}{2} \zeta_3 \zeta_4 \Bigr] \Bigr\}
\,, \label{ENMHVvgt1_4}
\eea
and
\bea
E_{3\to3}^{\prime\,(4)}(v>1) &=& \frac{1}{8} \Bigl\{
13440 h_{0}^{[8]} + 2880 ( h_{2}^{[8]} + h_{4}^{[8]} ) - 160 h_{6}^{[8]}
+ 2880 h_{8}^{[8]} + 744 h_{10}^{[8]} + 56 h_{12}^{[8]}
\nn\\ &&\hskip0.4cm\null
+ 140 h_{14}^{[8]}
+ 2880 h_{16}^{[8]} + 744 ( h_{18}^{[8]} + h_{20}^{[8]} ) - 12 h_{22}^{[8]}
+ 80 h_{24}^{[8]} + 60 h_{26}^{[8]} + 80 h_{28}^{[8]}
\nn\\ &&\hskip0.4cm\null
+ 60 h_{30}^{[8]}
+ 2880 h_{32}^{[8]} + 736 ( h_{34}^{[8]} + h_{36}^{[8]} ) - 16 h_{38}^{[8]}
+ 736 h_{40}^{[8]} + 196 h_{42}^{[8]} + 32 h_{44}^{[8]}
\nn\\ &&\hskip0.4cm\null
+ 56 h_{46}^{[8]}
+ 96 h_{48}^{[8]} + 56 ( h_{50}^{[8]} + h_{52}^{[8]} ) + 32 h_{54}^{[8]}
+ 64 h_{56}^{[8]} + 32 ( h_{58}^{[8]} + h_{60}^{[8]} )
\nn\\ &&\hskip0.4cm\null
- 12 h_{62}^{[8]}
+ 2880 h_{64}^{[8]} + 720 ( h_{66}^{[8]} + h_{68}^{[8]} ) - 16 h_{70}^{[8]}
+ 720 h_{72}^{[8]} + 192 h_{74}^{[8]} + 32 h_{76}^{[8]}
\nn\\ &&\hskip0.4cm\null
+ 60 h_{78}^{[8]}
+ 720 h_{80}^{[8]} + 192 ( h_{82}^{[8]} + h_{84}^{[8]} ) + 28 h_{86}^{[8]}
+ 32 h_{88}^{[8]} + 40 h_{90}^{[8]} + 48 h_{92}^{[8]}
\nn\\ &&\hskip0.4cm\null
+ 42 h_{94}^{[8]}
+ 96 h_{96}^{[8]} + 48 ( h_{98}^{[8]} + h_{100}^{[8]} ) + 32 h_{102}^{[8]}
+ 48 h_{104}^{[8]} + 36 h_{106}^{[8]}
\nn\\ &&\hskip0.4cm\null
+ 32 ( h_{108}^{[8]} + h_{110}^{[8]} )
+ 48 h_{112}^{[8]} + 32 ( h_{114}^{[8]} + h_{116}^{[8]} + h_{118}^{[8]}
+ h_{120}^{[8]} + h_{122}^{[8]} )
\nn\\ &&\hskip0.4cm\null
+ 24 h_{124}^{[8]} + 120 h_{126}^{[8]}
+ 2880 h_{128}^{[8]} + 624 ( h_{130}^{[8]} + h_{132}^{[8]} ) - 48 h_{134}^{[8]}
+ 624 h_{136}^{[8]}
\nn
\eea
\bea
&&\null
 + 152 h_{138}^{[8]} + 28 h_{142}^{[8]}
+ 624 h_{144}^{[8]} + 152 ( h_{146}^{[8]} + h_{148}^{[8]} ) - 4 h_{150}^{[8]}
+ 8 h_{154}^{[8]}
\nn\\ &&\hskip0.0cm\null
+ 16 h_{156}^{[8]} + 6 h_{158}^{[8]}
+ 624 h_{160}^{[8]} + 152 ( h_{162}^{[8]} + h_{164}^{[8]} ) - 4 h_{166}^{[8]}
+ 152 h_{168}^{[8]}
\nn\\ &&\hskip0.0cm\null
+ 42 h_{170}^{[8]} + 8 h_{172}^{[8]} + 3 h_{174}^{[8]}
+ 8 h_{178}^{[8]} + 8 h_{180}^{[8]} + 16 h_{184}^{[8]}
- 18 h_{190}^{[8]} 
\nn\\ &&\hskip0.0cm\null
+ \zeta_2 \Bigl(
   - 30720 h_{0}^{[6]} - 160 h_{1}^{[6]} - 6456 h_{2}^{[6]} + 140 h_{3}^{[6]}
   - 6456 h_{4}^{[6]} - 12 h_{5}^{[6]} - 140 h_{6}^{[6]} + 60 h_{7}^{[6]}
\nn\\ &&\hskip0.8cm\null
   - 6464 h_{8}^{[6]} - 16 h_{9}^{[6]} - 1644 h_{10}^{[6]} + 56 h_{11}^{[6]}
   - 184 h_{12}^{[6]} + 32 h_{13}^{[6]} - 128 h_{14}^{[6]} - 12 h_{15}^{[6]}
\nn\\ &&\hskip0.8cm\null
   - 6480 h_{16}^{[6]} - 16 h_{17}^{[6]} - 1608 h_{18}^{[6]} + 60 h_{19}^{[6]}
   - 1608 h_{20}^{[6]} + 28 h_{21}^{[6]} - 40 h_{22}^{[6]} + 42 h_{23}^{[6]}
\nn\\ &&\hskip0.8cm\null
   - 192 h_{24}^{[6]} + 32 h_{25}^{[6]} - 84 h_{26}^{[6]} + 32 h_{27}^{[6]}
   - 88 h_{28}^{[6]} + 32 h_{29}^{[6]} - 48 h_{30}^{[6]} + 120 h_{31}^{[6]}
\nn\\ &&\hskip0.8cm\null
   - 6576 h_{32}^{[6]} - 48 h_{33}^{[6]} - 1408 h_{34}^{[6]} + 28 h_{35}^{[6]}
   - 1408 h_{36}^{[6]} - 4 h_{37}^{[6]} + 8 h_{38}^{[6]} + 6 h_{39}^{[6]}
\nn\\ &&\hskip0.8cm\null
   - 1408 h_{40}^{[6]} - 4 h_{41}^{[6]} - 338 h_{42}^{[6]} + 3 h_{43}^{[6]}
   + 8 h_{44}^{[6]} - 40 h_{46}^{[6]} - 18 h_{47}^{[6]}  \Bigr)
\nn\\ &&\hskip0.0cm\null
+ \zeta_3 \Bigl( 3040 h_{0}^{[5]} + 1292 h_{1}^{[5]} + 756 h_{2}^{[5]} - 20 h_{3}^{[5]}
   + 752 h_{4}^{[5]} + 304 h_{5}^{[5]} + 24 h_{6}^{[5]} + 44 h_{7}^{[5]}
\nn\\ &&\hskip0.8cm\null
   + 736 h_{8}^{[5]} + 292 h_{9}^{[5]} + 164 h_{10}^{[5]} - 10 h_{11}^{[5]}
   + 16 h_{12}^{[5]} + 8 h_{13}^{[5]} - 80 h_{15}^{[5]}
   + 672 h_{16}^{[5]} + 276 h_{17}^{[5]}
\nn\\ &&\hskip0.8cm\null
   + 156 h_{18}^{[5]} - 6 h_{19}^{[5]}
   + 156 h_{20}^{[5]} + 73 h_{21}^{[5]} + 8 h_{22}^{[5]} + 18 h_{23}^{[5]} \Bigr)
\nn\\ &&\hskip0.0cm\null
- \zeta_4 \Bigl( 9060 h_{0}^{[4]} + 280 h_{1}^{[4]} + 2545 h_{2}^{[4]} + 356 h_{3}^{[4]}
   + 2460 h_{4}^{[4]} + 119 h_{5}^{[4]} + 185 h_{6}^{[4]} + 58 h_{7}^{[4]}
\nn\\ &&\hskip0.8cm\null
+ 1986 h_{8}^{[4]} - 21 h_{9}^{[4]}
+ \frac{1053}{2} h_{10}^{[4]} + 102 h_{11}^{[4]} \Bigr)
\nn\\ &&\hskip0.0cm\null
+ \zeta_5 \Bigl( 2208 h_{0}^{[3]} + 1068 h_{1}^{[3]} + 548 h_{2}^{[3]} - 64 h_{3}^{[3]}
   + 496 h_{4}^{[3]} + 225 h_{5}^{[3]} \Bigr)
\nn\\ &&\hskip0.0cm\null
- \zeta_2 \zeta_3 \Bigl( 6880 h_{0}^{[3]} + 2920 h_{1}^{[3]}
   + 1676 h_{2}^{[3]} + 40 h_{3}^{[3]}
   + 1524 h_{4}^{[3]} + 627 h_{5}^{[3]} \Bigr)
\nn\\ &&\hskip0.0cm\null
- \zeta_6 \Bigl( \frac{28465}{3} h_{0}^{[2]}
  + \frac{1023}{4} h_{1}^{[2]} + \frac{16151}{8} h_{2}^{[2]} \Bigr)
+ (\zeta_3)^2 \Bigl( 392 h_{0}^{[2]} + 39 h_{1}^{[2]} + 86 h_{2}^{[2]} \Bigr)
\nn\\ &&\hskip0.0cm\null
+ \frac{1}{16} \zeta_7 \Bigl( 25120 h_{0}^{[1]} + 12135 h_{1}^{[1]} \Bigr)
- \frac{1}{2} \zeta_2 \zeta_5 \Bigl( 9968 h_{0}^{[1]} + 4483 h_{1}^{[1]} \Bigr)
- \zeta_3 \zeta_4 \Bigl( 2435 h_{0}^{[1]} + 957 h_{1}^{[1]} \Bigr) 
\nn\\ &&\hskip0.0cm\null
- \frac{2461055}{288} \zeta_8 + 578 \zeta_3 \zeta_5 - 886 \zeta_2 (\zeta_3)^2
\nn\\ &&\hskip0.0cm\null
+ 2 \pi i \Bigr[ 6720 h_{0}^{[7]} + 1440 h_{2}^{[7]} + 1440 h_{4}^{[7]} + 28 h_{6}^{[7]}
   + 1440 h_{8}^{[7]} + 372 h_{10}^{[7]} + 40 h_{12}^{[7]} + 40 h_{14}^{[7]}
\nn\\ &&\hskip0.8cm\null
   + 1440 h_{16}^{[7]} + 368 h_{18}^{[7]} + 368 h_{20}^{[7]} + 16 h_{22}^{[7]}
   + 48 h_{24}^{[7]} + 28 h_{26}^{[7]} + 32 h_{28}^{[7]} + 16 h_{30}^{[7]}
\nn\\ &&\hskip0.8cm\null
   + 1440 h_{32}^{[7]} + 360 h_{34}^{[7]} + 360 h_{36}^{[7]} + 16 h_{38}^{[7]}
   + 360 h_{40}^{[7]} + 96 h_{42}^{[7]} + 16 h_{44}^{[7]} + 24 h_{46}^{[7]}
\nn\\ &&\hskip0.8cm\null
   + 48 h_{48}^{[7]} + 24 h_{50}^{[7]} + 24 h_{52}^{[7]} + 16 h_{54}^{[7]}
   + 24 h_{56}^{[7]} + 16 h_{58}^{[7]} + 16 h_{60}^{[7]} + 12 h_{62}^{[7]}
\nn\\ &&\hskip0.8cm\null
   + 1440 h_{64}^{[7]} + 312 h_{66}^{[7]} + 312 h_{68}^{[7]} + 312 h_{72}^{[7]}
   + 76 h_{74}^{[7]} + 8 h_{78}^{[7]} + 312 h_{80}^{[7]} + 76 h_{82}^{[7]}
\nn\\ &&\hskip0.8cm\null
   + 76 h_{84}^{[7]} + 4 h_{86}^{[7]} + 4 h_{90}^{[7]} + 8 h_{92}^{[7]}
\nn
\eea
\bea
&&\null
+ \zeta_2 \Bigl(
   - 1920 h_{0}^{[5]} + 28 h_{1}^{[5]} - 348 h_{2}^{[5]} + 40 h_{3}^{[5]}
            - 352 h_{4}^{[5]} + 16 h_{5}^{[5]} + 4 h_{6}^{[5]} + 16 h_{7}^{[5]}
\nn\\ &&\hskip0.9cm\null
	    - 360 h_{8}^{[5]} + 16 h_{9}^{[5]} - 84 h_{10}^{[5]} + 24 h_{11}^{[5]}
	                   + 16 h_{13}^{[5]} + 4 h_{14}^{[5]} + 12 h_{15}^{[5]}
	                   - 408 h_{16}^{[5]} - 80 h_{18}^{[5]}
\nn\\ &&\hskip0.9cm\null
                           + 8 h_{19}^{[5]}
	    - 80 h_{20}^{[5]} + 4 h_{21}^{[5]} + 4 h_{22}^{[5]} \Bigr)
\nn\\ &&\hskip0.0cm\null
   + \zeta_3 \Bigl( 1520 h_{0}^{[4]} + 664 h_{1}^{[4]} + 376 h_{2}^{[4]} + 8 h_{3}^{[4]}
           + 368 h_{4}^{[4]} + 152 h_{5}^{[4]} + 8 h_{6}^{[4]} + 4 h_{7}^{[4]}
	   + 336 h_{8}^{[4]}
 \nn\\ &&\hskip0.9cm\null
          + 144 h_{9}^{[4]} + 78 h_{10}^{[4]} \Bigr)
\nn\\ &&\hskip0.0cm\null
- \zeta_4 \Bigl( 690 h_{0}^{[3]} - 13 h_{1}^{[3]}
+ 150 h_{2}^{[3]} - 2 h_{3}^{[3]} + 153 h_{4}^{[3]} \Bigr)
   + \zeta_5 \Bigl( 1104 h_{0}^{[2]} + 516 h_{1}^{[2]} + 248 h_{2}^{[2]} \Bigr)
\nn\\ &&\hskip0.0cm\null
   - \zeta_2 \zeta_3 \Bigl( 400 h_{0}^{[2]} + 172 h_{1}^{[2]} + 90 h_{2}^{[2]} \Bigr)
   - \frac{2635}{6} \zeta_6 h_{0}^{[1]}
   + (\zeta_3)^2 \Bigl( 196 h_{0}^{[1]} + 12 h_{1}^{[1]} \Bigr) 
\nn\\ &&\hskip0.0cm\null
   + 785 \zeta_7 - 284 \zeta_2 \zeta_5 - \frac{355}{2} \zeta_3 \zeta_4
   \Bigr] \Bigr\}
\,. \label{ENMHVprimevgt1_4}
\eea

The results for $E' = E(v,1+|\de|,v)$ are more complicated than
those for $E=E(1+|\de|,v,v)$, because only the latter orientation obeys
a final entry condition that is compatible with the self-crossing
line.  Both sets of functions are smooth as $v\to1^+$.  While 
$E(1+|\de|,v,v)$ is also smooth as $v\to\infty$, $E(v,1+|\de|,v)$
has logarithmic divergences there.  These are precisely
the logarithms associated with the NMHV multi-particle factorization pole
discussed in refs.~\cite{Dixon2014iba,Dixon2015iva}, where the
function $U(u,v,w) = \ln E(u,v,w)$ was studied in the limit $u,w\to\infty$
with $v$ fixed.

In the remainder of this appendix we discuss the behavior of
the full NMHV amplitude in the overlap region between the $3\to3$
multi-Regge limit
and the self-crossing limit, $w\to-1$, where we can work to higher order
in the expansion around the self-crossing limit.  This is necessary
because the rational prefactors, or $R$-invariants, ``$(i)$'' in \eqn{Eform},
blow up in the self-crossing limit.  The denominators of some of the
$R$-invariants contain spinor strings such as
$\langle3^-|(1+2)|6^-\rangle = \langle3|(1+2)|6]$.  The square of this
factor is
\bea
\langle3^-|(1+2)|6^-\rangle \langle6^-|(1+2)|3^-\rangle
&=& \frac{1}{2} {\rm tr}[(1-\gamma_5)3(1+2)6(1+2)]\nn\\
&=& s_{123} s_{345} - s_{12} s_{45}
= (1-u) s_{123} s_{345}\nn\\
&=& \de s_{123} s_{345} \,.
\label{spinorpresing}
\eea
This behavior leads some of the $R$-invariants to blow up like
$1/\langle3|(1+2)|6] \propto 1/\sqrt{\de}$.

On the other hand, in the Euclidean region this ``spurious pole'' power-law
singularity is completely cancelled by a relation between the transcendental
functions $E$ and $\Et$~\cite{Dixon2011nj,Dixon2014iba,Dixon2015iva}.
Here we will see that in Minkowski kinematics a logarithmic singularity
survives.  Actually we will not do so for generic $3\to3$
self-crossing kinematics,
but only for $v\to0^-$, making use of the overlap with the $w\to-1$ multi-Regge
limit.  For the generic multi-Regge limit, for the helicity configuration
\be
3^+ 6^+ \, \to \, 2^+ 4^- 5^+ 1^+ \,,
\label{NMHVconfig}
\ee
an analysis of the behavior of the $R$-invariants~\cite{Dixon2014iba}
leads to the following formula for the BDS-like normalized amplitude:
\bea
\rho \equiv \frac{\mathcal{A}_6^{\rm NMHV}}{\mathcal{A}_6^{\rm BDS-like}} &=&
 \frac{1}{2(1+w^*)} \Bigl[ E(u,v,w) + \Et(u,v,w) + E(w,u,v) - \Et(w,u,v) \Bigr]
\nn\\
&&\null + \frac{w^*}{2(1+w^*)}
 \Bigl[ E(v,w,u) + \Et(v,w,u) + E(w,u,v) - \Et(w,u,v) \Bigr] \,.~~~
\label{NMHVBDSlikeMRK}
\eea
The $2\to4$ MRK behavior of the ratio function was provided through four loops
in refs.~\cite{Dixon2014iba,Dixon2015iva} in terms of
SVHPLs~\cite{BrownSVHPLs}.  Converting to the BDS-like normalized
functions $E$ and $\Et$, analytically continuing to $3\to3$ kinematics
with $v<0$, and taking the limit $w\to-1$, we find through four loops:
\bea
\rho^{(1)} &=& - \frac{1}{2} \lnvne{2}\, +\, 2 \zeta_2
+ \pi i \biggl[ \frac{1+w}{1+w^*} + 1 \biggr] \,,
\label{rho1}\\
\rho^{(2)} &=& \frac{1}{8} \lnvne{4}\, -\, \frac{1}{2} \zeta_2 \lnvne{2}
\nn\\ &&\null
+ \pi i \biggl[ \frac{1+w}{1+w^*}
   \biggl( - \frac{1}{2} \lndene{2}\, +\, \lnden\, +\, \zeta_2\, -\, 1 \biggr) 
  - \frac{1}{2} \lnvne{2}\, -\, \lnvn\,
  -\, \zeta_3\, +\, \zeta_2\, - 1 \biggr] \,,~~~~
\label{rho2}\\
\rho^{(3)} &=& 
- \frac{1}{48} \lnvne{6}\, -\, \frac{1}{4} \zeta_4 \lnvne{2}\, 
+\, \frac{91}{12} \zeta_6
\nn\\ &&\null
+ \pi i \biggl[ \frac{1+w}{1+w^*} \biggl(
  \frac{1}{8} \lndene{4}\, -\, \frac{1}{2} \lndene{3}\,
 +\, \frac{3}{2} \lndene{2}\, -\, (\zeta_3+3) \lnden\,
 +\, \frac{1}{2} \zeta_4\, +\, \zeta_3\, +\, 3 \biggr) 
\nn\\ &&\null\hskip1cm
  + \frac{1}{8} \lnvne{4}\, +\, \frac{1}{2} \lnvne{3}\,
  -\, \frac{1}{2} (\zeta_3-3) \lnvne{2}\,
  +\, (\zeta_3+3) \lnvn
\nn\\ &&\null\hskip1cm
  +\, 7 \zeta_5\, -\, 3 \zeta_2 \zeta_3\,
  +\, \frac{1}{2} \zeta_4\, +\, \zeta_3\, + 3 \biggr] \,,
\label{rho3}
\eea
\bea
\rho^{(4)} &=&
\frac{1}{384} \lnvne{8}\, +\, \frac{1}{48} \zeta_2 \lnvne{6}\,
 +\, \frac{7}{16} \zeta_4 \lnvne{4}\,
 +\, \Bigl( \frac{13}{48} \zeta_6 + \frac{1}{2} (\zeta_3)^2 \Bigr) \lnvne{2}\,
 -\, \frac{1325}{36} \zeta_8\, -\, 2 \zeta_2 (\zeta_3)^2
\nn\\ &&\null
+ \pi i \biggl[ \frac{1+w}{1+w^*} \biggl(
 - \frac{1}{48} \lndene{6}\, +\, \frac{1}{8} \lndene{5}\,
 -\, \frac{1}{8} ( \zeta_2 + 5 ) \lndene{4}\,
 +\, \Bigl( - \frac{1}{6} \zeta_3 + \frac{1}{2} \zeta_2 + \frac{5}{2} \Bigr)
      \lndene{3}\,
\nn\\ &&\null\hskip1.3cm
 - \Bigl( \frac{7}{4} \zeta_4 - \frac{1}{2} \zeta_3
        + \frac{3}{2} \zeta_2 + \frac{15}{2} \Bigr) \lndene{2}\,
 +\, \Bigl( 4 \zeta_5 - 3 \zeta_2 \zeta_3 + \frac{7}{2} \zeta_4
           - \zeta_3 + 3 \zeta_2 + 15 \Bigr) \lnden\,
\nn\\ &&\null\hskip1.3cm
 -\, \frac{13}{24} \zeta_6\, -\, 2 (\zeta_3)^2\,
 -\, 4 \zeta_5\, +\, 3 \zeta_2 \zeta_3\,
 -\, \frac{7}{2} \zeta_4\, +\, \zeta_3\, -\, 3 \zeta_2\, - 15 \biggr)
\nn\\ &&\null\hskip1cm
 - \frac{1}{48} \lnvne{6}\, -\, \frac{1}{8} \lnvne{5}\,
 +\, \Bigl( \frac{3}{8} \zeta_3 - \frac{1}{8} \zeta_2 - \frac{5}{8} \Bigr)
    \lnvne{4}\,
 +\, \Bigl( \frac{1}{12} \zeta_3 - \frac{1}{2} \zeta_2 - \frac{5}{2} \Bigr)
    \lnvne{3}\,
\nn\\ &&\null\hskip1cm
 +\, \Bigl( \frac{25}{8} \zeta_5 + 2 \zeta_2 \zeta_3 - \frac{7}{4} \zeta_4
       + \frac{1}{4} \zeta_3 - \frac{3}{2} \zeta_2 - \frac{15}{2} \Bigr)
   \lnvne{2}\,
\nn\\ &&\null\hskip1cm
 + \Bigl( - \frac{3}{2} (\zeta_3)^2 - 5 \zeta_5 + \zeta_2 \zeta_3 
       - \frac{7}{2} \zeta_4 + \frac{1}{2} \zeta_3 - 3 \zeta_2 - 15 \Bigl)
          \lnvn\,
\nn
\eea
\bea
&&\null\hskip1cm
 -\, \frac{1381}{32} \zeta_7\, +\, \frac{43}{2} \zeta_2 \zeta_5\,
 +\, 4 \zeta_3 \zeta_4 \, -\, \frac{13}{24} \zeta_6\,
 -\, \frac{5}{2} (\zeta_3)^2\, -\, 5 \zeta_5\, +\, \zeta_2 \zeta_3\,
\nn\\ &&\null\hskip1cm
 -\, \frac{7}{2} \zeta_4\, +\, \frac{1}{2} \zeta_3\,
 -\, 3 \zeta_2\, -\, 15 \biggr] \,.
\label{rho4}
\eea

In these expressions we let $(1+w)\to \xi(1+w)$, 
$(1+w^*)\to\xi(1+w^*)$, take $\xi\to0$, and drop terms that vanish
in this limit.  The only two
rational prefactors that can survive are then $1$ and $(1+w^*)/(1+w)$.
The results do not have uniform transcendentality because
the transcendental functions have been expanded to higher order
around the $w\to-1$ limit to cancel the $1/(1+w)$ factor in 
\eqn{NMHVBDSlikeMRK}.
We rewrite the logarithmic terms using the self-crossing variables,
i.e.~we let $\ln|1+w|^2 \to \lnden\ -\ \lnvn$. 
We see that there are indeed logarithmically singular $\lnden$ terms in
the imaginary part beginning at two loops.
However, the $\lnden$ terms appear only in the part with the $(1+w)/(1+w^*)$
prefactor; that is, the terms that are independent of the azimuthal component
of the vector $\vec{z}$ in \eqn{eq:del_eps} are finite.
Furthermore, the $\lnden$ terms contain no $v$ dependence
(which could only appear through powers of $\lnvn$ in the approximation
in which we are working).
This behavior is reminiscent of what we found for the
MHV configuration.  It would be very interesting to try to understand
these NMHV properties better, both in the MRK limit and more generally
along the full self-crossing line.



\section{Seven-point self-crossing kinematics}
\label{sckin7}

\begin{figure}
\begin{center}
\includegraphics[width=2.0in]{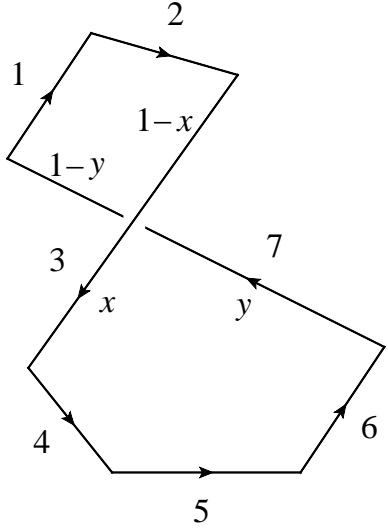}
\end{center}
\caption{The self-crossing configuration for $2\to5$ scattering,
where particles 3 and 7 are incoming, and 1, 2, 4, 5 and 6 are outgoing.}
\label{fig:sc7}
\end{figure}

In this section we briefly describe the self-crossing configuration
for the seven-point amplitude, or heptagonal Wilson loop shown in
\fig{fig:sc7}.  We consider
\bea
k_3\ +\ k_7\ &\to& k_1\ + k_2\ +\ k_4\ +\ k_5\ +\ k_6, \label{incout25}\\
(1-x) k_3\ +\ (1-y) k_7 &\to& k_1\ +\ k_2, \label{sub25A}\\
x k_3\ +\ y k_7 &\to& k_4\ +\ k_5\ +\ k_6.  \label{sub25B}
\eea
Incoming gluons 3 and 7 split into collinear pairs with
momentum fractions $x$ and $1-x$, and $y$ and $1-y$, respectively.
These pairs then undergo a $2\to2$ scattering into final state gluons
1, 2, and a $2\to3$ scattering into final state gluons 4, 5 and 6.

It is straightforward to derive the following relations
among the Mandelstam variables:
\bea
s_{12} &=& (1-x)(1-y) s_{37} \,, \\
(1-x) s_{23} &=& (1-y) s_{71} \,, \\
s_{456} &=& xy s_{37} \,, \\
s_{123} &=& -x(1-y) s_{37} \,, \\
s_{712} &=& -(1-x)y s_{37} \,, \\
x s_{34} &=& (1-y) s_{56} + y s_{567} \,, \label{eq567} \\
y s_{67} &=& (1-x) s_{45} + x s_{345} \,. \label{eq345} 
\eea
We would like to rewrite these constraints in terms of the seven dual
conformal cross ratios defined in ref.~\cite{Anastasiou2009kna},
\be
u_i \equiv u_{i,i+3} = \frac{x_{i,i+4}^2 x_{i+1,i+3}^2}
                          {x_{i,i+3}^2 x_{i+1,i+4}^2}
\ee
or
\be
u_1 = \frac{x_{15}^2 x_{24}^2}{x_{14}^2 x_{25}^2}
    = \frac{s_{23} s_{567}}{s_{123} s_{234}}
\ee
and cyclic permutations thereof (mod 7).

It is easy to see that
\be
u_7 = \frac{s_{12} s_{456}}{s_{712} s_{123}} = 1,
\label{u7eq1}
\ee
which is the analog of the $u=1$ constraint in the six-point case.

To find the second constraint on the cross ratios, analogous
to $v=w$ in the six-point case, we first look for combinations of
cross ratios from which the invariants $s_{234}$ and $s_{671}$
are absent:
\bea
\frac{u_1}{u_2 u_6}
&=& \frac{s_{23}}{s_{71}} \frac{s_{712}}{s_{123}} \frac{s_{567}}{s_{34}}
= \frac{y}{x} \frac{s_{567}}{s_{34}} \,, \label{u1overu2u6} \\
\frac{u_6}{u_1 u_5}
&=& \frac{s_{71}}{s_{23}} \frac{s_{123}}{s_{712}} \frac{s_{345}}{s_{67}}
= \frac{x}{y} \frac{s_{345}}{s_{67}} \,, \label{u6overu1u3} \\
\frac{u_3 u_4}{u_2 u_5}
&=& \frac{s_{45} s_{56} s_{712} s_{123}}{s_{34} s_{67} s_{456}^2}
= \frac{(1-y)s_{56}}{x s_{34}} \cdot \frac{(1-x) s_{45}}{y s_{67}} \,.
\label{u3u4overu2u5} 
\eea
Using \eqns{eq567}{eq345}, we can rewrite \eqn{u3u4overu2u5} as
\be
\frac{u_3 u_4}{u_2 u_5}
= \biggl[ 1 - \frac{y s_{567}}{x s_{34}} \biggr]
\biggl[ 1 - \frac{x s_{345}}{y s_{67}} \biggr]
= \biggl[ 1 - \frac{u_1}{u_2 u_6} \biggr]
\biggl[ 1 - \frac{u_6}{u_1 u_5} \biggr] \,,
\label{creqn2a}
\ee
or
\be
(u_1 u_5 - u_6) (u_2 u_6 - u_1) = u_1 u_3 u_4 u_6 \,.
\label{creqn2b}
\ee
We solve \eqn{creqn2b} for $u_4$, and insert that solution and
$u_7=1$ into the Gram determinant constraint that is obeyed by the seven
cross ratios to have four-dimensional kinematics.  We find that the Gram
determinant vanishing condition then contains a simple factor,
$u_6 - u_1 u_5 - u_3 u_6$.  Setting this factor to zero, we solve for
$u_3$ and plug the solution back into \eqn{creqn2b}.  We obtain
\be
u_7 = 1 \,, \qquad
u_3 = 1 - \frac{u_1 u_5}{u_6} \,, \qquad
u_4 = 1 - \frac{u_2 u_6}{u_1} \,.
\label{creqnfinal}
\ee
The self-crossing solution is then parametrized by the four remaining
cross ratios, $u_1$, $u_2$, $u_5$ and $u_6$. 

\newpage

\end{document}